\documentclass[floatfix,prx,twocolumn,letterpaper,lengthcheck,superscriptaddress,showpacs,amssymb,amsmath,amsfonts,aps,altaffilletter,nofootinbib,nopreprintnumbers,showpacs,longbibliography]{revtex4-1}

\usepackage{color}
\usepackage{hyperref}
\usepackage{amsmath}
\usepackage{amsthm}
\usepackage{multirow}
\usepackage{graphicx}
\usepackage{caption}
\usepackage{subcaption}
\usepackage{placeins}

\usepackage{mathrsfs}
\DeclareSymbolFontAlphabet{\mathrsfs}{rsfs}
\newcommand{\scrM}{\mathrsfs{M}}

\newcommand{\Surf}{\ensuremath{\mathcal{S}}}
\def\insubscript{\rm inner}
\def\outsubscript{\rm outer}
\newcommand{\Sout}{\Surf_{\outsubscript}}
\newcommand{\Sin}{\Surf_{\insubscript}}
\newcommand{\Sone}{\Surf_{1}}
\newcommand{\Stwo}{\Surf_{2}}

\newcommand{\HH}{\ensuremath{\mathcal{H}}}
\newcommand{\Hout}{\HH_{\outsubscript}}
\newcommand{\Hin}{\HH_{\insubscript}}
\newcommand{\Hone}{\HH_{1}}
\newcommand{\Htwo}{\HH_{2}}

\def\tname{T}
\def\ttouch{\tname_{\rm touch}}
\def\tbifurcate{\tname_{\rm bifurcate}}
\def\tmin{\tname_{\rm min}}

\def\twoq{\widetilde{q}}
\def\twoepsilon{\widetilde{\epsilon}}

\def\MM{\mathcal{M}}
\def\MADM{M_\text{ADM}}

\newcommand{\Mirr}{M_{\rm irr}}
\newcommand{\Mone}{\Mirr^{(1)}}
\newcommand{\Mtwo}{\Mirr^{(2)}}
\newcommand{\Mout}{\Mirr^{\rm \outsubscript}}

\newcommand{\Rout}{R_{\rm \outsubscript}}
\newcommand{\Rin}{R_{\rm \insubscript}}

\def\sI{\mathit{I}}
\def\sII{\mathit{II}}
\def\sIII{\mathit{III}}

\def\qI{\mathit{I}}
\def\qII{\mathit{II}}
\def\qIII{\mathit{III}}
\def\qIV{\mathit{IV}}

\newtheorem{defn}{Definition}

\begin{document}

\title[]{Horizons in a binary black hole merger I: Geometry and area increase}

\author{Daniel Pook-Kolb} 
\affiliation{Max-Planck-Institut f\"ur Gravitationsphysik (Albert
  Einstein Institute), Callinstr. 38, 30167 Hannover, Germany}
\affiliation{Leibniz Universit\"at Hannover, 30167 Hannover, Germany}

\author{Ofek Birnholtz}
\affiliation{Department of Physics, Bar-Ilan University, Ramat-Gan
  5290002, Israel}

\author{Jos\'e Luis Jaramillo}
\affiliation{Institut de Math\'ematiques de Bourgogne (IMB), UMR 5584,
  CNRS,Universit\'e de Bourgogne Franche-Comt\'e, F-21000 Dijon,
  France}

\author{Badri Krishnan} 
\affiliation{Max-Planck-Institut f\"ur Gravitationsphysik (Albert
  Einstein Institute), Callinstr. 38, 30167 Hannover, Germany}
\affiliation{Leibniz Universit\"at Hannover, 30167 Hannover, Germany}

\author{Erik Schnetter}
\affiliation{Perimeter Institute for Theoretical Physics, Waterloo, 
  ON N2L 2Y5, Canada}
\affiliation{Physics \& Astronomy Department, University of Waterloo,
  Waterloo, ON N2L 3G1, Canada}
\affiliation{Center for Computation \& Technology, Louisiana State
  University, Baton Rouge, LA 70803, USA}


\begin{abstract}

  Recent advances in numerical relativity have revealed how marginally
  trapped surfaces behave when black holes merge.  It is now known
  that interesting topological features emerge during the merger, and
  marginally trapped surfaces can have self-intersections.  This paper
  presents the most detailed study yet of the physical and geometric
  aspects of this scenario.  For the case of a head-on collision of
  non-spinning black holes, we study in detail the world tube formed
  by the evolution of marginally trapped surfaces.  In the first of
  this two-part study, we focus on geometrical properties of the
  dynamical horizons, i.e. the world tube traced out by the time
  evolution of marginally outer trapped surfaces.  We show that even
  the simple case of a head-on collision of non-spinning black holes
  contains a rich variety of geometric and topological properties and
  is generally more complex than considered previously in the
  literature.  The dynamical horizons are shown to have mixed
  signature and are not future marginally trapped everywhere.  We
  analyze the area increase of the marginal surfaces along a sequence
  which connects the two initially disjoint horizons with the final
  common horizon.  While the area does increase overall along this
  sequence, it is not monotonic.  We find short durations of anomalous
  area change which, given the connection of area with entropy, might
  have interesting physical consequences. We investigate the possible
  reasons for this effect and show that it is consistent with existing
  proofs of the area increase law.

\end{abstract}

\maketitle

\section{Introduction}
\label{sec:intro}

One of the remarkable predictions of general relativity is the
existence of black holes, purely geometric objects in a curved
spacetime which behave like compact physical objects in numerous
physical situations, and can power the most energetic phenomena in our
universe.  The properties of spacetime near black holes are unlike
anything observed in flat space, even for very large black holes where
the curvature near the horizon is not necessarily large.  One of these
unusual phenomena is the existence of \emph{closed trapped surfaces}.
These are 2-dimensional closed surfaces which have the unusual
property that even outgoing light rays emanating from the surface are
convergent. Such surfaces cannot exist completely contained in flat
spacetime regions. In fact, within classical general relativity, they
can only exist in geodesically incomplete spacetimes, usually taken to
indicate the presence of a singularity
\cite{Penrose:1964wq,Hawking:1969sw}.  \emph{Marginally outer trapped
  surfaces} (MOTS) are limiting cases of trapped surfaces where the
outgoing light rays have vanishing convergence.  The outermost MOTS on
a given constant time hypersurface, also known as an apparent horizon,
can be shown locally to have the property of a one-way membrane,
i.e. any material particle having fallen into it cannot cross it
again.  The outermost MOTS can also be shown to have non-decreasing
area and furthermore these are found to satisfy the laws of black hole
mechanics. Black hole spacetimes however contain a much wider variety
of MOTSs which have interesting physical and geometric properties.

Numerous merger events where two black holes merge to form a larger
remnant black hole have now been observed by gravitational wave
detectors
\cite{LIGOScientific:2018mvr,TheLIGOScientific:2016pea,Nitz:2018imz,Nitz:2019hdf,Venumadhav:2019lyq,Zackay:2019tzo}.
The number of detections will increase by orders of magnitude in the
next years as the detectors become more sensitive and new generations
of detectors are built.  It is common to understand such mergers using
event horizons.  A well known example is \cite{Matzner:1995ib}, the
``pair of pants picture'', which clearly shows the merger of two
disjoint surfaces to yield a final horizon. How should one think of
the merger process in terms of marginally trapped surfaces, and does
this yield a picture analogous to the ``pair of pants''?

\begin{figure*}
  \centering    
  \includegraphics[trim=46 36 33 16,clip,width=\columnwidth]{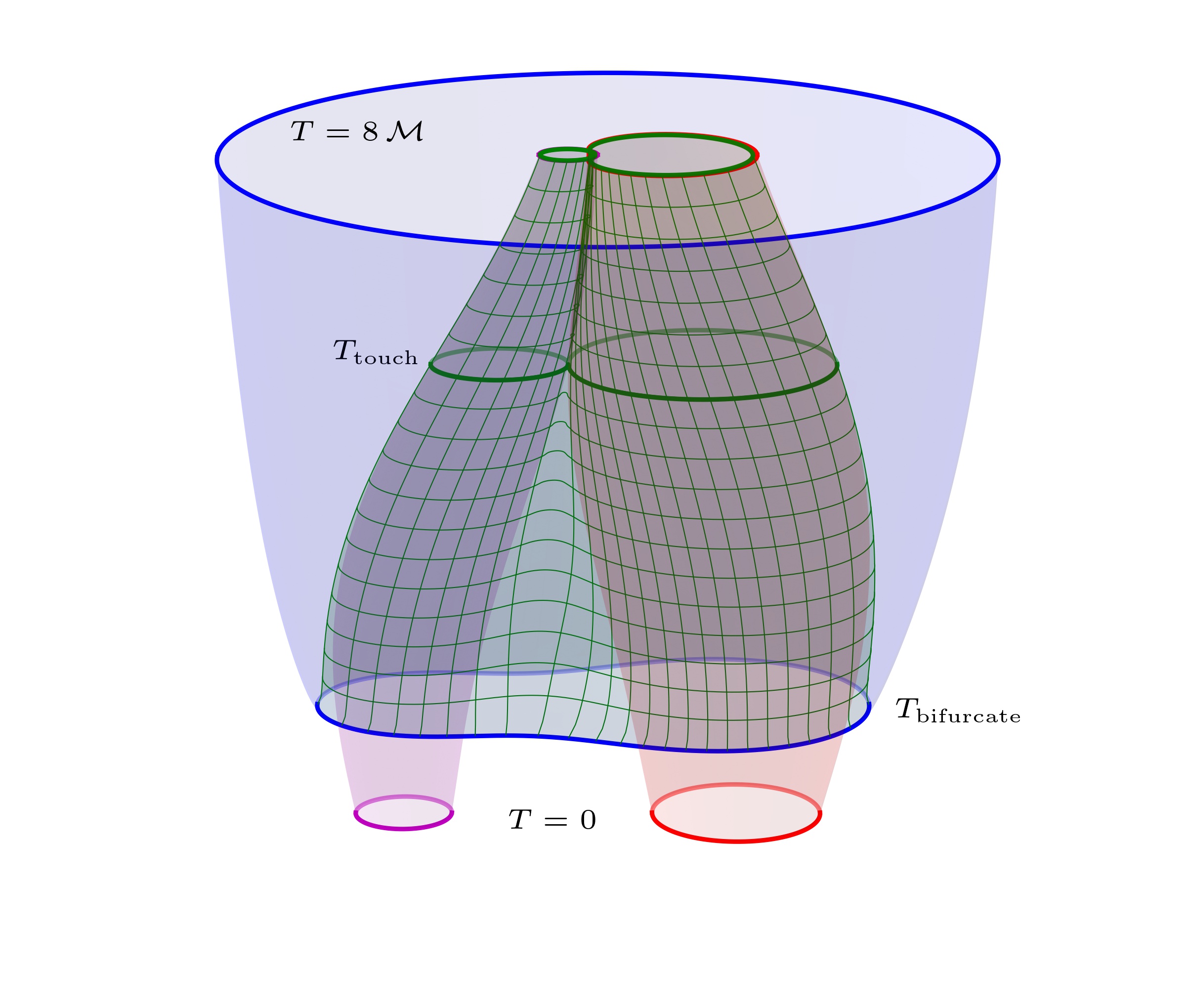}
  \includegraphics[trim=0 0 0 0,clip,width=\columnwidth]{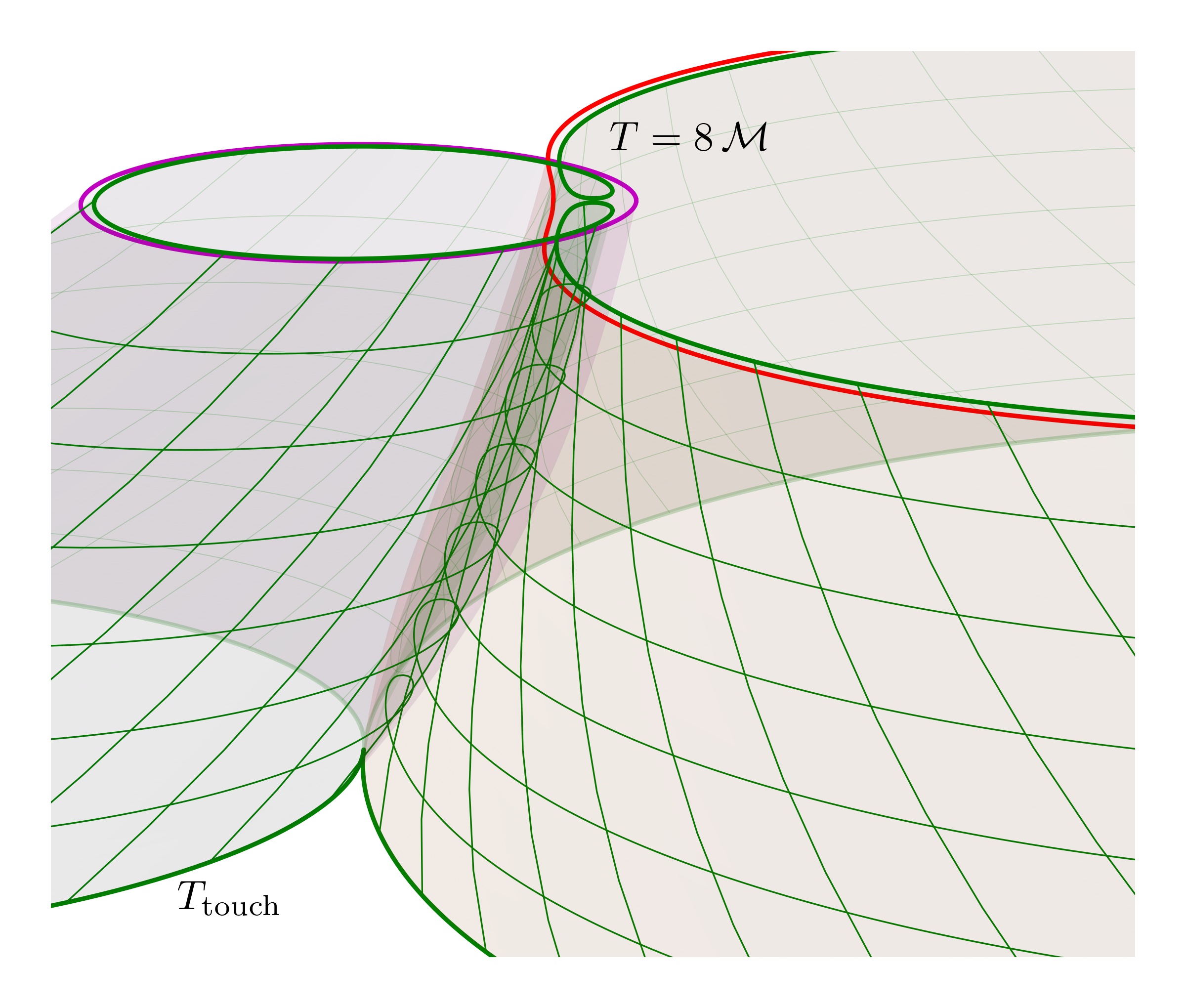}
  \caption{The behavior of MOTSs in a binary black hole merger
    obtained from a numerical simulation of the full vacuum Einstein
    equations.  The left panel shows an overview up to
    $T=8\,\MM > \ttouch$
    and the right panel a close-up of the end section of the two
    individual horizons (purple, red) and the inner common horizon (green) with
    the self-intersections visible.
    The blue contour describes the outer common horizon.
    See text for details. }
  \label{fig:3d}
\end{figure*}

The scenario of how two MOTSs merge has been recently established for
the first time numerically, and is summarized in Fig.~\ref{fig:3d}.
This figure is obtained from a numerical solution of the full vacuum
Einstein equations for the head-on collision of two unequal mass black
holes.  We start with simplest puncture initial data where the black
holes have no spin or initial linear momentum, namely Brill-Lindquist
initial data \cite{Brill:1963yv}.  The initial data is prescribed on
Euclidean space with two points (the ``punctures'') removed.  The data
is time symmetric, i.e. the extrinsic curvature vanishes.  The
3-metric $h_{ab}$ is conformally flat: $h_{ab}=\Phi^4\delta_{ab}$.
The conformal factor at a point $\mathbf{r}$ is
\begin{equation}
  \Phi(\mathbf{r}) = 1 + \frac{m_1}{2r_1} + \frac{m_2}{2r_2}\,,
\end{equation}
where $r_1$ and $r_2$ are the distances from $\mathbf{r}$ to the two
punctures, $d$ the distance between the punctures, and $m_1,m_2$ are
the bare masses associated with the punctures. The ADM mass is seen to
be $\MADM = m_1+m_2$.  There turns out be a rich variety of MOTSs
even in this simple initial data; see \cite{Pook-Kolb:2018igu} for a
detailed study.  We choose a particular configuration $m_1= 0.5$ and
$m_2 = 0.8$, and the initial coordinate separation between the two
punctures is $d_0/\MADM=1$.  For these parameters, there are initially
only two disjoint MOTSs (representing the two black holes) surrounding
the two punctures.  Throughout, we state times in units of
$\MM := \MADM/1.3$.

As we evolve this initial data using the Einstein equations, the
result is shown in Fig.~\ref{fig:3d}.  This is the analog of the
``pair of pants picture''.  In this figure time goes vertically
upwards and horizontal sections of the tubes yield sections of the
MOTSs at a given time (the MOTSs are axially symmetric, and thus the
full MOTS can be obtained by revolving these sections around their
respective axes).  The tubes in red and purple are the world tubes of
the two individual MOTSs. These get closer to each other and
eventually touch at a time labeled $\ttouch$, and go through each
other after $\ttouch$.  At a time $\tbifurcate$ which is somewhat
earlier than $\ttouch$, when the two horizons get sufficiently close
to each other, a common MOTS is formed outside the individual ones.
This common MOTS immediately bifurcates into an inner and outer branch
shown respectively as a green mesh and in blue.  The outer branch (in
blue) becomes more symmetric and reaches an equilibrium state
corresponding to a MOTS of a Schwarzschild black hole horizon.  The
inner branch on the other hand becomes increasingly distorted.
Eventually it merges with the two individual MOTSs precisely at the
time $\ttouch$.  Finally, the inner MOTS develops self-intersections
immediately after $\ttouch$.  For reference, we find
$\ttouch \approx 5.53781\,\MM$ ($\approx 4.25985\,\MADM$) and
$\tbifurcate \approx 1.37460\,\MM$ ($\approx 1.05738\,\MADM$).

It is useful to rephrase the above in terms of sections of the world
tubes of Fig.~\ref{fig:3d}.  Before time $\tbifurcate$ the
intersection of the world tubes with a Cauchy surface will consist of
two disjoint spherical surfaces which we shall denote $\Sone$ and
$\Stwo$.  The 3-dimensional world tubes generated by them will be
denoted $\Hone$ and $\Htwo$ respectively.  Between $\tbifurcate$ and
$\ttouch$, $\Sone$ and $\Stwo$ continue to exist separately, but are
now surrounded by a pair of marginally trapped surfaces which enclose
both $\Sone$ and $\Stwo$.  The inner of these is denoted $\Sin$ and
the outer one (the apparent horizon) by $\Sout$.  The world tubes
generated by them are $\Hin$ and $\Hout$.  At $\ttouch$, $\Sone$ and
$\Stwo$ touch and at later times they go through each other while
remaining spherical. Also at $\ttouch$, $\Sin$ coincides with
$\Sone\cup\Stwo$, i.e. it forms a cusp.  After $\ttouch$, the cusps of
$\Sin$ develop into knots, i.e.  self-intersections, which become
larger with time.  The eventual fate of $\Sone$, $\Stwo$ and $\Sin$ is
not yet known \cite{Mosta:2015sga,Evans:2020lbq}; these get closer to
the punctures whence they become difficult to track numerically
(though constraints on their possible dynamics follow from general
results precluding the change of topology of Cauchy hypersurfaces
during evolution~\cite{Gan75,Gan76}).  $\Sout$ continues moving
outwards becoming ever more symmetric as it reaches its final
equilibrium fate.

These results were first reported in \cite{PhysRevLett.123.171102},
and the numerical method is detailed in
\cite{PhysRevLett.123.171102,Pook-Kolb:2018igu}.  In particular, the
existence of self-intersecting MOTSs has been proven in detail with
high accuracy (cf. also the recent \cite{Booth:2020qhb}).  A number of
questions still remain to be answered about this scenario.  These
include understanding geometric properties of the world tubes such as
the status of the area increase law, physical properties such as
multipole moments, fluxes of energy across the horizons, and the
stability properties. The goal of the present series of papers is to
discuss these physical properties in detail for this same initial
configuration.  In this paper, the first of two parts, we shall
discuss the geometric properties of the world tubes shown in
Fig.~\ref{fig:3d}. This includes the signature of the world tubes, the
expansion of the ingoing null rays, and most importantly, the area
increase law.  The area of the final apparent horizon $\Sout$ at late
times is certainly larger than the sum of the areas of $\Sone$ and
$\Stwo$ at early times.  Moreover, we can trace a sequence of surfaces
which takes us from the initially disjoint surfaces $\Sone$ and
$\Stwo$ to the single final horizon $\Sout$.  However, the area does
not increase monotonically along this sequence; there are in fact
short durations of area \emph{decrease} along this sequence.  This
fact needs an explanation, and might have important physical
implications.  In short, the geometrical quantities studied in this
first paper show \emph{how} the area increases.  The second paper
(henceforth paper II) addresses the question of \emph{why} the area
increases, i.e. the energy fluxes across the horizon.

The plan for the rest of the paper is the
following. Sec.~\ref{sec:quasilocal} sets up notation and briefly
summarizes the basic notions and results that we shall use later. The
behavior of MOTSs under time evolution, even for the simple case of a
head-on collision, exhibits a rich variety of geometric and physical
properties, and we will need the full machinery of quasi-local
horizons to describe these features. For this reason we summarize
different notions of quasi-local horizons, and we shall find it
appropriate to modify existing terminology in some cases.
Sec.~\ref{sec:areaincrease} discusses a very basic geometric and
physical aspect, namely the area of the MOTSs, and the status of the
area increase law.  This will involve issues such as the signature of
the world tube and whether the MOTSs are future or past trapped.  The
classic area increase law for event horizons might lead us to believe
that area should always increase to the future.  The situation will be
somewhat more complicated for us.  While the area of the MOTS does
increase overall, we find that there are small durations where this
does not hold.  Sec.~\ref{sec:theta_n} studies in detail the expansion
of the ingoing null normal.  Future trapped surfaces have negative
ingoing expansion everywhere, indicating the presence of a singularity
to the future.  We shall see that the dynamical horizons are not
everywhere future trapped, and have portions with positive ingoing
expansion.  Sec.~\ref{sec:signature} studies the signature of the
dynamical horizons and it shows that the horizons have both timelike
and spacelike portions.  Sec.~\ref{sec:anomalous} revisits the area
increase, and considers the correspondence of geometric fields on the
horizon with properties of a 2-dimensional fluid, first suggested
within the membrane paradigm.  Sec.~\ref{sec:conclusions} concludes by
discussing open questions and possible directions for future work.
Appendix~\ref{appendix:bousso-engelhardt} presents a detailed
comparison of our results with the proof of area monotonicity on a
dynamical horizon.  Appendix~\ref{appendix:Damour-Navier-Stokes}
extends the membrane paradigm analogy to spinning black holes, and
finally Appendix~\ref{sec:contact_structures} speculates on a possible
geometric interpretation of self-intersecting MOTSs.

\section{Basic Notions}
\label{sec:quasilocal} 

\subsection{The optical scalars and marginally trapped surfaces}
\label{subsec:basics}

We collect here the basic notions and definitions related to
quasi-local black hole horizons we shall need later.  While we will
try to be as self contained as possible, our goal in this section is
not to provide a comprehensive overview of the subject, but rather to
summarize the connections between the different notions with suitable
references to the literature. Reviews with diverse viewpoints can be
found in
e.g. \cite{Hayward:2000ca,Booth:2005qc,Ashtekar:2004cn,Gourgoulhon:2005ng,Visser:2009xp,Jaramillo:2011zw,hayward2013black,Faraoni:2015pmn}.

Spacetime is a 4-dimensional manifold $\scrM$ with a Lorentzian metric
$g_{ab}$ with signature $(-,+,+,+)$.  The derivative operator
compatible with $g_{ab}$ will be denoted $\nabla_a$, and the Riemann
tensor $R_{abcd}$ is defined according to
$(\nabla_a\nabla_b - \nabla_b\nabla_a)X_c = {R_{abc}}^dX_d$ for an
arbitrary 1-form $X_c$.  Let $\Surf \subset \scrM$ be a smooth,
orientable, closed 2-dimensional spacelike surface.  While it is
possible to consider higher genus surfaces, these are generically
expected to be unstable \cite{Newman:1987} and we restrict ourselves
to spherical topology in this paper.  $\Surf$ is naturally endowed
with two null-normal fields denoted $\ell^a$ and $n^a$.  These vector
fields are required to be future-directed, null and orthogonal to
$\Surf$.  We are free to rescale them by positive-definite functions.
This rescaling freedom can be reduced by fixing the inner-product
$\ell\cdot n = -1$ which ties the rescalings of $\ell$ and $n$:
\begin{equation}
\label{eq:ln-rescaling}
  \ell \rightarrow f\ell\,,\quad n\rightarrow f^{-1}n\,, \quad f>0\,.  
\end{equation}
Since $\Surf$ is spacelike, the spacetime metric $g_{ab}$, when
restricted to the tangent space of $\Surf$, yields a Riemannian metric
\begin{equation}
  \twoq_{ab} = g_{ab} + \ell_an_b + n_a\ell_b\,. 
\end{equation}
We denote the volume 2-form on $\Surf$ by $\twoepsilon_{ab}$, and the
derivative operator on $\Surf$ compatible with $\twoq_{ab}$ is denoted
$\mathcal{D}$.  Integrals over $\Surf$ will be written with $dA$ as
the measure.

Of special importance for us will be the so called optical scalars,
i.e. the expansion, shear, and twist of $\ell^a$ and $n^a$.  The
derivative $\nabla_a\ell_b$ projected on $\Surf$ can be separated into
a symmetric and anti-symmetric part, and the symmetric part can in
turn be decomposed into a trace and trace-free part:
\begin{equation}
  \label{eq:dl}
  \twoq_a^c\twoq_b^d\nabla_c\ell_d = \frac{1}{2}\Theta_{(\ell)}\twoq_{ab} + \sigma^{(\ell)}_{ab} + \omega^{(\ell)}_{ab}\,.
\end{equation}
Here $\Theta_{(\ell)}$ is the \emph{expansion} of $\ell^a$, the
symmetric tracefree tensor $\sigma^{(\ell)}_{ab}$ is the \emph{shear},
and the anti-symmetric tensor $\omega^{(\ell)}_{ab}$ is the
\emph{twist} of $\ell^a$.  The expansion, shear, twist of $n^a$ are
defined analogously:
\begin{equation}
  \label{eq:dn}
  \twoq_a^c\twoq_b^d\nabla_cn_d = \frac{1}{2}\Theta_{(n)}\twoq_{ab} + \sigma^{(n)}_{ab} + \omega^{(n)}_{ab}\,.
\end{equation}
The most important objects for us will be the two expansions, and
$\sigma_{ab}^{(\ell)}$.  In general, whenever we refer to shear or
$\sigma_{ab}$, we shall mean $\sigma^{(\ell)}_{ab}$ unless indicated
otherwise.  The twist will vanish because, by construction, $\ell$ and
$n$ are orthogonal to a smooth surface $\Surf$.

It will often be useful to complete $(\ell,n)$ to a null-tetrad
$(\ell, n, m,\bar{m})$ where $m^a$ is a complex null vector orthogonal
to both $\ell$ and $n$.  Thus $m^a$ is tangent to $\Surf$ and
satisfies $m\cdot \bar{m} = 1$ ($\bar{m}$ is the complex conjugate of
$m$).  Given the null-tetrad, the information contained in the
symmetric-tracefree tensor $\sigma_{ab}$ on $\Surf$ can be reduced to
a single complex field:
\begin{equation}
  \sigma := \sigma_{ab}m^a m^b = m^am^b\nabla_a\ell_b\,.
\end{equation}
The choice of $m$ can be changed by a phase:
$m\rightarrow e^{i\psi}m$.  Under this change, the shear transforms as
$\sigma \rightarrow e^{2i\psi}\sigma$ which means that $\sigma$ has
spin weight $+2$. This implies that $\sigma$ can be expanded into
angular modes on $\Surf$ using spin-weighted spherical harmonics of
spin-weight $+2$.  This will play a very important role in paper II.

The projections of $\nabla_a\ell_b$ and $\nabla_an_b$ given in
Eqs.~(\ref{eq:dl}) and (\ref{eq:dn}) are the two extrinsic curvatures
of $\Surf$ embedded in spacetime $\scrM$.  The other important
quantity is the connection on the normal bundle of $\Surf$.  Since
$\Surf$ has co-dimension $2$, the connection on the normal bundle is
given by a single 1-form $\omega_a$ defined as
\begin{equation}
  \omega_a = -n_b\twoq_{a}^c\nabla_c\ell^b\,.
\end{equation}
This 1-form determines the angular momentum associated with the
horizon; we will always deal with non-spinning black holes and will
have $\omega_a=0$ in this paper.

$\Surf$ is said to be a \emph{future-marginally-outer-trapped} surface
if $\Theta_{(\ell)}=0$ and $\Theta_{(n)}<0$.  Note that the expansions
$\Theta_{(\ell)}$ and $\Theta_{(n)}$ are also rescaled under the
transformation of Eq.~(\ref{eq:ln-rescaling}), but these conditions
remain unchanged since $\ell^a$ and $n^a$ remain future directed.  If
instead $\Theta_{(n)}>0$ (and still requiring $\Theta_{(\ell)}=0$),
then $\Surf$ is said to be \emph{past-marginally-outer-trapped}.  We
shall often just refer to \emph{marginally trapped surfaces} with the
understanding that we are referring to future-marginally-outer-trapped
surfaces.  Surfaces satisfying only $\Theta_{(\ell)}=0$ (with no
condition on $\Theta_{(n)}$) are the marginally outer trapped
surfaces, or MOTS in short -- these are the basic objects that we
shall study in this paper.  As we shall see, $\Theta_{(n)}< 0$ will
not always be satisfied; it is therefore necessary to keep track of
this condition in the various definitions and results.

A MOTS is a geometric concept in the full 4-dimensional spacetime
independent of any spatial slices.  In numerical simulations however,
they are connected to Cauchy surfaces because in order to locate them,
we only require the Cauchy data, i.e. the 3-metric and extrinsic
curvature.  For a closed 2-surface $\Surf$ in $\Sigma$, let $R^a$ be
the spacelike outward pointing normal to $\Surf$, and $T^a$ the
unit-timelike normal to $\Sigma$.  A convenient choice for the null
normals is
\begin{equation}
  \label{eq:normals}
  \ell^a = \frac{1}{\sqrt{2}}\left( T^a + R^a\right)\,,\quad
  n^a = \frac{1}{\sqrt{2}}\left( T^a - R^a\right)\,.
\end{equation}
The expansion condition $\Theta_{(\ell)}=0$ can be written in terms of
the extrinsic curvature $K_{ab}$ of $\Sigma\subset \scrM$:
\begin{equation}
  D_aR^a + K_{ab}R^aR^b - K = 0
\end{equation}
where $D_a$ is the derivative operator on $\Sigma$. Taking $\Surf$ to
be the level set of a suitable function $h$, this equation can, in
turn, be expressed as a second order non-linear differential equation
for $h$.  Our choice for $h$ is based on choosing a reference surface
sufficiently close to $\Surf$. Details on how the reference surface is
chosen, the associated coordinate system, and the numerical method for
solving the above equation can be found in
\cite{Pook-Kolb:2018igu,PhysRevD.100.084044} and our implementation is
available at \cite{pook_kolb_daniel_2019_2591105}.  This is an
extension of the widely used method developed in
\cite{Thornburg:2003sf,Thornburg:2006zb,Thornburg:1995cp,Thornburg:1995cp,Shoemaker:2000ye,Lin:2007cd,Jaramillo:2009zz}.
Our numerical calculations use the \emph{Einstein Toolkit}
\cite{Loffler:2011ay, EinsteinToolkit:web}. We use \emph{TwoPunctures}
\cite{Ansorg:2004ds,Ansorg:2005bp} to set up initial conditions and an axisymmetric
version of \emph{McLachlan} \cite{Brown:2008sb} to solve the Einstein
equations, which uses \emph{Kranc} \cite{Husa:2004ip, Kranc:web} to
generate efficient C++ code.
For the results in the current series of papers, we performed
simulations with three spatial resolutions
$1/\Delta x = 960$, $480$, $60$, running, respectively, until
$\tname_\text{max} = 7\,\MM, 20\,\MM, 50\,\MM$.
Further details of the simulation
specific to our problem are detailed in \cite{PhysRevD.100.084044}.

We now collect basic definitions and results pertaining to quasi-local
horizons that we shall use while presenting our results. The goal here
is not a detailed review of the subject, but mainly to orient the
reader and to set up notation and terminology.

\subsection{The MOTS stability operator}
\label{subsec:stability}

Starting with a MOTS on a Cauchy surface, it is natural to ask how it
behaves under time evolution.  It is not \emph{a priori} obvious that
a MOTS should evolve smoothly.  It is now known that the behavior of a
MOTS under time evolution is controlled by a second order
non-self-adjoint elliptic operator $L_\Sigma$ known as the stability
operator
\cite{Andersson:2005gq,Andersson:2007fh,Andersson:2008up,Pook-Kolb:2018igu}.
$L_\Sigma$ is constructed from variations of $\Surf$
\cite{Newman:1987}.  Given a surface $\Surf$, let $\Surf_\lambda$ be a
family of surfaces parameterized by a real variable $\lambda$; the
surfaces $\Surf_\lambda$ depend smoothly on $\lambda$.  We take
$\Surf_{\lambda=0}$ to coincide with $\Surf$, and $\lambda$ can take
values in an interval $(-\epsilon,\epsilon)$.  The variation
$\Surf_\lambda$ is assumed to be smooth which implies that if we pick
a point $p$ on $\Surf$, the variation produces a smooth curve passing
through $p$, and the tangent vector to these curves at $\lambda=0$
defines a vector field $X$ on $\Surf$.  With this structure, one can
naturally define the variation of geometric quantities on $\Surf$
\cite{Booth:2006bn,Gao:2008jv}.  Of particular importance is the
variation of $\Theta_{(\ell)}$ denoted $\delta_X\Theta_{(\ell)}$.  On
each $\Surf_\lambda$ construct the null normals $\ell^{(\lambda)}$ and
$n^{(\lambda)}$ as for $\Surf$ itself.  This defines the expansion
$\Theta_{(\ell)}^{(\lambda)}$ for all $\lambda$, and allows us to
differentiate it:
\begin{equation}
  \delta_X\Theta_{(\ell)} := \left(\frac{d\Theta_{(\ell)}^{(\lambda)}}{d\lambda}\right)_{\lambda=0}\,.
\end{equation}
This variation should not be confused with usual derivatives of
$\Theta_{(\ell)}$.  In particular, while
$\delta_{cX}\Theta_{(\ell)} = c\delta_X\Theta_{(\ell)}$ for constants
$c$, it turns out that
$\delta_{\psi X}\Theta_{(\ell)} \neq \psi\delta_X\Theta_{(\ell)}$ if
$\psi$ is a non-constant function.  If $X$ is tangent to $\Surf$ and
$\Theta_{(\ell)}=0$ on $\Surf$, then it is obvious that
$\delta_X\Theta_{(\ell)}=0$.  Thus, we only need to consider variation
fields $X$ normal to $\Surf$.  One could consider $X$ to be
proportional to $\ell$ or $n$, but in the context of a Cauchy
evolution, it is natural to take $X$ along the normal $R^a$,
$X^a=\psi R^a$. Thus, we define the stability operator associated with
$\Sigma\supset\Surf$ as
\begin{equation}
  L_\Sigma\left[\psi\right] :=  \sqrt{2}\;\delta_{\psi R}\Theta_{(\ell)}\,, 
\end{equation}
where a global constant positive factor can be arbitrarily chosen, and
$\sqrt{2}$ is chosen to simplify later expressions.  An explicit
calculation shows that $L_\Sigma$ is a second order elliptic operator
but it is not necessarily self-adjoint.  In vacuum spacetimes, the
expression for $L_\Sigma$ is the following:
\begin{eqnarray}
  L_\Sigma\left[\psi\right] &=& -\Delta_\Surf \psi + 2\omega^a\mathcal{D}_a \psi \nonumber \\
                  &+& \left( \frac{1}{2}\mathcal{R} + \mathcal{D}_a\omega^a - \omega_a\omega^a - \sigma_{ab}\sigma^{ab} \right) \psi \,.
\end{eqnarray}
Here $\Delta_\Surf$ is the Laplacian on $\Surf$ and $\mathcal{D}_a$ is
the derivative operator on $\Surf$. In the present case, we shall deal
with the head-on collision of non-spinning black holes so that
$\omega_a=0$, whence $L_\Sigma$ will be self-adjoint and will have
real eigenvalues.

In the dynamical evolution setting, the importance of $L_\Sigma$ lies in the following result
\cite{Andersson:2005gq,Andersson:2007fh,Andersson:2008up,Pook-Kolb:2018igu}:
\begin{itemize}
\item \emph{A MOTS evolves smoothly as long as $L_\Sigma$ is invertible,
    i.e. as long as none of its eigenvalues vanish.}
\end{itemize}
In simple cases when the smallest eigenvalue $\Lambda_0$ is strictly
positive, then $L_\Sigma$ is obviously invertible and the MOTS evolves
smoothly. In a binary black hole merger, this is what happens for the
outermost MOTS and the two individual MOTSs.  However, as shown in
\cite{Pook-Kolb:2018igu,PhysRevLett.123.171102,PhysRevD.100.084044},
the inner common MOTS is more complicated.  It is born at
$T_{\rm bifurcate}$ with $\Lambda_0=0$ which immediately becomes
negative. None of the other eigenvalues cross $0$ and the MOTS
continues to evolve smoothly.  It is clear from this that the complete
spectrum of $L_\Sigma$, and not just its principal eigenvalue, is
potentially of interest.  This is especially true for spinning black
holes when the eigenvalues can be complex, thus leading to the full
MOTS-spectral problem formulated in \cite{Jaramillo:2013rda} and
initiated in \cite{Jaramillo:2014oha,Jaramillo:2015twa}.  We shall
explore the spectrum of $L_\Sigma$ in paper II.

\subsection{Marginally trapped tubes and dynamical horizons}
\label{subsec:dhbasics}

With the time evolution understood, we consider the three-dimensional
world tube traced out by a MOTS.  This world tube is known as a
\emph{marginally outer trapped tube}. More formally (following
\cite{Andersson:2007fh}):
\begin{defn}[Marginally Outer Trapped Tube]
  A smooth 3-dimensional surface $\HH$ in a spacetime is said
  to be a marginally outer trapped tube (MOTT) if
  \begin{itemize}
  \item it has topology $S^2\times \mathbb{R}$, i.e. it admits a
    foliation by 2-spheres,
  \item each leaf of the foliation is a MOTS.
  \end{itemize}
\end{defn}
Note that $\HH$ is allowed to have arbitrary signature and no
restrictions are placed on the ingoing expansion $\Theta_{(n)}$ for
any of the MOTSs which constitute $\HH$.  As we have seen, the results
involving the stability operator mentioned above do not assume any
condition on $\Theta_{(n)}$, and hold for any MOTT.  The classic
examples of MOTTs in spherical symmetry are the well known Vaidya
\cite{Vaidya:1999zz} and Oppenheimer-Snyder \cite{Oppenheimer:1939ue}
solutions.  Further examples in spherical symmetry can be found in
e.g. \cite{Booth:2005ng,Helou:2016xyu,Chatterjee:2020khj}.  These
examples already show the wide variety of cases that can appear even
in spherical symmetry.  See \cite{Booth:2012xm} for a construction of
the spacetime locally near such a horizon.
See e.g. \cite{Schnetter:2006yt,Jaramillo:2009zz,Mosta:2015sga,Gupta:2018znn}
for previous examples of numerical studies concerning dynamical horizons in
various physical situations.  The present paper shall provide the most
detailed numerical study yet of these horizons in a black hole merger.

Imposing additionally $\Theta_{(n)}<0$ leads to the definition of a
marginally trapped tube (following \cite{Ashtekar:2005ez}):
\begin{defn}[Marginally Trapped Tube]
  A MOTT is said to be a marginally trapped tube (MTT) if it satisfies
  in addition $\Theta_{(n)}<0$ everywhere.
\end{defn}
As we shall discuss below in Sec.~\ref{sec:areaincrease}, the
condition $\Theta_{(n)}<0$ is employed in different proofs of the area
increase law.  The proof of the area increase law
\cite{Bousso:2015qqa,Bousso:2015mqa,Sanches:2016pga} holds for an MTT
of arbitrary signature, though with additional technical assumptions
we shall discuss later (MTTs are referred to as holographic screens in
this work).

Additional restrictions can be placed on a MOTT depending on the
physical situation one is interested in.  When the marginally trapped
tube is in equilibrium, i.e. there is no energy flux across
$\HH$, we need the notion of a non-expanding horizon
\cite{Ashtekar:2000hw}:
\begin{defn}[Non-expanding horizon]
  A smooth 3-dimensional surface $\Delta$ in a spacetime is said to be
  a non-expanding-horizon if
  \begin{itemize}
  \item $\Delta$ has topology $S^2\times\mathbb{R}$
  \item $\Delta$ is null
  \item Any null normal to $\Delta$, denoted $\ell^a$, has vanishing
    expansion ($\Theta_{(\ell)}=0$).
  \item The Einstein field equations hold at $\Delta$ and, if $T_{ab}$
    is the stress energy tensor, $-T^a_{b}\ell^b$ is future directed
    and causal when $\ell^a$ is future directed.
  \end{itemize}
\end{defn}
From the properties of a null surface, it can be shown that every
complete cross-section of $\Delta$ is a MOTS.  Thus, a non-expanding
horizon is, in essence, a MOTT with null signature.  The last
condition is an energy condition and is implied by, for example, the
dominant energy condition.  It can also be shown that each
cross-section of $\Delta$ has the same area; the black hole here is in
equilibrium in an otherwise dynamical spacetime.  Not all geometric
fields on a non-expanding horizon are time independent.  Further
physical restrictions requiring the derivative operator to be time
independent lead to the notion of an isolated horizon
\cite{Krishnan:2012bt,Ashtekar:2001is,Ashtekar:2001jb}.  It is
interesting to note that a version of the stability operator also
appears in going from non-expanding to isolated horizons
\cite{Ashtekar:2001jb}, and again, the invertibility of the stability
operator turns out to be the relevant condition.

Local constructions of spacetime neighborhoods near non-expanding
horizons is given in
\cite{Lewandowski:1999zs,Krishnan:2012bt,Gunasekaran:2019jrq,Booth:2012xm,Gurlebeck:2015xpa,Flandera:2016qwg,Lewandowski:2014nta,Lewandowski:2018khe}.
All stationary black holes and Killing horizons, including of course
the Schwarzschild and Kerr black holes, are non-expanding horizons. A
detailed study of the Kerr-Newman black hole viewed as a non-expanding
horizon can be found in \cite{Scholtz:2017ttf}.  Apart from these,
there are also the so-called distorted black holes representing
stationary black holes in the presence of external fields
\cite{Geroch:1982bv} (see also \cite{Fairhurst:2000xh} for exact
solutions representing charged distorted black holes).  Distorted
black holes can potentially have positive $\Theta_{(n)}$
\cite{Pilkington:2011aj}.

Moving now to the general dynamical case, we will work with a general
MOTT.  As we shall see, a MOTT can be spacelike or timelike, or even
have sections of mixed signature. In addition it can have positive or
negative $\Theta_{(n)}$.  In principle we could add qualifiers in
front of MOTT and refer to, for example, spacelike or timelike MOTTs.
However, to minimize the number of acronyms and to perhaps make it
easier to remember:
\begin{defn}[Dynamical Horizons]
  We shall refer to a generic MOTT as a dynamical horizon. Additional
  qualifiers will be added as appropriate.  Thus we can have spacelike
  or timelike dynamical horizons depending on the signature, and
  future or past depending on whether $\Theta_{(n)}<0$ or $>0$
  respectively.
\end{defn}
The reader might be aware that previously, dynamical horizons referred
to spacelike MTTs \cite{Ashtekar:2002ag,Ashtekar:2003hk} (this is
closely related to but not the same as a future outer trapping horizon
\cite{Hayward:1993wb,Hayward:1994yy,Hayward:2004fz,Hayward:2006ss}).
However, already in \cite{Ashtekar:2003hk} (Appendix B), timelike
cases were considered and referred to as timelike dynamical horizons.
Dynamical horizons were always meant to refer to a general MOTT and
the spacelike case was initially thought to be the most relevant
case. We shall therefore use different terminology in this paper.  A
general MOTT will be called a dynamical horizon and qualifiers will be
added as appropriate.  

We conclude this section by a short discussion of the area increase
and fluxes across dynamical horizons.  Consider a portion
$\Delta \HH$ between two MOTSs with initial area $A_i$ and
final area $A_f$.  As shown in
\cite{Ashtekar:2002ag,Booth:2003ji,Booth:2006bn}, the area change
$A_f-A_i$ can be written as an integral over $\Delta \HH$,
with the integrand being local fields on $\Delta\HH$.  The
integrand can be viewed as a flux, whence the area is seen to change
due to the flux of radiation across the horizon.  We shall discuss the
fluxes in great detail in paper II, but here we just mention two
points: i) the dominant contribution is due to the shear
$\sigma_{ab}^{(\ell)}$, which was recently seen to be closely
correlated with the outgoing flux measured at infinity
\cite{Prasad:2020xgr}.  Thus, the fluxes provide a critical link
between horizon dynamics and observations of gravitational waves. ii)
the fluxes are manifestly positive definite for spacelike dynamical
horizons, but not so for timelike cases \cite{Ashtekar:2003hk}.

Besides these flux laws, there is an alternate formulation of the area
change. The starting point is the membrane paradigm for black hole
event horizons
\cite{Damour:membrane,damour1982surface,Price:1986yy,Thorne:1986}.  By
applying the Einstein equations to an event horizon, Damour showed a
close analogy between evolution equations on the horizon and the
Navier-Stokes equation for a 2-dimensional fluid
\cite{Damour:membrane,damour1982surface}.  In this way, it is possible
to relate fields on the event horizon and physical properties of a
corresponding 2-dimensional fluid such as energy density, pressure,
bulk and shear viscosity.  An interesting feature of this
correspondence is that one obtains a negative bulk viscosity for the
fluid, suggesting an instability.  As shown in
\cite{Gourgoulhon:2005ch,Gourgoulhon:2006uc,Gourgoulhon:2008pu}, this
correspondence also holds for dynamical horizons and one can similarly
obtain counterparts to the various physical quantities listed above.
In particular, the bulk viscosity now turns out to be positive as
expected.  We shall explore certain aspects of this analogy later in
Sec.~\ref{sec:anomalous}.

\section{The area increase law}
\label{sec:areaincrease}

The laws of black hole thermodynamics can be satisfactorily formulated
using quasi-local horizons
\cite{Ashtekar:1998sp,Ashtekar:1999yj,Ashtekar:2001is,Ashtekar:1999sn,Ashtekar:2000hw,Ashtekar:2002ag,Ashtekar:2003hk}.
For example, formulations of the first law of black hole mechanics
based on event horizons \cite{Bekenstein:1973ur,Bardeen:1973gs} use a
mixture of quantities defined at the horizon (such as the area) and
infinity (such as the ADM mass, and also surface gravity which uses
the timelike Killing vector normalized at infinity).  The quasi-local
formulation of the first law satisfactorily addresses this problem,
and coherently uses quantities defined only at the horizon.  Here we
shall not review all aspects of black hole thermodynamics, and instead
focus on one aspect, namely the area increase law.

\begin{figure*}
  \centering    
  \includegraphics[width=0.45\linewidth]{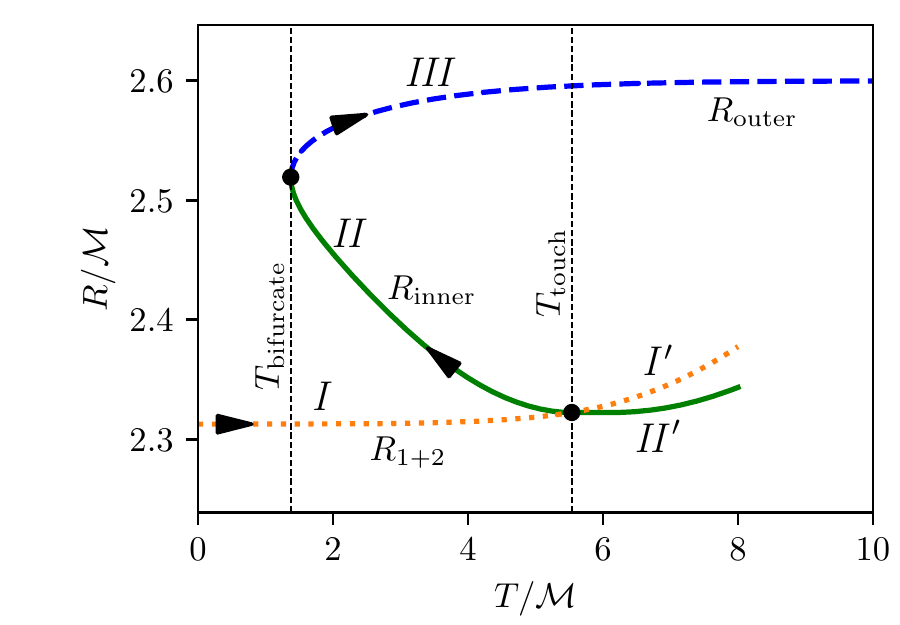}
  \includegraphics[width=0.45\linewidth]{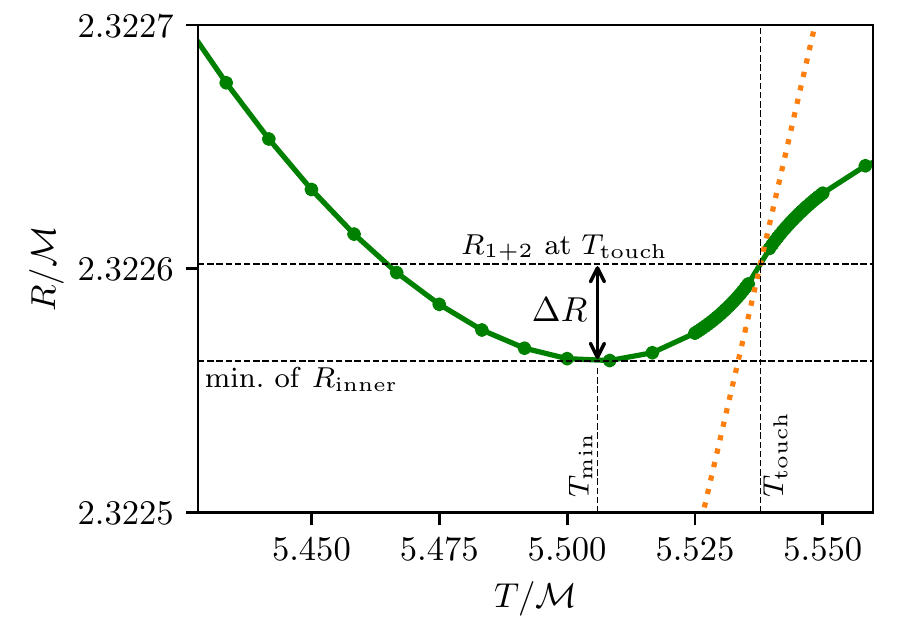}
  \caption{The first panel shows the radii of the various horizons as
    functions of time.  The orange dotted line is an ``effective''
    area-radius for the two individual horizons, the blue curve is the
    radius of the apparent horizon, while the solid green curve is the
    radius of the inner common MOTS. It is possible to connect the
    initial radii with the final one by following the segment $\sI+\sII+\sIII$
    (the segment $\sII$ is followed backwards in time).  The second panel
    shows a close-up near $\ttouch$ where an anomalous increase in
    the area of the inner common horizon is observed.}
  \label{fig:radii}
\end{figure*}

The areas of the various horizons for our particular configuration are
straightforward to calculate and have been presented previously
\cite{PhysRevLett.123.171102,PhysRevD.100.084044}.  Here we present
the same data first in terms of the radii of the black holes; see
Fig.~\ref{fig:radii}.  For a spherical surface with area $A$, one can
define a radius $R$ according to $\sqrt{A/4\pi}$, known as the
area-radius. This is straightforward to define for $\Sin$ and $\Sout$,
and their radii will be denoted $\Rin$ and $\Rout$ respectively.  For
the two individual horizons $\Sone$ and $\Stwo$, we can similarly
define an effective area radius $R_{1+2}$ as $\sqrt{(A_1+A_2)/4\pi}$.
Fig.~\ref{fig:radii} plots these radii as functions of time.  To
connect the two initial horizons to the final one, one can follow the
curves along the segment $\sI$, then follow segment $\sII$ backwards
in time, and then segment $\sIII$ which takes us to the final remnant
black hole.  While the overall area change is of course positive, the
area does not change monotonically along $\sI+\sII+\sIII$.  There is a
small duration of anomalous area increase on segment $\sII$ just prior
to $\ttouch$.  The second panel shows a close-up near $\ttouch$ and
the anomalous area increase of $\Sin$.  For reference, the local
minimum of the area occurs at $\tmin = 5.50594\,\MM$.

\begin{figure}
  \centering    
  \includegraphics[width=0.9\columnwidth]{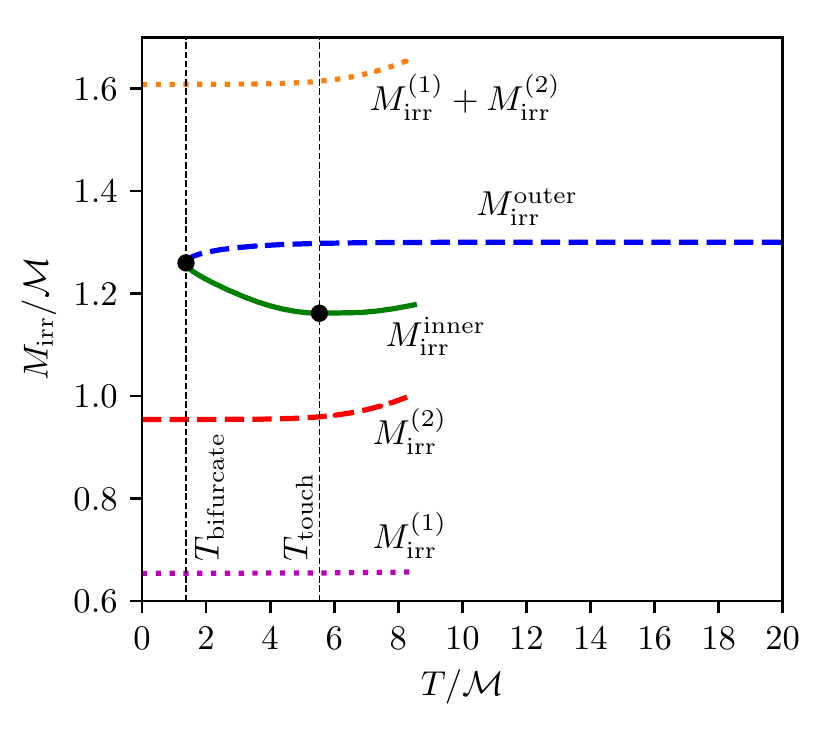}
  \caption{Irreducible masses of the various horizons as functions of
    time.}
  \label{fig:irreducible-mass}
\end{figure}

Fig.~\ref{fig:irreducible-mass} shows the irreducible masses
$\Mirr=\sqrt{A/16\pi}$ of the various horizons.  For the two
individual horizons $\Sone$ and $\Stwo$, we show them separately and
also the sum
\begin{equation}
  \Mone + \Mtwo = \sqrt{A_1/16\pi} +
  \sqrt{A_2/16\pi}\,. 
\end{equation}
This measure takes into account the interaction energy between the two
black holes, which in fact is quite significant given that the
separation between the black holes is small.  Thus unlike
Fig.~\ref{fig:radii}, in Fig.~\ref{fig:irreducible-mass}, the curve
for $\Mone + \Mtwo$ lies above the curves for the outer horizons.  For
the apparent horizon, the value of $\Mout$ at late times is a good
approximation to the Bondi mass, i.e. the mass left in the spacetime
after all the gravitational radiation has left the system.  Since this
is a very short simulation, the amount of radiation is small and the
difference between this estimate of the Bondi mass and the ADM mass is
smaller than $0.1\%$.

In the next sections we will go deeper into the various ingredients
which determine the area change.  Let us therefore conclude this
section by outlining why the expansion $\Theta_{(n)}$, and the
signature of the dynamical horizons, are the main ingredients we
should be looking at.  There are some simplified cases when the area
increase law can be easily proved, namely for purely future-spacelike
and purely future-timelike dynamical horizons, i.e. we assume
everywhere $\Theta_{(n)}<0$ and fix the signature.  Let us start with
the spacelike case, and let $\HH$ be the dynamical horizon.  The null
normal choices of Eq.~(\ref{eq:normals}) are tied to a Cauchy surface
intersecting the dynamical horizon. We need instead null-normals
entirely determined by $\HH$.  Since $\HH$ is spacelike, it has a
unit-timelike normal $\widehat{\tau}^a$.  The foliation on $\HH$
determines a unit-spacelike vector field $\widehat{r}^a$, which we
take to be outward pointing.  Then, analogous to
Eq.~(\ref{eq:normals}), a suitable choice of null normals is
\begin{equation}
  \label{eq:normals-dh}
  \widehat{\ell}^a = \frac{\widehat{\tau}^a + \widehat{r}^a}{\sqrt{2}}\,,\quad   \widehat{n}^a = \frac{\widehat{\tau}^a - \widehat{r}^a}{\sqrt{2}}\,.
\end{equation}
Then, with $\Theta_{(\widehat{\ell})}=0$, we easily get
\begin{equation}
  D_a\widehat{r}^a = -\frac{\Theta_{(\widehat{n})}}{\sqrt{2}} > 0\,,
\end{equation}
where $D_a$ is the derivative operator on $\HH$.  This shows
that the area of the cross-sections of $\HH$ increases along
$\widehat{r}^a$.  Similarly, for a timelike dynamical horizon, the
roles of $\widehat{r}^a$ and $\widehat{\tau}^a$ are interchanged and
it is now $\widehat{\tau}^a$ which is tangent to $\HH$.  We
get for the divergence of $\widehat{\tau}^a$:
\begin{equation}
  D_a\widehat{\tau}^a =  +\frac{\Theta_{(\widehat{n})}}{\sqrt{2}} < 0\,,
\end{equation}
whence the area decreases along $\widehat{\tau}^a$.  The reader should
bear in mind that the null normals $\widehat{n}^a$ and $n^a$ are
related by a scaling as in Eq.~(\ref{eq:ln-rescaling}). Thus the
expansions of $\widehat{n}^a$ and $n^a$ are also related by a scaling,
but the sign of the expansion remains unchanged.

These simple calculations illustrate the importance of the sign of
$\Theta_{(n)}$ and the signature.  The proof of the area increase law
by Bousso \& Engelhardt \cite{Bousso:2015mqa,Bousso:2015qqa} which we
will discuss below, does not make any assumption on the signature but
does assume $\Theta_{(n)}<0$.  With this assumption, and additional
technical requirements which will be important, it can be shown that
the area must be monotonic on a future dynamical horizon.  Now,
viewing $\Hin$ and $\Hout$ as a single dynamical horizon
(i.e. consider the union of the segments $\qIII$, $\qII$, and
$\qII^\prime$ shown in Fig.~\ref{fig:radii}), the area is in fact
monotonic \emph{except for the anomalous area increase starting from
  $\tmin$}.  We need to explain which of the conditions assumed in the
proof are violated.  Some obvious questions for us are thus: Do the
various dynamical horizons behave as one might have naively expected?
For example, is the outer common horizon everywhere spacelike and does
it have $\Theta_{(n)}<0$?  Similarly, is the inner horizon always
timelike?  What happens during the anomalous area increase shown in
the second panel of Fig.~\ref{fig:radii}?  We shall now proceed to
address these questions.

In the next sections we shall examine the signature and the behavior
of $\Theta_{(n)}$ for all the horizons.  Keep in mind the different
role played by these two aspects: $\Theta_{(n)}$ is part of the
extrinsic curvature of a MOTS, i.e. it is determined by how $\Surf$ is
embedded in a spacetime manifold.  The signature of the dynamical
horizons $\HH$, on the other hand, necessarily involves MOTSs
at different time steps; we need to obtain at least a small portion of
$\HH$ in order to evaluate its signature.  Thus, we first
investigate $\Theta_{(n)}$ followed by the signature.

\section{The expansion of the ingoing null rays}
\label{sec:theta_n}

\begin{figure}
  \centering    
  \includegraphics[width=\columnwidth]{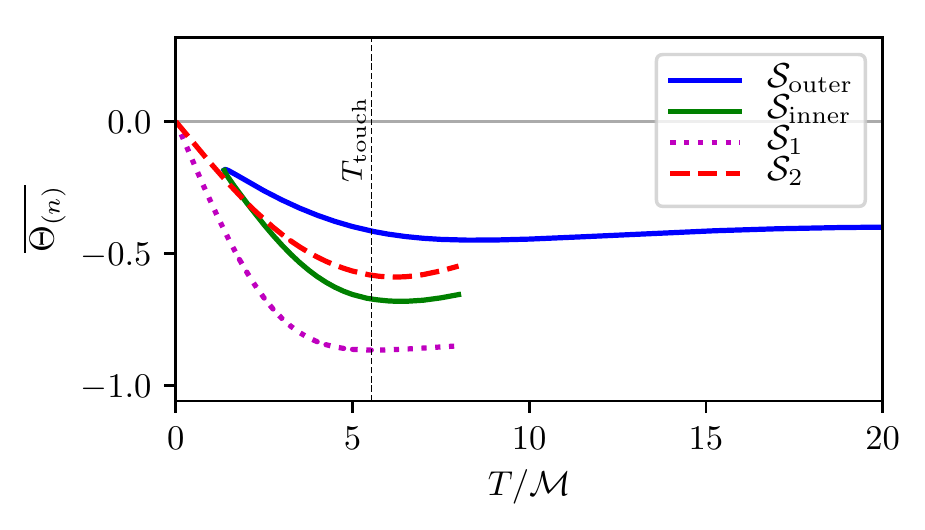}
  \caption{The average ingoing expansion}
  \label{fig:expansion-avg}
\end{figure}

As we have seen, the expansion of the inward pointing null rays,
$\Theta_{(n)}$, is of great importance for the area increase law.  The
average $\overline{\Theta_{(n)}}$ over a closed 2-surface $\mathcal{S}$
with area $A_{\mathcal{S}}$ and area 2-form $\twoepsilon$ is
\begin{equation}
  \overline{\Theta_{(n)}} = \frac{1}{A_{\mathcal{S}}}\oint_{\mathcal{S}} \Theta_{(n)} \twoepsilon\,.  
\end{equation}
The average of $\Theta_{(n)}$ is shown in Fig.~\ref{fig:expansion-avg}
for all four horizons as functions of time.  The initial data is time
symmetric, which implies that $\Theta_{(n)}=0$ for $\mathcal{S}_{1,2}$
initially.  The average becomes negative and remains negative at all
times.  The common horizons are born with negative
$\overline{\Theta_{(n)}}$ and they remain negative at all times.  

\begin{figure*}
  \centering    
  \includegraphics[width=\columnwidth]{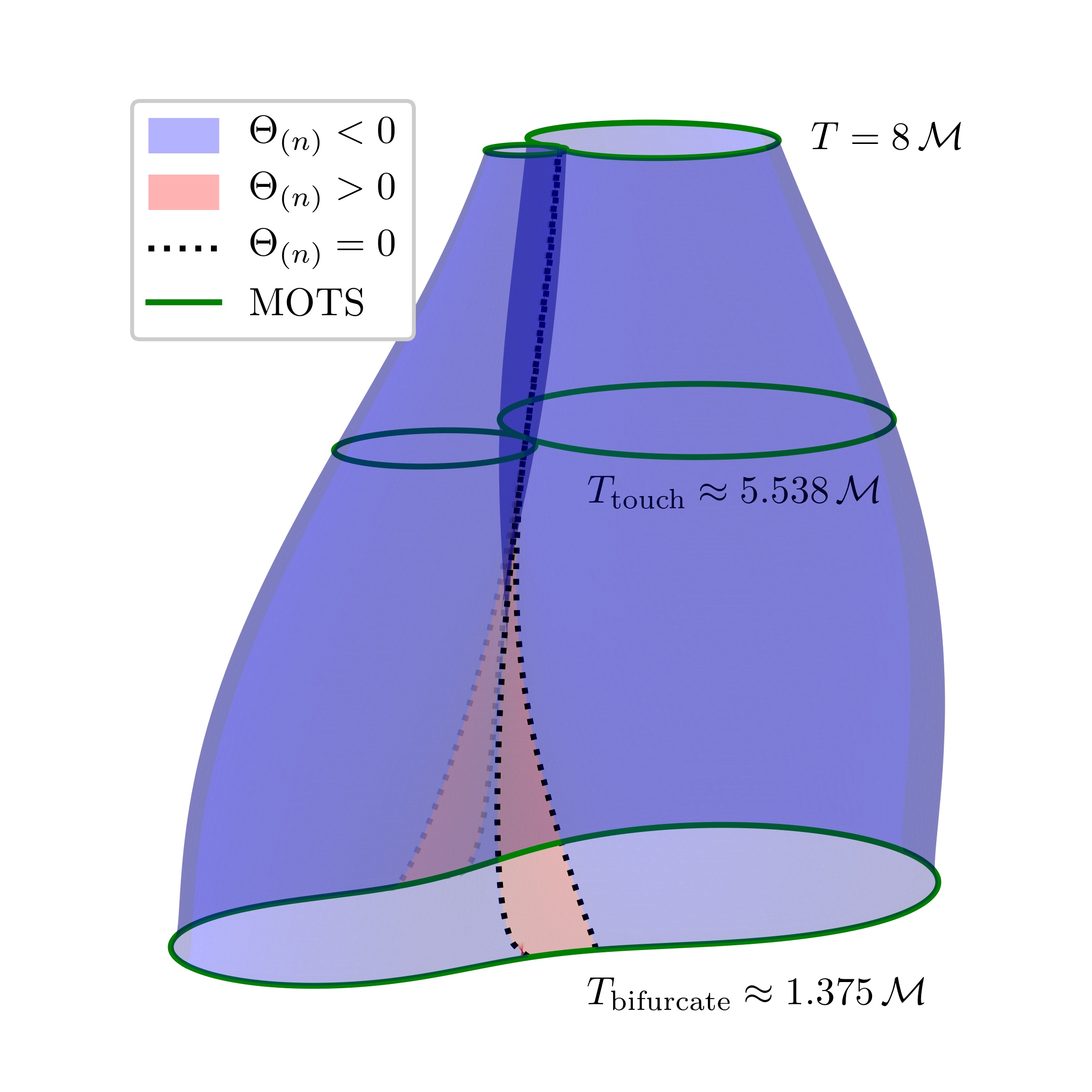}
  \includegraphics[width=\columnwidth]{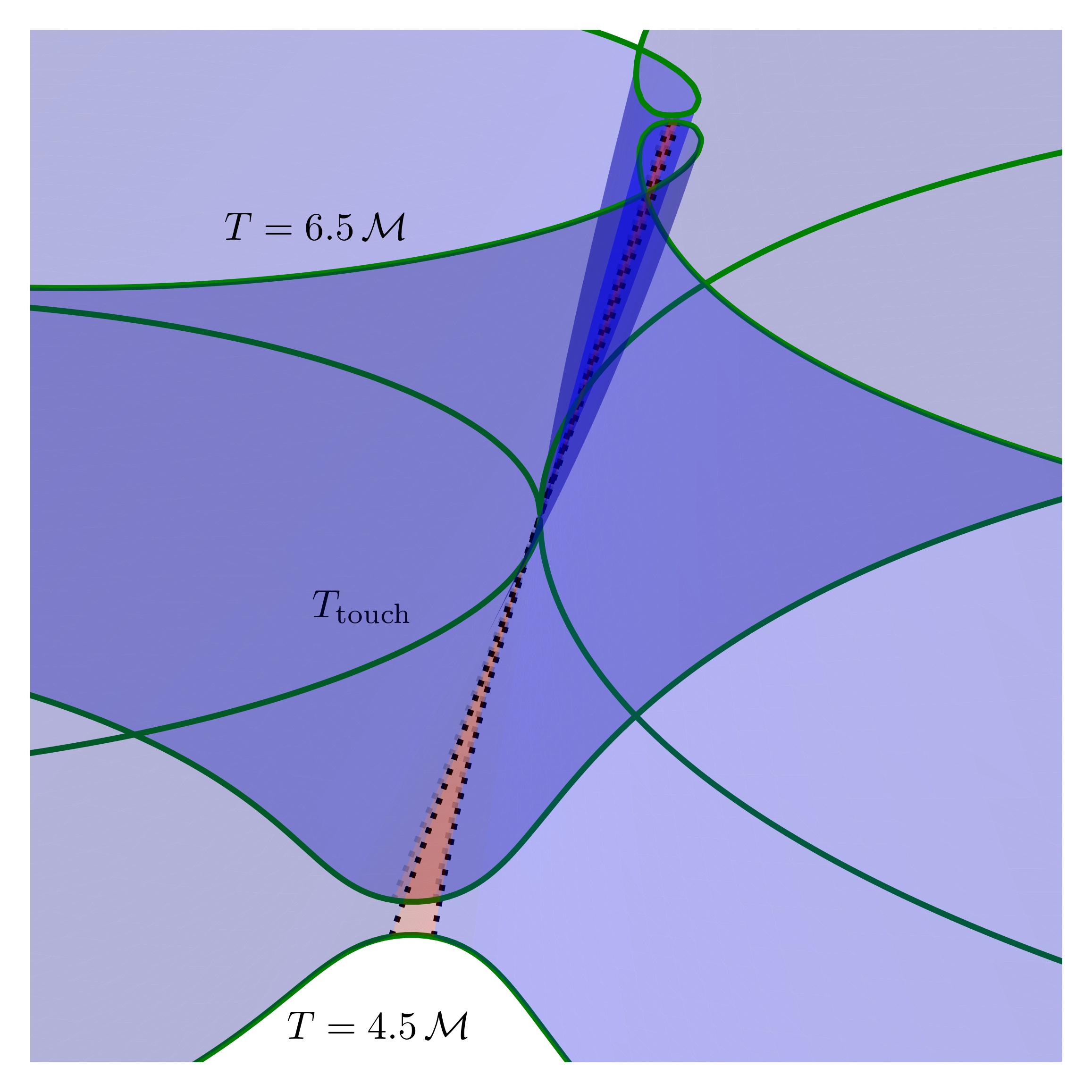}
  \caption{The sign of $\Theta_{(n)}$ for $\Sin$.  As for the apparent
    horizon, $\Sin$ has a portion around its waist with
    $\Theta_{(n)}>0$, and this portion becomes smaller over time but
    does not disappear.  The second panel shows details near
    $\ttouch$.  We see that the portion with positive $\Theta_{(n)}$
    momentarily vanishes at $\ttouch$, but reappears again immediately
    afterwards. }
  \label{fig:expansion-inner}
\end{figure*}

The behavior of $\Theta_{(n)}$ beyond the averages is more
interesting.  The individual horizons however remain boring:
$\Theta_{(n)}\leq 0$ for the individual horizons at all times.  Since
the initial data is time symmetric, the individual black holes
initially have $\Theta_{(n)}=0$ but it is strictly negative
thereafter.  The individual horizons are therefore conventional
future-dynamical horizons.  The outer dynamical horizon generated by
the apparent horizons are somewhat more interesting: they do not
always have $\Theta_{(n)}<0$ as one might have expected.  Upon
formation, there is a small region with $\Theta_{(n)}>0$ around the
``waist'' defined as follows.  Each MOTS is axisymmetric, and thus has
an axial symmetry vector $\varphi^a$.  This vanishes at two points
which defines the two poles.  We can calculate the proper length of
each orbit of $\varphi^a$.  The proper length vanishes at the poles
and, for a regular round sphere, it is maximum at the equator.
However, for some of the MOTSs in our problem, we find that the proper
circumference has a local minimum around the equator. This is most
obvious for $\Hin$ near $\ttouch$ where it looks like a figure eight
(see second panel of Fig.~\ref{fig:expansion-inner}), but it is also
true for the apparent horizon just when it is born at $\tbifurcate$.

\begin{figure*}
  \centering    
  \includegraphics[width=\columnwidth]{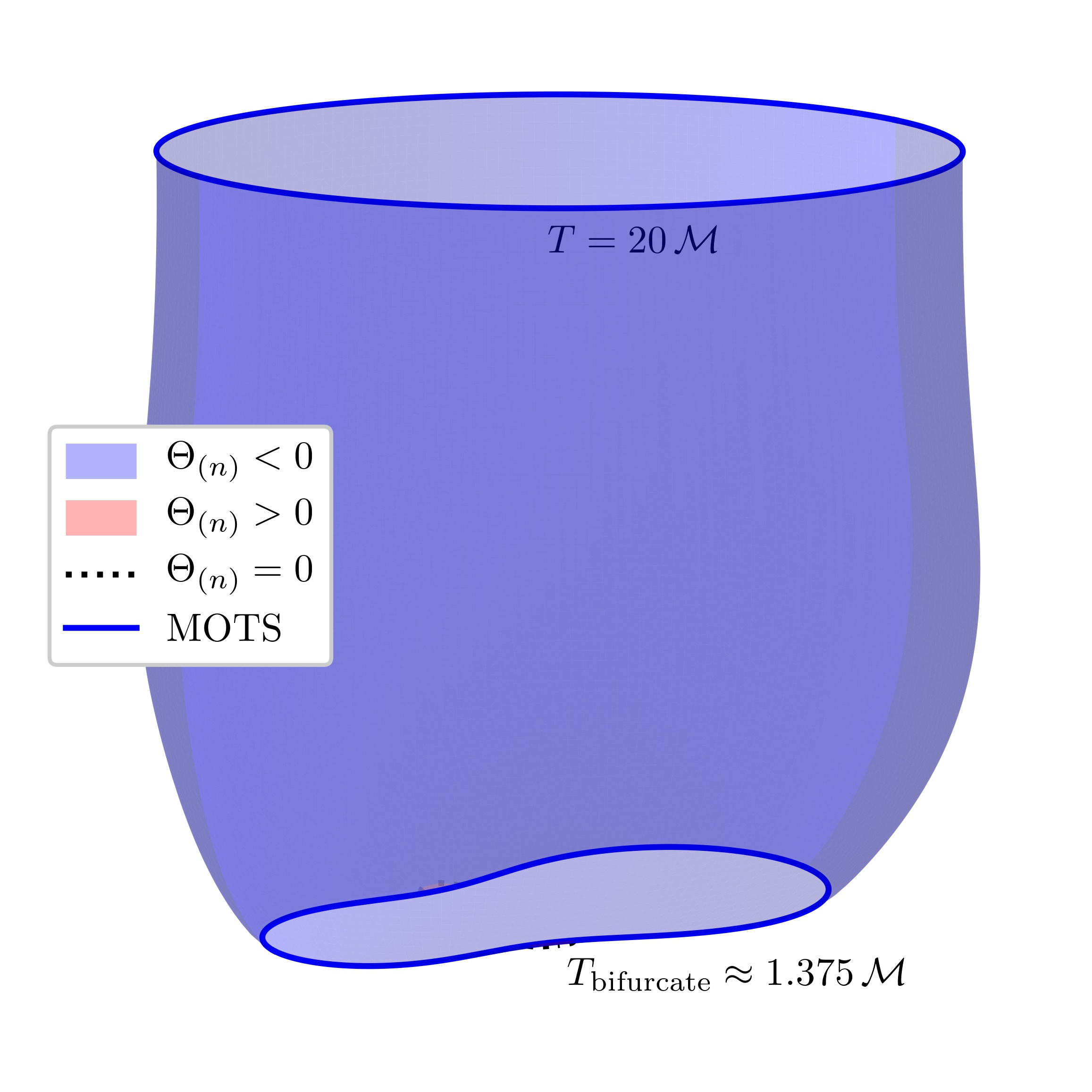}
  \includegraphics[width=\columnwidth]{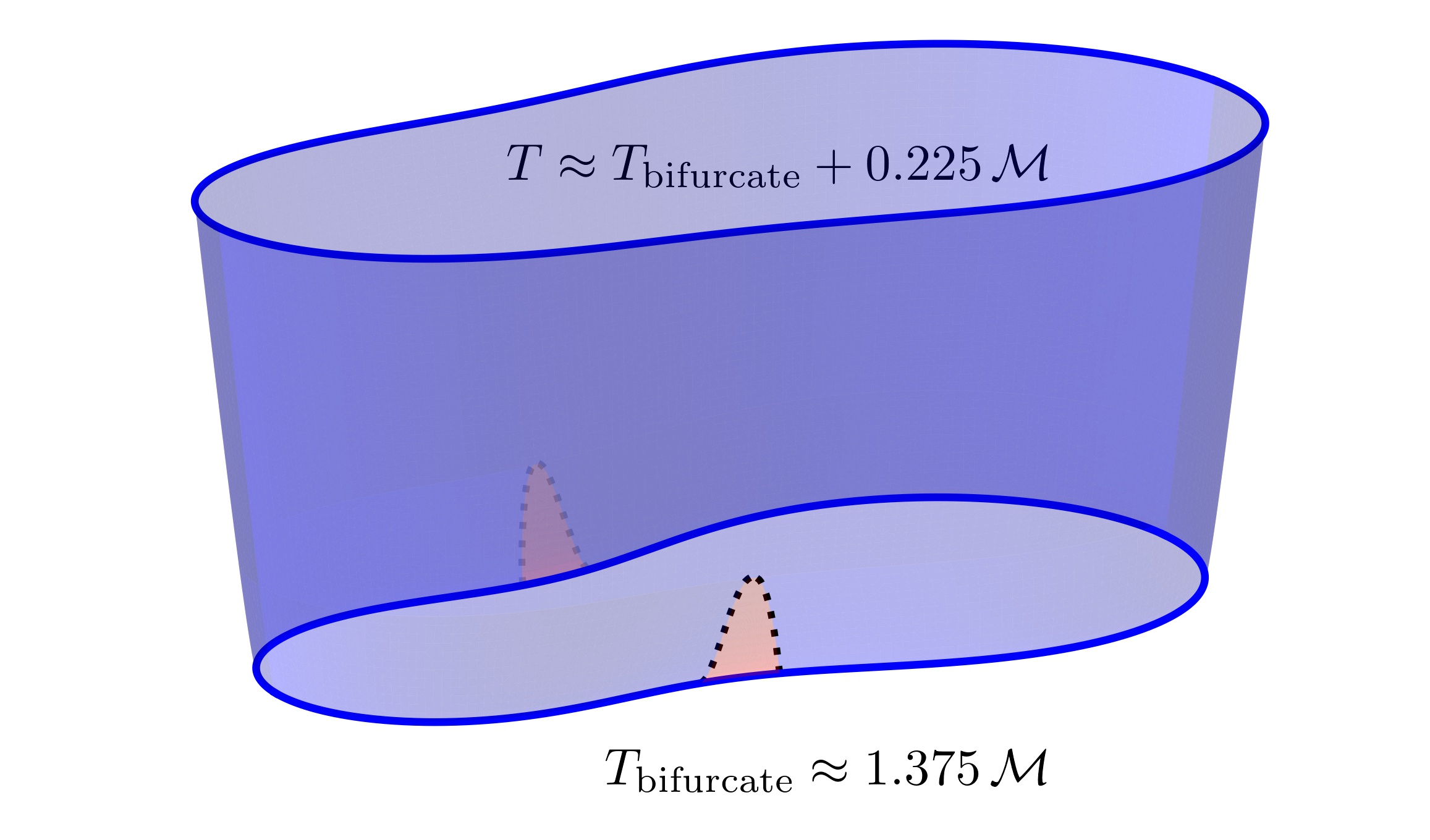}
  \caption{The sign of $\Theta_{(n)}$ for $\Hout$.  Upon formation,
    there is a small portion around the waist that has
    $\Theta_{(n)}>0$.  This portion soon disappears after which time
    the outer dynamical horizon is a conventional spacelike
    future-trapped dynamical horizon.  The second panel shows a zoom
    of $\Hout$ just after it is formed.}
  \label{fig:expansion-AH}
\end{figure*}

For $\Hout$, this portion soon disappears and we have the conventional
$\Theta_{(n)}<0$ at all points on the apparent horizons after
this. See Fig.~\ref{fig:expansion-AH}.  A similar feature was also
seen in \cite{Gupta:2018znn}.  Evidently, the small portion with the
``wrong'' sign of $\Theta_{(n)}$ does not affect the area increase law
for $\Hout$; the relevant portion of the horizon is too small to have
an overall effect and the apparent horizon area is monotonically
increasing.

As might be expected, the inner horizon $\Sin$ is yet more
interesting.  As shown in Fig.~\ref{fig:expansion-inner}, $\Sin$ never
truly becomes future marginally trapped, i.e. it always has a portion
(around its ``waist'') with positive $\Theta_{(n)}$.  This shrinks
with time and eventually vanishes momentarily at $\ttouch$, but
reappears immediately afterwards.  Thus we see that $\Hin$ is never
truly future marginally trapped.  However, just after $\tbifurcate$,
$\Sin$ is decreasing in area (as it should) despite this effect.

We see then that \emph{dynamical horizons generally do not have
  $\Theta_{(n)}<0$ everywhere} -- this is the takeaway message from
this section.  This condition has been widely used in previous
literature on quasi-local horizons to prove the area increase law.  In
particular, it is used in the proof of the area increase law for
future dynamical horizons of arbitrary signature
\cite{Bousso:2015mqa,Bousso:2015qqa}.  Thus, these proofs are not
directly applicable for $\Hin$ and there is, strictly speaking, no
contradiction.  However, there is a more subtle reason why the proof
of \cite{Bousso:2015mqa,Bousso:2015qqa} does not apply to $\Hin$, and
we shall return to this point shortly.

\section{The signature of the dynamical horizons}
\label{sec:signature}

Given a MOTS and a dynamical horizon $\HH$ obtained by evolving it,
the signature of $\HH$ can be computed in a straightforward way.  For
any point $p$ on the world tube, we can construct three linearly
independent vectors tangent to $\HH$, say $\mathbf{e}_A$ ($A=1,2,3$)
and find their inner-products
$q_{AB} := \mathbf{e}_A\cdot\mathbf{e}_B$.  The eigenvalues of
$q_{AB}$ then yield the signature; if it has a negative eigenvalue it
is timelike, and it is spacelike if all eigenvalues are positive.  If
the matrix is degenerate then $\HH$ is a null surface.

\begin{figure}
  \centering    
  \includegraphics[width=0.7\columnwidth]{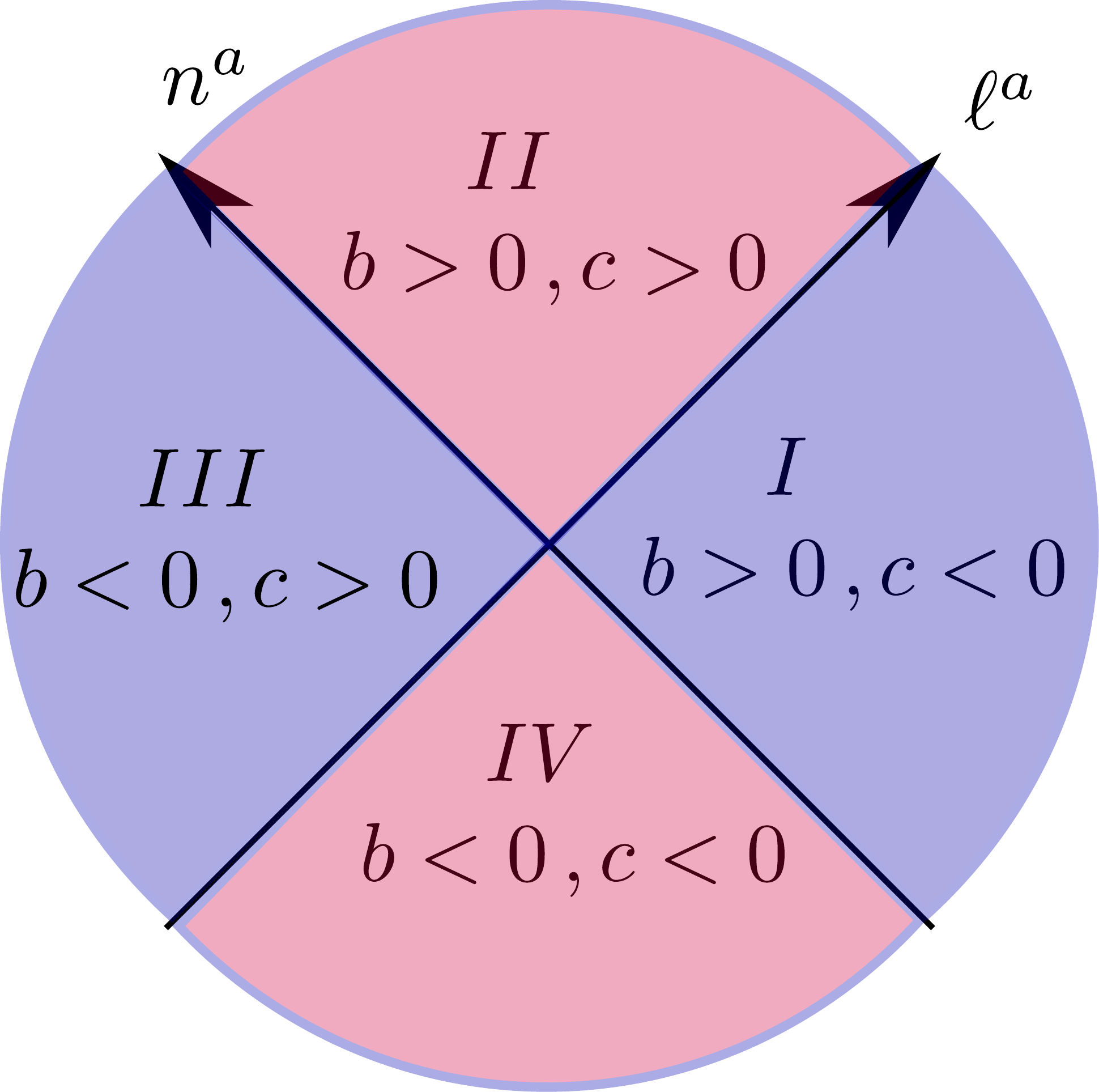}
  \caption{The four different types of time evolution vector fields on
    a generic dynamical horizon. The right and left quadrants, $\qI$ and
    $\qIII$ respectively, refer to spacelike dynamical horizons. The top
    and the bottom quadrants, $\qII$ and $\qIV$ respectively, are timelike.
    Horizons in the right quadrant are moving outwards while those in
    the left quadrant move inwards. We do not have any horizons in the
    bottom quadrant ($\qIV$) since in our simulations $V^a$ can never be
    past directed.}
  \label{fig:signature-types}
\end{figure}

Alternatively, we can consider properties of a ``time evolution''
vector on the dynamical horizon.  Consider a dynamical horizon
$\HH$ of arbitrary signature and arbitrary $\Theta_{(n)}$.
Let $V^a$ be a vector field on $\HH$ such that it is
orthogonal to the leaves of the MOTSs which constitute $\HH$,
and it maps one foliation to the next. Thus, if on a dynamical horizon
the MOTSs are labeled by a parameter $\lambda$, then we can choose
$\lambda$ such that $V^a\partial_a\lambda=1$. Each MOTS $\Surf$ is
taken to lie on a given Cauchy surface and thus equipped with null
normals $(\ell^a,n^a)$ according to Eq.~(\ref{eq:normals}).  Since
$V^a$ is orthogonal to $\Surf$, there must exist functions $b$ and $c$
on $\Surf$ such that
\begin{equation}\label{eq:tev_expanded}
  V^a = b\ell^a + cn^a\,.
\end{equation}
Since $\ell\cdot n = -1$, we have $V\cdot V = -2bc$.  Thus, the
signature of $\HH$ is controlled by the sign of $bc$;
$\HH$ is spacelike if $b$ and $c$ have different signs, and
timelike if they have the same signs. Readers more familiar with $T^a$
and $R^a$ might find the following expression for $V^a$ more
illuminating:
\begin{equation}
  \sqrt{2}V^a = (b+c)T^a + (b-c)R^a\,. 
\end{equation}
We identify the 4 cases shown in Fig.~\ref{fig:signature-types}
depending on the signs of $b$ and $c$.
If $b$ and $c$ are both positive, then intuitively, the term
along $T^a$ dominates and is positive and thus $V^a$ is timelike.
Similarly, when $b$ and $c$ have opposite signs, the term along $R^a$
dominates. When $b>c$, i.e. when $b>0$ and $c<0$, $V^a$ points
outwards (i.e. along $R^a$ by definition), and in the opposite case
when $b<c$ then it points inwards.  We note that in our simulations,
by construction the MOTS is found on Cauchy surfaces referring to a
given time, and we essentially construct $V^a$ by connecting a MOTS at
a given time with another MOTS at a later time.  This means that when
it is timelike, $V^a$ can never be past directed and thus case $\qIV$
cannot occur in our simulations.  

The null case corresponds to either of $b$ or $c$ vanishing, and it is
usually assumed that this does not occur on open sets (this is borne
out in our numerical results).  The null portions arise when
$\HH$ transitions between any two of the four cases listed
above.  Furthermore, only one of the signs can change in a transition;
the vanishing of both $b$ and $c$ means that $V^a$ vanishes which
cannot happen as long as the foliation of the MOTT is regular.  If
$\twoepsilon$ is the area 2-form on $\Surf$, and when $\Surf$ is a
MOTS so that $\Theta_{(\ell)}=0$, then
\begin{equation}
  \label{eq:ctheta_n}
  \mathcal{L}_V\twoepsilon = c\Theta_{(n)}\twoepsilon\,.
\end{equation}
Locally, the area increase is determined by the product
$c\Theta_{(n)}$.  Thus, $c$ plays a double role: the product $bc$
determines the signature while the product $c\Theta_{(n)}$ determines
the change in area.  $V^a$ is strictly null only on a set of measure
zero.  As discussed earlier, the null case is nevertheless very
important for conceptual reasons. All of the well known stationary
Kerr and Schwarzschild horizons are null. Moreover, $\Hout$ has $b$
large and positive, and $c$ small but negative in the limit of late
times as it reaches equilibrium, so that
$V\cdot V = -2bc \gtrapprox 0$.  The same holds for $\Hone$ ad $\Htwo$
at early times.

\begin{figure*}
  \centering    
  \includegraphics[width=\columnwidth]{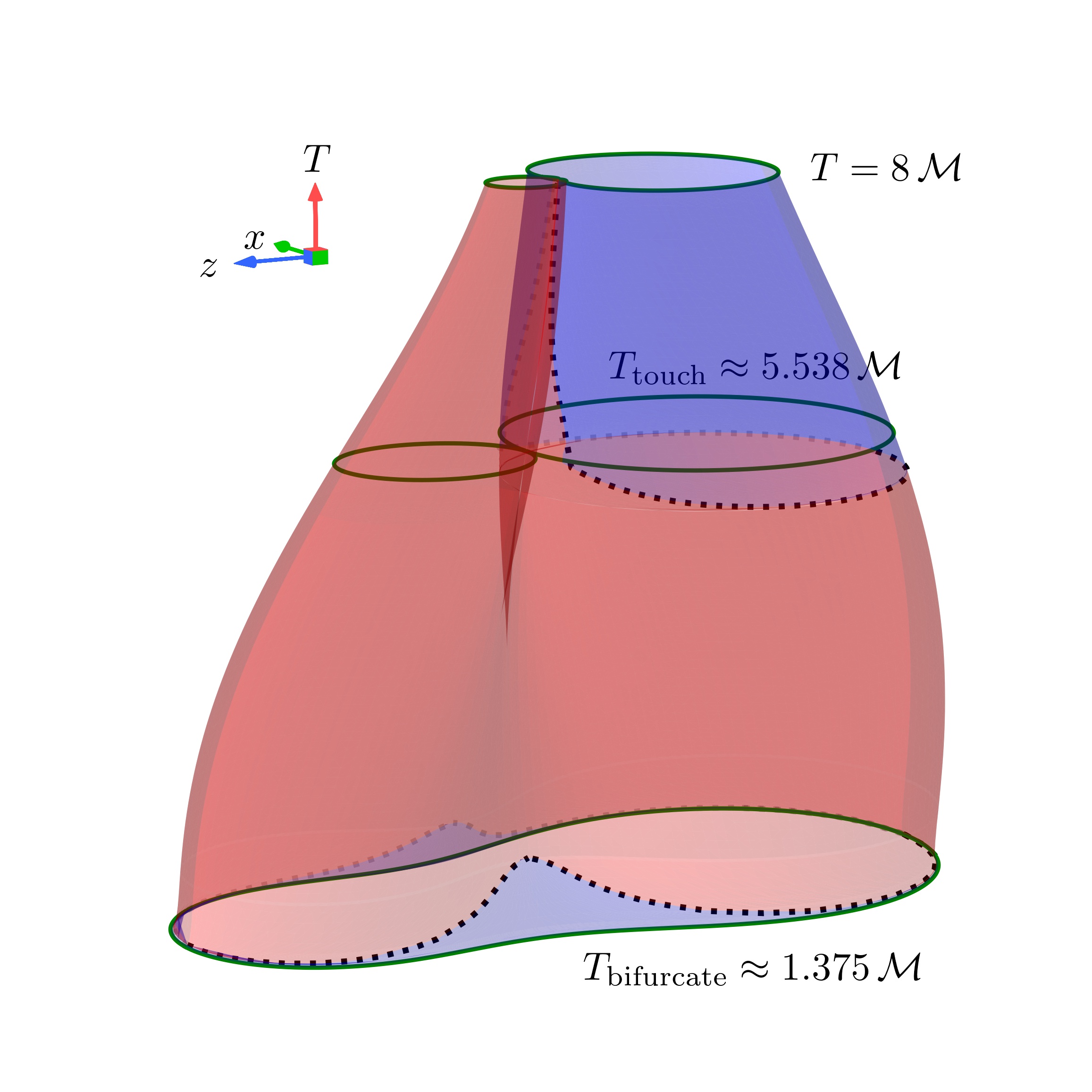}
  \includegraphics[width=\columnwidth]{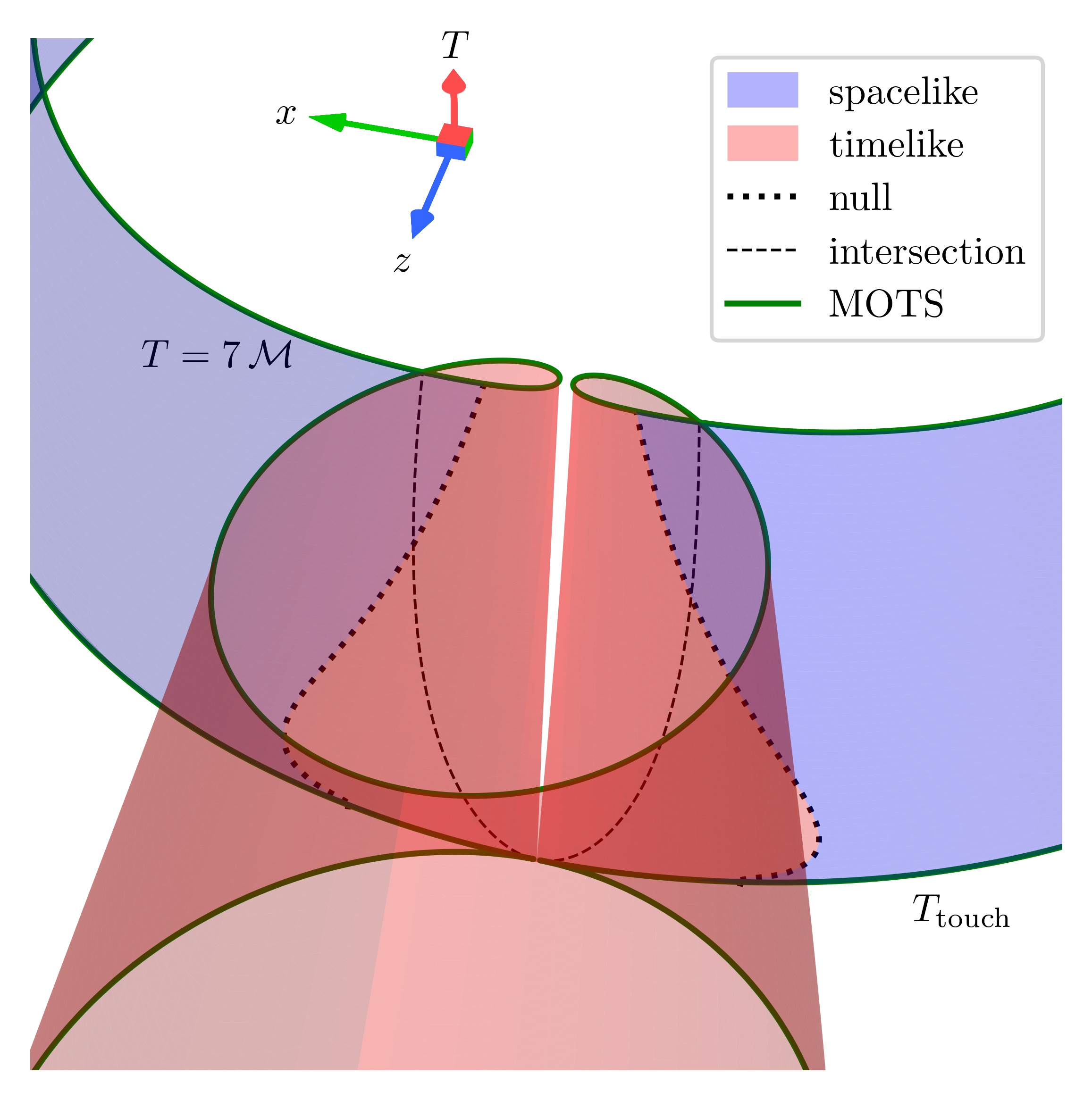}
  \caption{The signature of the inner-horizon $\Hin$.  Blue shows the
    spacelike portions and red is timelike. The null portions where
    the horizon transitions between spacelike and timelike are shown
    as dotted lines.  Upon formation (at $\tbifurcate$) $\Hin$ is
    entirely spacelike. This phase however lasts for a very short time
    (and is not easy to make out in the figure).  It develops timelike
    portions and soon becomes fully timelike.  The portions around the
    ``waist'' persist in remaining spacelike for the longest.  At a
    later time, a little bit before $\ttouch$, the part of $\Hin$
    surrounding the larger black hole reverts to being spacelike.  The
    right panel shows details near the self-intersection.  The thin
    dashed curve is the location of the self-intersection. }
  \label{fig:signature}
\end{figure*}

\begin{figure}
  \centering
  \includegraphics[width=\columnwidth]{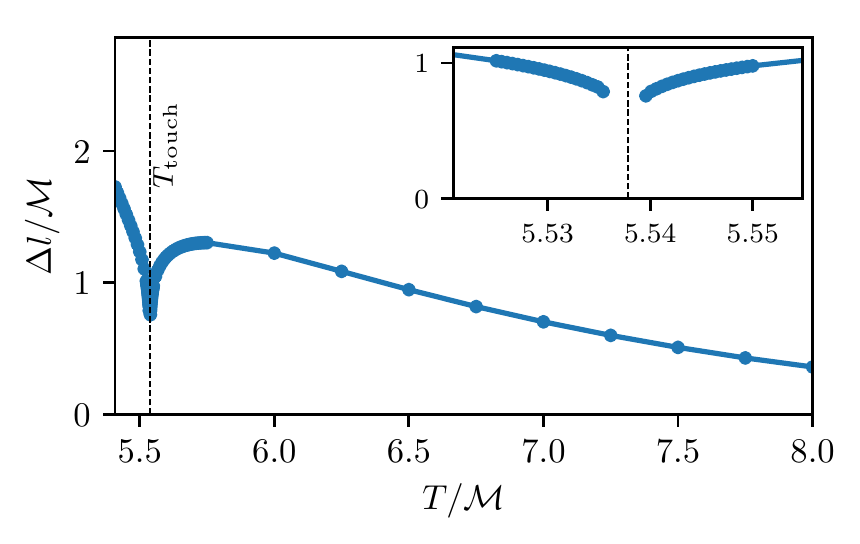}
  \caption{%
    Proper length $\Delta l$ of the curve segment connecting the waist of
    $\Sin$ to the point where the signature changes from
    timelike to spacelike.
    This corresponds to the proper distance measured along the MOTS
    when going
    from the waist to the dotted line in Fig.~\ref{fig:signature}.
    We cannot numerically resolve whether this signature change happens
    precisely at the cusp (i.e. $\Delta l = 0$) when $\tname = \ttouch$.
  }
  \label{fig:signature-null-to-neck}
\end{figure}

With this understanding, we can now present our results regarding the
signature.  $\Hone$, $\Htwo$ and $\Hout$ always turn out to be
spacelike; this is consistent with them being stable in the sense of
the principal eigenvalue of the stability operator being positive
\cite{Andersson:2007fh} (this will be discussed further in paper II).
Only $\Hin$ shows interesting behavior in this regard.  The signature
of this world tube is shown in Figs.~\ref{fig:signature}.  As
discussed earlier, this horizon develops cusps and self intersections.
From Fig.~\ref{fig:signature} we see clearly that $\Hin$ is mostly
timelike, but there are interesting and non-negligible portions which
are spacelike.  When it is initially formed at $T_{\rm bifurcate}$, it
is completely spacelike; it must of course agree there with $\Hout$
which is always spacelike.  However, it remains fully spacelike for
only a few time-steps after which most portions become timelike; the
region around the ``waist'' remains spacelike for the longest.  After
this, $\Hin$ remains entirely timelike until just before $\ttouch$
when the portion around the \emph{larger} black hole develops
spacelike portions.  The second panel of Fig.~\ref{fig:signature}
shows a close-up of a portion around the self-intersecting knot.  We
see there is only one change of signature as we traverse each knot.
Fig.~\ref{fig:signature-null-to-neck} shows the distance of the waist
to the point where this change happens.  The region around the knot is
becoming increasingly spacelike.

We can dissect this behavior further in terms of the functions $b$ and
$c$ introduced earlier.  The apparent horizon at late times is easiest
to understand: $V^a$ is spacelike and outward pointing, but strongly
tilted towards $\ell^a$. Thus, $c$ is small and negative, and $b$ is
positive which means that, at late times, the dynamical horizon
$\Hout$ generated by $\Sout$ must be of type $\qI$.  The same holds
for $\Sone$ and $\Stwo$ at early times. Closer to $\tbifurcate$ when
$\Sout$ is growing rapidly, we must have $b>0$ and $c<0$, but $c$ will
not be small.  The inner horizon is radically different.  As it is
born, it moves \emph{inwards} rapidly and it is spacelike: it has
$b<0$ and $c>0$ and is of type $\qIII$.  The spacelike portions at
early times shown in Fig.~\ref{fig:signature} (at the bottom of the
world tube) are of type $\qIII$.  This world tube soon becomes
timelike of type $\qII$ wherein $b>0$ and $c>0$.  This continues till
we approach $\ttouch$.  Shortly before $\ttouch$, a part of $\Sin$
again becomes spacelike: this is the portion which envelops the larger
MOTS $\Stwo$.  However, in this spacelike portion it turns out that we
have $b>0$ and $c<0$, i.e. it is of type $\qI$.  The inner horizon
thus shows the following transitions:
$\qIII\rightarrow \qII \rightarrow \qI\, ({\rm partially}$).  It is in
fact this spacelike portion of type $\qI$ which is responsible for the
anomalous area increase shown in the right panel of
Fig.~\ref{fig:radii}.

To explain this, we need to go back to the Bousso-Engelhardt proof of
the area increase law \cite{Bousso:2015mqa,Bousso:2015qqa}.  A key
intermediate result in this work is Theorem IV.2 of
\cite{Bousso:2015qqa} which shows that $c$ cannot change sign.  This
would seem to rule out the transition $\qII \rightarrow \qI$ described
in the previous paragraph.  However, this proof requires the existence
of a MOTS which has $c<0$ everywhere, i.e. it requires that the
spacelike portion contains at least one complete MOTS.  We see that
around $\ttouch$, $\Hin$ has complete MOTSs in the timelike portion,
but none in the spacelike portion.  Moreover, after $\ttouch$, $\Hin$
violates another requirement assumed in
\cite{Bousso:2015mqa,Bousso:2015qqa}, namely that each MOTS should
have disjoint ``inside'' and ``outside'' regions.  Thus, again, there
is no contradiction with the proof.  Further details can be found in
Appendix \ref{appendix:bousso-engelhardt}.  The appendix also shows
that a straightforward extension of the proof to our case does not
work.

\section{The anomalous area increase and the membrane paradigm}
\label{sec:anomalous}

In this final section, we indulge in some speculations on the
anomalous area increase.  We have mentioned briefly earlier that
because of the relation between horizon area and entropy, the
anomalous area increase of $\Hin$ might be physically significant.
One approach where this might play a role is in the fluid-gravity
correspondence.  For black holes in $d+1$-dimensional anti-deSitter
space, it is suggested in \cite{Bhattacharyya:2008jc} that the area
increase law for the event horizon has a dual description in terms of
an ``entropy current'' defined for a relativistic fluid living on the
$d$-dimensional boundary.  The calculation presented in
\cite{Bhattacharyya:2008jc} is perturbative, where many of the
complications of full non-linear general relativity, such as those we
have studied, do not arise.  In order to extend this correspondence to
non-perturbative situations, it has been argued that the event horizon
might not be the appropriate concept, and one should consider
dynamical horizons instead \cite{Booth:2011qy,Figueras:2009iu}.
Indeed, because of the teleological and non-local nature of the event
horizon, it would be unusual if its properties could be mapped to a local
hydrodynamics description (except in situations where it can be
treated perturbatively).

\begin{figure*}
  \centering
  \includegraphics[width=0.45\linewidth]{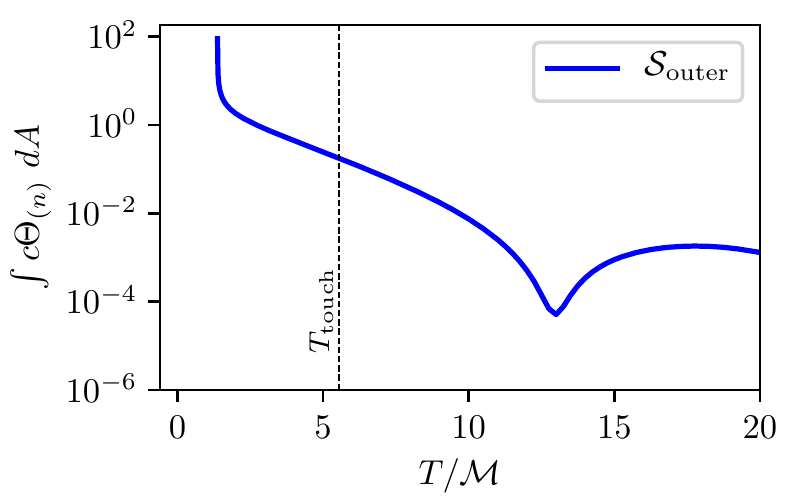}%
  \includegraphics[width=0.45\linewidth]{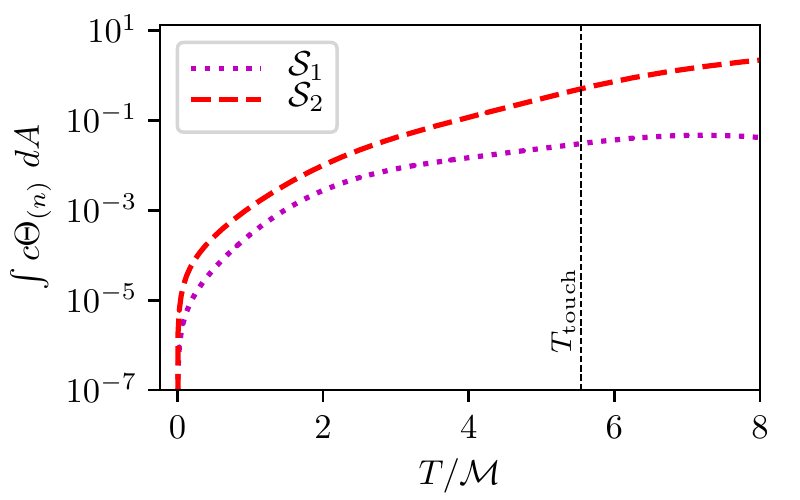}
  \caption{%
    Plot of the integral of $c\Theta_{(n)}$ as a function of
    time for $\Sout$ (left panel) and $\Sone$, $\Stwo$ (right panel),
    on a logarithmic scale. As explained in the text, this quantity is
    essentially the rate of change of the area.  As expected, and
    consistent with Fig.~\ref{fig:radii}, this shows that the area of
    $\Sout$ increases rapidly on formation (just after $\tbifurcate$)
    and then settles down.  The dip near $\sim 13\,\MM$ will be
    explained in paper II.  The individual horizons $\Sone$ and $\Stwo$
    show the opposite behavior, i.e. increasing rapidly only as
    $\ttouch$ is approached. }
  \label{fig:ctheta_n_outer}
\end{figure*}

\begin{figure*}
  \centering    
  \includegraphics[width=0.45\linewidth]{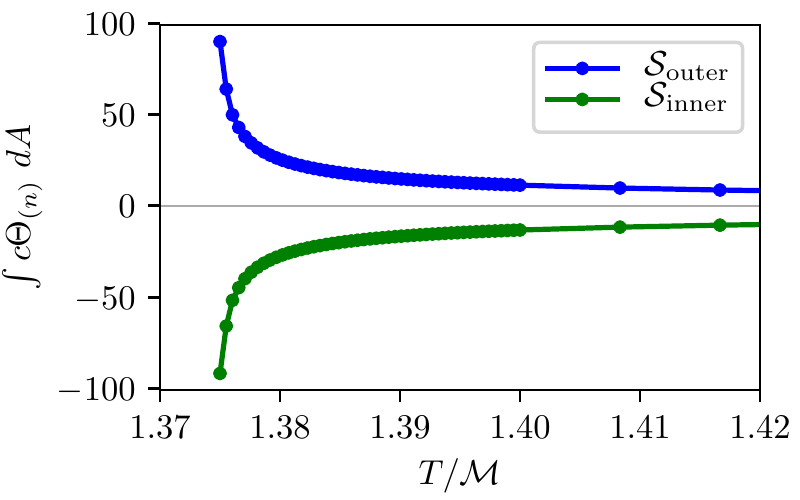}%
  \includegraphics[width=0.45\linewidth]{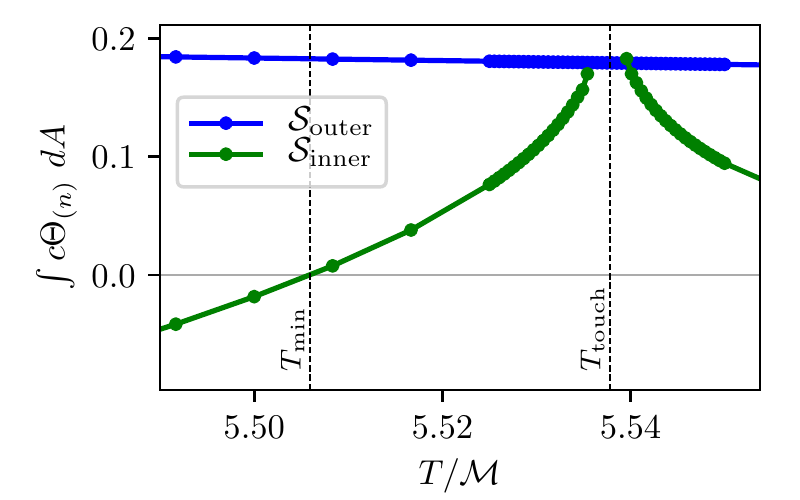}
  \caption{ Plot of the integral of $c\Theta_{(n)}$ for the inner
    horizon $\Sin$.  The left panel shows the values just after
    $\tbifurcate$ when $\Sin$ is formed and rapidly decreases in area,
    while the right panel shows values near $\ttouch$.  The results are
    consistent with the area results. In particular, as we see from
    the second panel of Fig.~\ref{fig:radii}, the area of $\Sin$ has a
    minimum at $\tmin$.  In the second panel of this plot, we get a
    zero at $\tmin$, consistent with it being the rate of change of
    the area.  }
  \label{fig:ctheta_n_inner}
\end{figure*}

At present, a viable proposal for the dual entropy current for
dynamical horizons is lacking.  We suggest that binary mergers might
provide an interesting test case.  If each of the horizons $\Hone$ and
$\Htwo$ at early times have a dual hydrodynamics description and so
does the final horizon $\Hout$ at late times, then the overall
increase in area might be viewed as the increase in entropy due to the
interaction and mixing between the two fluids.  As we have detailed in
this paper, the inner horizon $\Hin$ provides the link between the
initial and the final states.  Thus, if such a dual description is
generally viable then $\Hin$, and in particular the quantity
$c\Theta_{(n)}$ appearing in Eq.~(\ref{eq:ctheta_n}), is likely to play
an important role.  The product $c\Theta_{(n)}$ yields the time
derivative of the area according to Eq.~(\ref{eq:ctheta_n}) averaged
over the horizons:
\begin{equation}\label{eq:dotA}
  \oint_{\Surf}\mathcal{L}_V\twoepsilon = \dot{A}_{\Surf} = \oint_{\Surf}c\Theta_{(n)}\,dA \,.
\end{equation}
The integrals of $c\Theta_{(n)}$ as functions of time for $\Hone$,
$\Htwo$ and $\Hout$ are shown in Fig.~\ref{fig:ctheta_n_outer} and
these are consistent with Fig.~\ref{fig:radii}.  Similarly,
Fig.~\ref{fig:ctheta_n_inner} for the inner horizon is consistent with
the right panel of Fig.~\ref{fig:radii}; the minimum of the area in
the right panel of Fig.~\ref{fig:radii} is consistent with the zero of
$\int c\Theta_{(n)}$ in Fig.~\ref{fig:ctheta_n_inner}.  While these
results are guaranteed mathematically, it is still a useful numerical
check since Eq.~(\ref{eq:dotA}) is an independent calculation of
$\dot{A}_\Surf$.

There exists in fact a different description of a black hole horizon
in terms of fluids, namely that arising in the membrane paradigm
mentioned in Sec.~\ref{subsec:dhbasics}.  As mentioned there, the
analogy between fields on black hole horizons and a 2-dimensional
fluid also works for $\HH$.  One of these quantities is the energy
density which, it turns out, is proportional to the expansion of
$V^a$.  Since $\Theta_{(V)} = c\Theta_{(n)}$, we see that the energy
density is proportional to $c\Theta_{(n)}$.  The interpretation of
$c\Theta_{(n)}$ as an energy density means that $\Sin$ has large
negative energy when it is formed, and its energy becomes positive
after $\tmin$.

To explain this more fully, we revisit the discussion of the
quasi-local membrane paradigm in
\cite{Gourgoulhon:2005ch,Gourgoulhon:2006uc,Gourgoulhon:2008pu},
adapting it to MOTTs of arbitrary signature. Given a hypersurface
$\HH$, we introduce the orthogonal vector $W^a$ (compare with $V^a$ in
Eq.~(\ref{eq:tev_expanded}))
\begin{equation}\label{eq:W}
  W^a = b\ell^a - cn^a\,,
\end{equation}
satisfying $V\cdot W=0$, $W\cdot W =2bc=-V\cdot V$.  The evolution of
its expansion $\Theta_{(W)} = b\Theta_{(\ell)} - c\Theta_{(n)}$ along
$\HH$ is written as (cf. e.g.  \cite{Booth:2006bn,Cao:2010vj})
\begin{eqnarray}
  \mathcal{L}_V\Theta_{(W)} &=& \kappa^{(V)} \Theta_{(V)} -\frac{1}{2}\Theta_{(V)}\Theta_{(W)}
  -\sigma^{(V)}_{ab}\sigma_{(W)}^{ab} \nonumber \\
  &&- G_{ab}V^aW^b
  + (\mathcal{L}_Vb) \Theta_{(\ell)} -  (\mathcal{L}_Vc) \Theta_{(n)}  \nonumber \\
  &&+  \mathcal{D}^a\left(b\mathcal{D}_ac - c\mathcal{D}_ab - 2bc\;\omega_a \right)\,,
\end{eqnarray}
where $\kappa^{(V)}=-n^bV^a\nabla_a\ell_b$,
$\sigma^{(V)}_{ab}= b\sigma_{ab}^{(\ell)} + c\sigma_{ab}^{(n)}$,
$\sigma^{(W)}_{ab}= b\sigma_{ab}^{(\ell)} - c\sigma_{ab}^{(n)}$ and
$G_{ab}$ is the Einstein tensor. First we note that, if ${\cal H}$ is
a smooth event horizon (so in particular a null hypersurface), by
making $b=1$, $c=0$ so that $V^a=W^a=\ell^a$, we immediately recover
the null Raychaudhuri equation.  This equation was interpreted by
Price and Thorne in \cite{Price:1986yy,Thorne:1986} as an energy
balance law by introducing the surface energy density
$\varepsilon =-\Theta_{(W)}/8\pi= -\Theta_{(\ell)}/8\pi$.  For a
dynamical horizon, namely with $\Theta_{(\ell)}=0$ and thus
$\Theta_{(W)}=-\Theta_{(V)}$, we can write
\begin{eqnarray}\label{eq:evol_theta_V}
  \mathcal{L}_V\Theta_{(V)} + \Theta_{(V)}^2 &=& -\kappa^{(V)} \Theta_{(V)} +\frac{1}{2}\Theta_{(V)}^2
  +\sigma^{(V)}_{ab}\sigma_{(W)}^{ab} \nonumber \\
  &&+ G_{ab}V^aW^b
   +  (\mathcal{L}_V\ln c) \Theta_{(V)}  \nonumber \\
  &&+  \mathcal{D}^a\left(c\mathcal{D}_ab - b\mathcal{D}_ac + 2bc\;\omega_a \right)\,.
\end{eqnarray}
Identifying again $\varepsilon =-\Theta_{(W)}/8\pi$ as a formal
surface energy density, that in the MOTT case translates into
$\varepsilon = \Theta_{(V)}/8\pi= c\Theta_{(n)}/8\pi$, we can
interpret $\dot{A}$ in Eq.~(\ref{eq:dotA}) in terms of a total surface
energy $\mathcal{E}$
\begin{equation}
  \mathcal{E} = \oint_{\Surf}\varepsilon \;\twoepsilon  = \frac{1}{8\pi}\oint_{\Surf}\Theta_{(V)}\twoepsilon
  = \frac{1}{8\pi}\oint_{\Surf}c\Theta_{(n)}\twoepsilon
=\frac{\dot{A}_{\Surf}}{8\pi}\,.
\end{equation}
The rate of change of $\mathcal{E}$
\begin{equation}
  \dot{\mathcal{E}} = \frac{\ddot{A}_{\Surf}}{8\pi} =
  \oint_{\Surf}\mathcal{L}_V(\varepsilon\twoepsilon) =\oint_{\Surf}\left(\mathcal{L}_V\varepsilon
  + \Theta_{(V)}\varepsilon\right)\twoepsilon\,,
\end{equation}
is controlled by Eq.~(\ref{eq:evol_theta_V}). This can be cast as an
energy balance law
\begin{eqnarray}\label{eq:energy_balance_equation}
  \mathcal{L}_V\varepsilon + \Theta_{(V)}\varepsilon &=& -\left(\frac{\kappa^{(V)}}{8\pi}\right) \Theta_{(V)}
  +\Theta_{(V)}\left(\frac{\Theta_{(V)}}{16\pi}\right) \nonumber \\
  &&+\sigma^{(V)}_{ab}\left(\frac{\sigma_{(W)}^{ab}}{8\pi}\right) 
  + \Pi - \mathcal{D}_aQ^a\,,
\end{eqnarray}
where $\kappa^{(V)}/8\pi$ is a surface tension (2-dimensional pressure
term), $\Theta_{(V)}$ is the fluid expansion (so that a bulk viscosity
coefficient $\zeta=1/16$ can be identified), $\sigma^{(V)}_{ab}$ and
$\sigma^{(W)}_{ab}/8\pi$ are, respectively, the shear strain and
stress tensors (in general not proportional, so MOTTs do not
correspond to Newtonian fluids and therefore a shear viscosity $\mu$
cannot be defined),
$\Pi:= T_{ab}V^aW^b + \frac{1}{8\pi}(\mathcal{L}_V\ln c) \Theta_{(V)}$
is an external energy production rate (enforced by the Einstein
equations) and
$Q^a:=(b\mathcal{D}^ac - c\mathcal{D}^ab - 2bc\;\omega^a)/8\pi$ is
heat flux.


This fluid description arising in the membrane paradigm is however
only a formal analogy. There is, unlike in the fluid-gravity
correspondence, no deeper interpretation in terms of any dual boundary
description or otherwise.  Nevertheless, it is still interesting that
the analogy goes through for the inner horizon as well. In
Appendix~\ref{appendix:Damour-Navier-Stokes}, we show that the
analogy also extends to spinning black holes, i.e. to include the
rotation 1-form $\omega_a$, which serves to define a momentum density
on $\HH$.  We shall make further use of this analogy in paper II.  In
particular following \cite{Jaramillo:2011rf,Jaramillo:2012rr} in this
viscous fluid picture, the evolution equation for $\Theta_{(V)}$ will
be employed to introduce decay and oscillation timescales leading to a
slowness parameter \cite{Price:2011fm} for the approach of a dynamical
horizon to equilibrium.

\section{Conclusions}
\label{sec:conclusions}

In this paper we have studied geometrical properties of the world tube
of marginally outer trapped surfaces in a binary black hole merger.
This includes the status of the area increase law, and the different
ingredients which go into the rate of change of the area, i.e. the
expansion of the ingoing null normal $\Theta_{(n)}$ and the signature
of the world tube.  We have seen that the horizons are of mixed
signature with various transitions between spacelike and timelike
portions, especially for the inner horizon.  Cross-sections of the
inner horizon can be of mixed signature.  Similarly, the condition
$\Theta_{(n)}<0$ is not satisfied everywhere for the inner horizon.
The anomalous area increase apparently contradicts existing proofs of
the area increase law, in particular the Bousso-Engelhardt result.  We
have argued that technical assumptions required for this proof do not
hold in our case.  We have briefly discussed the anomalous area
increase in terms of the membrane-paradigm analogy using the energy
density of a fictitious 2-dimensional fluid. The deeper physical
significance of the anomalous area increase, if any, is still not
understood. 

The second paper will continue this study and consider physical
quantities such as energy fluxes, multipole moments and the stability
operator on all of these horizons.

\begin{acknowledgments}
  We are indebted to Abhay Ashtekar, Ivan Booth and Ricardo Uribe-Vargas
  for valuable discussions and suggestions.  Research at Perimeter Institute is
  supported in part by the Government of Canada through the Department
  of Innovation, Science and Economic Development Canada and by the
  Province of Ontario through the Ministry of Colleges and
  Universities. We also thank the French EIPHI Graduate School (ANR-17-EURE-0002) and
  the Spanish FIS2017-86497-C2-1 project (with FEDER contribution) for support.
\end{acknowledgments}

\appendix

\section{Comparison with the Bousso-Engelhardt area increase law}
\label{appendix:bousso-engelhardt}

In this Appendix, we show that the anomalous area increase does not
violate any of the existing proofs of the area increase laws.  The
most general proof to-date is due to Bousso \& Engelhardt
\cite{Bousso:2015qqa,Bousso:2015mqa}, and an extension thereof due to
Sanches \& Weinberg \cite{Sanches:2016pga}.  These proofs formally
apply to future dynamical horizons, or holographic screens in the
terminology of \cite{Bousso:2015qqa,Bousso:2015mqa}. We note that this
proof is in fact an application of the maximum principle for elliptic
operators.  The application of the maximum principle to null-surfaces
was studied previously by Galloway \cite{Galloway:1999ny}. These
methods have previously been applied by Ashtekar \& Galloway to
spacelike dynamical horizons to show e.g. the uniqueness of the
foliation by MOTSs \cite{Ashtekar:2005ez}.  We shall discuss the proof
in more detail below, but roughly speaking, these results would
naively indicate that the inner horizon should have decreasing area
and would seem to rule out the anomalous area increase.  While these
results assume $\Theta_{(n)}<0$, this is only a sufficient but not
necessary condition.  The regions on the horizon with $\Theta_{(n)}>0$
are small near $\ttouch$ and do not, by themselves, explain the area
increase.  In other words, we can split the area (and similarly also
its rate of change) as the sum of two terms depending on the sign of
$\Theta_{(n)}$:
\begin{equation}
  A_{\Surf} = \int_{\Theta_{(n)}<0}\,dA + \int_{\Theta_{(n)}>0}\,dA \,.
\end{equation}
The area over the $\Theta_{(n)}<0$ portion turns out to be much larger
than the second term, and it has a correspondingly larger effect on
$\dot{A}_\Surf$.  We saw for example that the area of $\Sout$
increases as expected despite it not having $\Theta_{(n)}<0$
everywhere.

The difficulty, and apparent contradiction, lies elsewhere.  It is the
spacelike portion on the inner horizon near $\ttouch$ that leads to
the anomalous area increase.  A key ingredient of the
Bousso-Engelhardt results is an intermediate step showing that the
function $c$ is not allowed to change sign.  Thus, if we actually had
$\Theta_{(n)}<0$, then $c\Theta_{(n)}$ cannot change sign, and the
area increase law follows directly from Eq.~(\ref{eq:ctheta_n}) (it
must increase with time for regions $\qI$ and $\qIV$, and decrease
with time for regions $\qII$ and $\qIII$).  In the transition from
timelike to spacelike of the inner horizon near $\ttouch$, which we
have seen is a transition $\qII\rightarrow \qI$ of
Fig.~\ref{fig:signature-types}, this is precisely what goes wrong: $c$
goes from positive to negative. Moreover, this intermediate result
only uses the condition $\Theta_{(\ell)}=0$ and does not rely upon
$\Theta_{(n)}<0$.  So how can this intermediate result be consistent
with the results of this paper?  To understand this we need to delve
into some of the technical conditions required in the
Bousso-Engelhardt proof.  We shall not spell out the details of the
proof, but we instead offer a pictorial description which will make
the result plausible.

\begin{figure}
  \centering    
  \includegraphics[width=0.9\columnwidth]{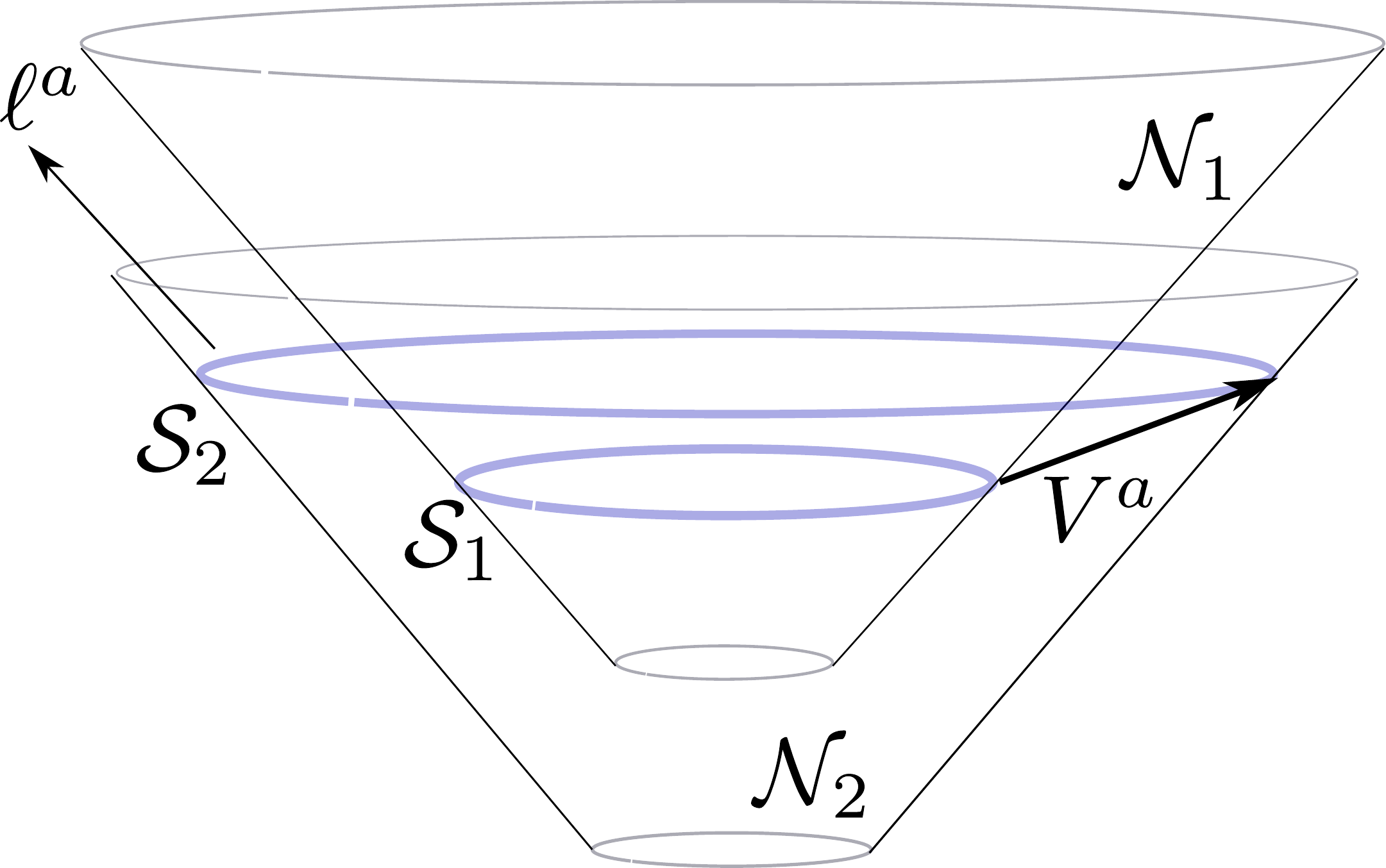}
  \caption{As a MOTS evolves in a spatially outward direction from
    $\Surf_1$ to $\Surf_2$, the null surface $\mathcal{N}_2$ is
    nowhere to the future of $\mathcal{N}_1$.  }
  \label{fig:bousso-proof-null}
\end{figure}

The simplest case which we sketch here is the proof that the
transition $\qI\rightarrow \qII$ is ruled out.  We thus start out with
a MOTS evolving spatially outward, which later partially switches over
to a future timelike direction.  For a MOTS $\Surf$, we construct the
outgoing null surface $\mathcal{N}$ generated by the null-rays
starting from $\ell^a$.  Denote the part of $\mathcal{N}$ to the
future of $\Surf$ by $\mathcal{N}^+$, and the portion to the past by
$\mathcal{N}^-$.  If we move $\Surf_1$ spatially outwards along $V^a$ 
to a new MOTS $\Surf_2$, the null surface $\mathcal{N}_2$ is easily
seen to be nowhere to the future of $\mathcal{N}_1$; see
Fig.~\ref{fig:bousso-proof-null}.  If $V^a$ partially changes to a
future timelike direction, this will eventually cease to hold.

Consider now the dynamical horizon, i.e. the world tube spanned by
$\Surf$ as it evolves along $V^a$.  Let the part of $\HH$ with
$c>0$ be denoted as $\HH^+$, the part with $c<0$ as
$\HH^-$, and the part with $c=0$ as $\HH^0$.  Assume
that $\HH$ has initially a leaf completely of type $\qI$,
i.e. we assume $c<0$ on a complete leaf.  Let $\lambda$ be the affine
parameter along $V^a$, i.e. $V^a\partial_a\lambda=1$, and let
$\lambda=\lambda_0$ at the initial MOTS $\Surf_{\lambda_0}$ (which has
$c<0$).

\begin{figure}
  \centering    
  \includegraphics[width=0.9\columnwidth]{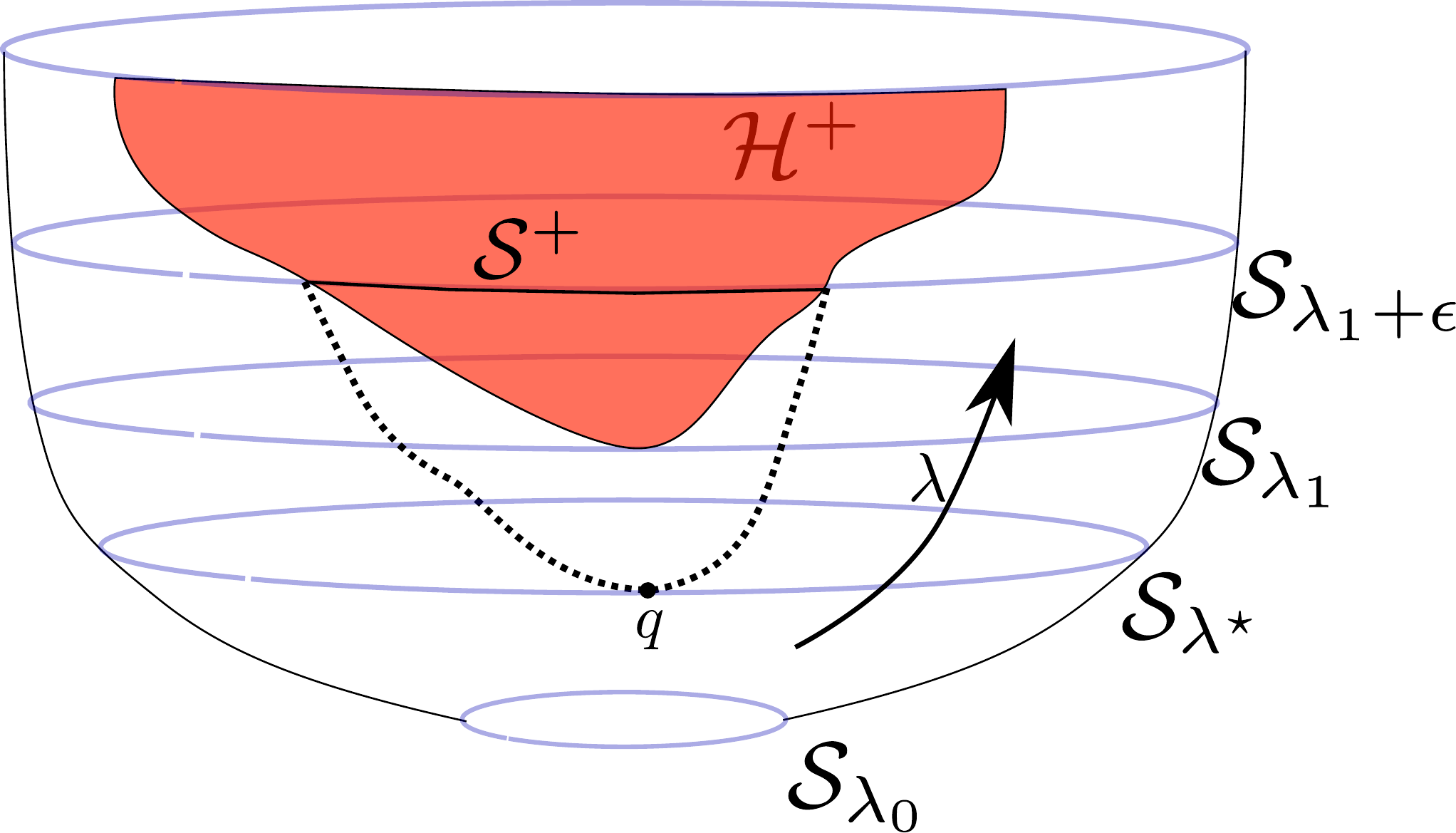}
  \caption{A pictorial sketch of the Bousso-Engelhardt proof. The
    portion of $\HH$ with the offending timelike region of
    type $\qII$ with $c>0$ is $\HH^+$. The rest of the
    horizon is of type $\qI$. The parameter $\lambda$ increases upwards.
    $\Surf_{\lambda_0}$ is the MOTS at $\lambda_0$, and $\lambda_1$ is
    the earliest that $\HH^+$ appears.  $\Surf^+$ is the
    portion of $\Surf_{\lambda_1+\epsilon}$ within
    $\HH^+$. The dashed curve depicts the intersection of
    $\HH$ with the null surface $\mathcal{N}^-$ (generated by
    the past-inward directed null curves along $-\ell^a$ starting from
    $\Surf^+$). The point $q$ is where $\lambda$ has a minimum over
    the dashed curve. At $q$, the null generator of the
    $\mathcal{N}^-$ coincides with the outgoing null normal of the
    MOTS $\Surf_{\lambda^\star}$ which is, by definition, supposed to
    have vanishing expansion.  }
  \label{fig:bousso-1}
\end{figure}

\begin{figure}
  \centering    
  \includegraphics[width=0.9\columnwidth]{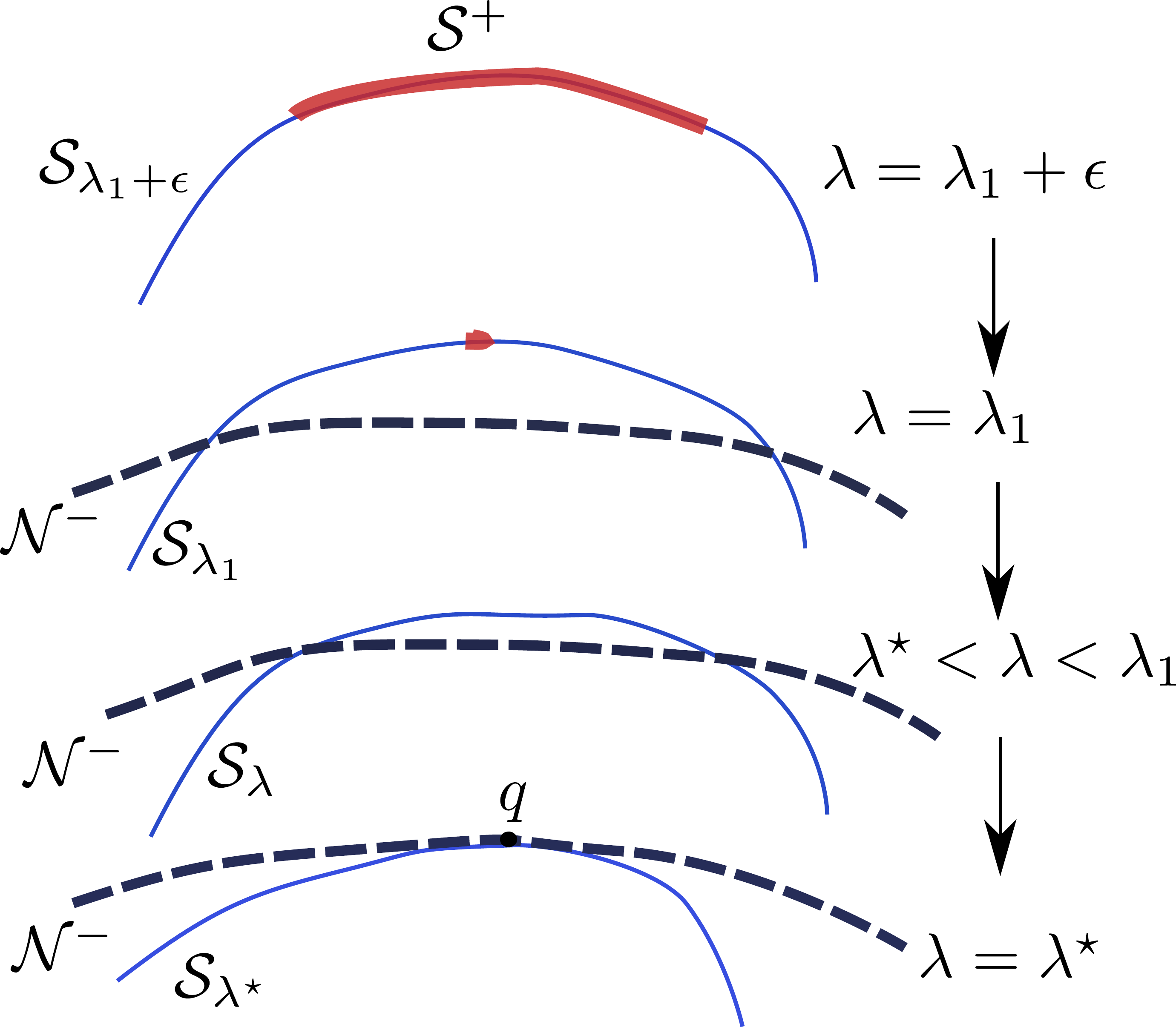}
  \caption{Same as Fig.~\ref{fig:bousso-1}, but now showing how the
    various surfaces would appear on Cauchy surfaces. We start with
    $\lambda_1+\epsilon$ where we start tracing back
    $\mathcal{N}^-$. The color scheme is consistent with
    Fig.~\ref{fig:bousso-1}: the portion in red is the timelike
    portion $\Surf^+$, the blue curve the MOTS, while the dashed curve
    shows the intersection of $\mathcal{N}^-$ with the Cauchy
    surfaces.  The timelike portion will contract slower than the null
    surface, and will thus stay outside the null portion.  Eventually,
    at $\lambda^\star$, the intersection is the point $q$.
    $\Surf_{\lambda^\star}$ is tangent to $\mathcal{N}^-$ at $q$, and
    the rest of it lies ``inside'' $\mathcal{N}^-$. The maximum
    principle is applied in a neighborhood of $q$. }
  \label{fig:bousso-proof-surfaces}
\end{figure}

Let us take $\HH^+$ to lie in the $\lambda>\lambda_0$ region;
this will be shown to lead to a contradiction.  Let
$\lambda_1>\lambda_0$ be the smallest value of $\lambda$ on
$\HH^+$, i.e. when the timelike portion first appears.
Clearly, $\mathcal{N}_{\lambda_1}$ lies nowhere to the future of
$\mathcal{N}_{\lambda_0}$.  We can move further infinitesimally to
$\lambda_1+\epsilon$ still without entering the future of
$\mathcal{N}_{\lambda_0}$.  Consider then the MOTS
$\Surf_{\lambda_1+\epsilon}$ and its subset
$\Surf^+ = \Surf_{\lambda_1+\epsilon}\cap \HH^+$.  Then trace
back the null rays to the past along $\mathcal{N}^-$, and let $k^a$
denote the null generator of $\mathcal{N}^-$.  Let us follow
$\mathcal{N}^-$ back in time and look at its intersection with Cauchy
surfaces corresponding to various values of $\lambda<\lambda_1$.  The
intersection $\mathcal{N}^-\cap \HH$ is generically a curve on
$\HH$.  Let the minimum value of $\lambda$ on this
intersection be denoted $\lambda^\star$.  Since $\epsilon$ is chosen
sufficiently small, $\mathcal{N}^-$ lies to the past of
$\Surf_{\lambda_0}$ so that $\lambda^\star>\lambda_0$. Let
$q\in\Surf_{\lambda^\star}$ be the point where the minimum is
achieved. This construction is shown pictorially in
Figs.~\ref{fig:bousso-1} and \ref{fig:bousso-proof-surfaces}.
Fig.~\ref{fig:bousso-1} depicts the world tube, its timelike portion,
and the intersection with $\mathcal{N}^-$.  The same situation is
shown in Fig.~\ref{fig:bousso-proof-surfaces} in terms of how these
surfaces would appear on Cauchy surfaces at various times.

Then, it can be shown that $\mathcal{N}^-$ is tangent to
$\Surf_{\lambda^\star}$; they share the same null normal at $q$ so
that by a suitable rescaling we can set $k^a=\ell^a$ at $q$.  However,
since $\Surf_{\lambda^\star}$ is ``inside'' $\mathcal{N}^-$, it has
larger curvature, and thus we must have
$\Theta_{(\ell)}(q) \geq \Theta_{(k)}(q)$ where $\Theta_{(k)}(q)$ is
the expansion of $k^a$ at $q$.  From the Raychaudhuri equation, and
assuming the null energy condition and the positivity of
$|\sigma^{(k)}|^2$, it follows that $\Theta_{(k)}(q) >0$.  This
implies then that $\Theta_{(\ell)}(q) \geq \Theta_{(k)}(q) > 0$ which
contradicts $\Theta_{(\ell)}(q) = 0$ (the defining condition for a
MOTS).  Thus, $\HH^+$ cannot exist and the transition
$\qI\rightarrow \qII$ is ruled out.  

A similar argument, works for the transition from region
$\qII \rightarrow \qI$, except we follow the null surface to the
future, and it is required to have a complete MOTS with $c<0$.  The
other forbidden transitions where $c$ changes sign are
$\qIV\rightarrow\qIII$ and $\qIII\rightarrow\qIV$.  In both of these
cases, a direct attempt at applying the above argument does not work.
Instead, one needs to reverse the direction of $V$ so that these are
reduced to the previous two cases.

The details of this proof are given in
\cite{Bousso:2015qqa,Bousso:2015mqa}; see also
\cite{Galloway:1999ny,Ashtekar:2005ez}.  Here we list the technical
conditions and Lemmas which must hold for the above argument to go
through:
\begin{enumerate}
\item Each MOTS $\Surf_\lambda$ which foliate $\HH$ must have
  an ``inside'' and an ``outside'', i.e. if they lie on a Cauchy
  surface $\Sigma$, they must split it into two disjoint portions.
\item The existence of a MOTS $\Surf_0$ which has $c<0$ everywhere --
  required for the point $q$ which minimizes $\lambda$ to exist.  
\item Genericity condition on the zeros of $b$ and $c$,
  $\partial \HH^+ = \HH^0 = \partial \HH^-$ --
  this excludes non-generic zeros of $c$ or $b$ (e.g. the zeros must
  not coincide with extrema, and the functions cannot vanish in an
  open set).
\item On any $\Surf$, $|\sigma|^2$ must be positive definite (in the
  presence of matter, we would include the null energy condition) -
  this ensures the positivity of $\Theta_{(k)}(q)$.  
\item The proof of $\Theta_{(\ell)}(q) \geq \Theta_{(k)}(q)$ is given
  in \cite{Wall:2013uza}; see \cite{Galloway:1999ny} for a more
  general proof with, e.g. weaker smoothness assumptions, explicitly
  using the maximum principle.
 
\end{enumerate}

All of these conditions do hold on $\Sout$, $\Sone$ and $\Stwo$ but
not on $\Sin$ where the first two are violated.  (1) is violated by
the self intersecting MOTSs but this only happens after $\ttouch$ and
is not relevant for the anomalous area increase which occurs
\emph{before} $\ttouch$. The culprit is then condition (2): We see
that near $\ttouch$, we have $c<0$ only over a part of the MOTSs
before $\ttouch$.  Thus, the proof of
\cite{Bousso:2015qqa,Bousso:2015mqa} does not rule out the anomalous
area increase scenario presented above.  Repeating the proof by
dropping the requirement of a complete MOTS with $c<0$, and replacing
it with requiring that there should be a complete MOTS in region
$\qII$ (as is the case here) is seen to not work either.

To show this, let us look explicitly at the case of interest to us,
namely a transition $\qII\rightarrow\qI$ (future-timelike to
outward-spacelike). The transition is partial and we do not have any
section which is entirely spacelike.  We do however have a complete
MOTS in the timelike portion.  The picture is very similar to
Fig.~\ref{fig:bousso-1}, with the timelike and spacelike portions
interchanged.  Thus, $\HH^+$ becomes $\HH^-$ and
$\Surf^+$ can be labeled $\Surf^-$.  Similarly,
Fig.~\ref{fig:bousso-proof-surfaces} can be reused but with the curves
for $\mathcal{N}^-$ and the various $\Surf$ interchanged.  Since
$\Surf^-$ is spacelike, it contracts faster than $\mathcal{N}^-$, and
thus it goes inside $\mathcal{N}^-$.  At $\lambda^\star$, we again
have $\mathcal{N}^-$ tangent to $\Surf_{\lambda^\star}$, but
critically, $\Surf_{\lambda^\star}$ now lies ``\emph{outside}''
$\mathcal{N}^-$. Thus, while we still have $\Theta_{(k)}(q) >0$, but
now $\Theta_{(\ell)}(q) \leq \Theta_{(k)}(q)$, and no contradiction
arises with $\Theta_{(\ell)}(q)=0$.

\section{Damour-Navier-Stokes equation in MOTTs}
\label{appendix:Damour-Navier-Stokes}

For completeness, in the context of the quasi-local membrane paradigm
discussed in section \ref{sec:anomalous}, we present here the equation
for the evolution of the rotation form $\omega_a$, interpreted as a
(Damour)-Navier-Stokes equation for the momentum density of the
two-dimensional fluid.

As we did for $\Theta_{(V)}$, we evaluate now the evolution of
$\omega_a$ along a hypersurface $\HH$
\begin{eqnarray}\label{eq:evol_omega}
  \mathcal{L}_V\omega_a + \Theta_{(V)}\omega_a &=& \mathcal{D}_a\kappa^{(V)}
  + \frac{1}{2} \mathcal{D}_a \Theta_{(V)}
  - \mathcal{D}_b{\sigma^{(W)b}}_a  \\
    &&+ {q_a}^bG_{bc}W^c
  - \Theta_{(\ell)}\mathcal{D}_ab + \Theta_{(n)}\mathcal{D}_ac\,. \nonumber
\end{eqnarray}
Making first $b=1$ and $c=0$, i.e. $V^a=W^a=\ell^a$, we recover the evolution equation
of the rotation form on a null hypersurface \cite{Gourgoulhon:2005ng}, in particular the
one satisfied on a general (smooth) event horizon
\begin{eqnarray}
 \mathcal{L}_V\omega_a + \Theta_{(\ell)}\omega_a &=& \mathcal{D}_a\kappa^{(\ell)}
  + \frac{1}{2} \mathcal{D}_a \Theta_{(\ell)}
  - \mathcal{D}_b{\sigma^{(\ell)b}}_a \nonumber \\
    &&+ {q_a}^bG_{bc}W^c\,. 
\end{eqnarray}
Following Damour \cite{Damour:membrane,damour1982surface}, this equation can be interpreted
as a Navier-Stokes equation by defining a momentum surface density $\pi_a = -\omega_a/8\pi$,
leading to a Newtonian viscous fluid picture with 
negative bulk viscosity $\zeta = -1/16\pi$ and shear viscosity $\mu = 1/16\pi$.
In our MOTT case, making $\Theta_{(\ell)}=0$ in Eq.~(\ref{eq:evol_omega}), we
get
\begin{eqnarray}\label{eq:evol_omega_MOTT}
  \mathcal{L}_V\omega_a + \Theta_{(V)}\omega_a &=& \mathcal{D}_a\kappa^{(V)}
  + \frac{1}{2} \mathcal{D}_a \Theta_{(V)}
  - \mathcal{D}_b{\sigma^{(W)b}}_a  \nonumber \\
    &&+ {q_a}^bG_{bc}W^c
   + \Theta_{(V)}\mathcal{D}_a\ln c\,.
\end{eqnarray}
Defining, as in the event horizon case, a  momentum surface density $\pi_a = -\omega_a/8\pi$
we get \cite{Gourgoulhon:2005ch}
\begin{eqnarray}\label{eq:DNS_MOTT}
  \mathcal{L}_V\pi_a + \Theta_{(V)}\pi_a &=& -\mathcal{D}_a\left(\frac{\kappa^{(V)}}{8\pi}\right)
  + \mathcal{D}_a \left(\frac{\Theta_{(V)}}{16\pi}\right)  \nonumber \\
   && + \mathcal{D}_b\left(\frac{\sigma^{(W)b}_a}{8\pi}\right)
   + f_a \,,
\end{eqnarray}
that corresponds to a viscous fluid with the interpretation of the
terms given after Eq.~(\ref{eq:energy_balance_equation}), namely a
Newtonian fluid in the trace part of the viscous stress tensor (since
it is proportional to the trace of the strain tensor, i.e.  the
expansion $\Theta_{(V)}$) with positive bulk viscosity
$\zeta = 1/16\pi$, non-Newtonian in the shear stress tensor part
(since $\sigma^{(W)}_{ab}/8\pi$ is not in general proportional to the
strain shear tensor $\sigma^{(V)}_{ab}$). The term
$f_a := -{q_a}^bT_{bc}W^c - \frac{1}{8\pi}\Theta_{(V)}\mathcal{D}_a\ln
c$, where Einstein equations are imposed, corresponds to an external
force surface density\footnote{To complete this picture, it is interesting
  to note that the principal eigenvalue $\Lambda_o$ of the MOTS stability
  operator, that will be systematically
  studied in Paper II, also admits a membrane paradigm interpretation.
  Namely, it can be seen as the pressure difference at the interphase
between two fluids, in the spirit of a Young-Laplace law \cite{Jaramillo:2013rda}.}.  See \cite{PhysRevD.83.044048} for a critical
account of this viscous fluid interpretation.

Equations (\ref{eq:evol_omega}) and (\ref{eq:DNS_MOTT}) are valid for
general signature MOTTs, where $V^a$ can be proportional $n^a$.  In
contrast with the $\Theta_{(V)}$ evolution and the energy balance
equation discussed in section \ref{sec:anomalous}, where
the ``heat flux'' term $Q^a$ must be generalized,
Eqs.~(\ref{eq:evol_omega}) and (\ref{eq:DNS_MOTT}) coincide exactly
with those in
\cite{Gourgoulhon:2005ch,Gourgoulhon:2006uc,Gourgoulhon:2008pu}, that
were originally restricted to deviations around the outgoing null
vector $\ell^a$.

\section{Contact structures: a tentative bridge from immersed MOTS to
  wavefronts and caustics}
\label{sec:contact_structures}

As observed in Fig. \ref{fig:signature} and fully discussed in Section
VI of \cite{PhysRevD.100.084044}, the inner common horizon $\Sin$
transitions from i) a smooth ``embedded curve'' before time $\ttouch$,
into ii) a singular curve (vanishing of its differential) with cusps
at $\ttouch$, and then to iii) an ``immersed curve'' with
self-intersections (knot) after $\ttouch$.  In a complementary view to
this ``parametrised curve'' perspective, we could approach such a
sequence in terms of projections: the described transition corresponds
indeed to the generic curve metamorphoses (``perestroikas'') happening
when projecting a three-dimensional curve into a plane with the
projection direction changing with a parameter \cite{Uribe04}. Indeed,
as illustrated in \cite{Booth:2020qhb}, such self-intersections seem
to have a ``genericity'' flavor in trapped regions.

Such a projection view suggests a picture with the ``lifted'' curve  as the fundamental
object and self-intersections as an artifact of the projections. The natural question is:
what is the appropriate higher-dimensional space in which MOTSs ``truly'' live? 
A suggestive tentative answer is given in terms of contact structures \cite{Arnold89,Friedrich:1983vi}.

Specifically, and dwelling now beyond axisymmetry, let us consider a $n$-dimensional spacetime $\scrM$
and its cotangent bundle  $T^*\!\!\scrM$
with natural Liouville form $\Lambda = p_adx^a$.
After removing the vanishing one-forms, $T^*\!\!\scrM\backslash\{0^*_M\}$, we  
can take quotient by one-form rescalings with  non-vanishing real numbers ${\mathbb R}^*$. The resulting
space $PT^*\!\!\scrM = (T^*\!\!\scrM\backslash\{0^*_M\})/{\mathbb R}^*$ is the
``projectified cotangent bundle''. Crucially, at each point, the kernel of $\Lambda$ is invariant under such
${\mathbb R}^*$-rescalings and projects onto hyperplanes in the tangent space of $PT^*\!\!\scrM$:
such field of (non-integrable) hyperplanes defines a ``contact structure'' on $PT^*\!\!\scrM$. In
a local chart $(x^a,p_a)$, with say $p_n\neq0$, we can consider the ``affine'' chart $(x^i,p_i,z)$,
$i\in\{1,\ldots,n-1\}$ with $z=x^n, p_n=1$. Then $\Lambda'= p_idx^i+dz$ is the contact form
of the $(2n-1)$-contact manifold $PT^*\!\!\scrM$ (cf. \cite{Friedrich:1983vi} for a discussion in a
general relativistic setting; see also \cite{EhlNew00}).

The relevance of such contact structures is that they rule the properties of light propagation 
in the geometric optics approximation, in particular the geometry of light wavefronts and the formation of
caustics \cite{Arnold13}. In this sense, and given the constitutive relation between (marginally) trapped surfaces 
and light convergence, such a higher-dimensional geometry seems a promising setting in which
lifts of spacetime MOTSs could exist as embedded, properly not self-intersecting, surfaces.
Elucidating their ultimate relation to wavefronts and caustics could shed light onto the here discussed
patterns of MOTS dynamics and the understanding of the trapped region.

\bibliography{mtt}{}

\begin{thebibliography}{110}%
\makeatletter
\providecommand \@ifxundefined [1]{%
 \@ifx{#1\undefined}
}%
\providecommand \@ifnum [1]{%
 \ifnum #1\expandafter \@firstoftwo
 \else \expandafter \@secondoftwo
 \fi
}%
\providecommand \@ifx [1]{%
 \ifx #1\expandafter \@firstoftwo
 \else \expandafter \@secondoftwo
 \fi
}%
\providecommand \natexlab [1]{#1}%
\providecommand \enquote  [1]{``#1''}%
\providecommand \bibnamefont  [1]{#1}%
\providecommand \bibfnamefont [1]{#1}%
\providecommand \citenamefont [1]{#1}%
\providecommand \href@noop [0]{\@secondoftwo}%
\providecommand \href [0]{\begingroup \@sanitize@url \@href}%
\providecommand \@href[1]{\@@startlink{#1}\@@href}%
\providecommand \@@href[1]{\endgroup#1\@@endlink}%
\providecommand \@sanitize@url [0]{\catcode `\\12\catcode `\$12\catcode
  `\&12\catcode `\#12\catcode `\^12\catcode `\_12\catcode `\%12\relax}%
\providecommand \@@startlink[1]{}%
\providecommand \@@endlink[0]{}%
\providecommand \url  [0]{\begingroup\@sanitize@url \@url }%
\providecommand \@url [1]{\endgroup\@href {#1}{\urlprefix }}%
\providecommand \urlprefix  [0]{URL }%
\providecommand \Eprint [0]{\href }%
\providecommand \doibase [0]{http://dx.doi.org/}%
\providecommand \selectlanguage [0]{\@gobble}%
\providecommand \bibinfo  [0]{\@secondoftwo}%
\providecommand \bibfield  [0]{\@secondoftwo}%
\providecommand \translation [1]{[#1]}%
\providecommand \BibitemOpen [0]{}%
\providecommand \bibitemStop [0]{}%
\providecommand \bibitemNoStop [0]{.\EOS\space}%
\providecommand \EOS [0]{\spacefactor3000\relax}%
\providecommand \BibitemShut  [1]{\csname bibitem#1\endcsname}%
\let\auto@bib@innerbib\@empty
\bibitem [{\citenamefont {Penrose}(1965)}]{Penrose:1964wq}%
  \BibitemOpen
  \bibfield  {author} {\bibinfo {author} {\bibfnamefont {Roger}\ \bibnamefont
  {Penrose}},\ }\bibfield  {title} {\enquote {\bibinfo {title} {{Gravitational
  collapse and space-time singularities}},}\ }\href {\doibase
  10.1103/PhysRevLett.14.57} {\bibfield  {journal} {\bibinfo  {journal} {Phys.
  Rev. Lett.}\ }\textbf {\bibinfo {volume} {14}},\ \bibinfo {pages} {57--59}
  (\bibinfo {year} {1965})}\BibitemShut {NoStop}%
\bibitem [{\citenamefont {Hawking}\ and\ \citenamefont
  {Penrose}(1970)}]{Hawking:1969sw}%
  \BibitemOpen
  \bibfield  {author} {\bibinfo {author} {\bibfnamefont {S.~W.}\ \bibnamefont
  {Hawking}}\ and\ \bibinfo {author} {\bibfnamefont {R.}~\bibnamefont
  {Penrose}},\ }\bibfield  {title} {\enquote {\bibinfo {title} {{The
  Singularities of gravitational collapse and cosmology}},}\ }\href {\doibase
  10.1098/rspa.1970.0021} {\bibfield  {journal} {\bibinfo  {journal} {Proc.
  Roy. Soc. Lond.}\ }\textbf {\bibinfo {volume} {A314}},\ \bibinfo {pages}
  {529--548} (\bibinfo {year} {1970})}\BibitemShut {NoStop}%
\bibitem [{\citenamefont {Abbott}\ \emph {et~al.}(2019)\citenamefont {Abbott}
  \emph {et~al.}}]{LIGOScientific:2018mvr}%
  \BibitemOpen
  \bibfield  {author} {\bibinfo {author} {\bibfnamefont {B.P.}\ \bibnamefont
  {Abbott}} \emph {et~al.} (\bibinfo {collaboration} {LIGO Scientific,
  Virgo}),\ }\bibfield  {title} {\enquote {\bibinfo {title} {{GWTC-1: A
  Gravitational-Wave Transient Catalog of Compact Binary Mergers Observed by
  LIGO and Virgo during the First and Second Observing Runs}},}\ }\href
  {\doibase 10.1103/PhysRevX.9.031040} {\bibfield  {journal} {\bibinfo
  {journal} {Phys. Rev. X}\ }\textbf {\bibinfo {volume} {9}},\ \bibinfo {pages}
  {031040} (\bibinfo {year} {2019})},\ \Eprint
  {http://arxiv.org/abs/1811.12907} {arXiv:1811.12907 [astro-ph.HE]}
  \BibitemShut {NoStop}%
\bibitem [{\citenamefont {Abbott}\ \emph {et~al.}(2016)\citenamefont {Abbott}
  \emph {et~al.}}]{TheLIGOScientific:2016pea}%
  \BibitemOpen
  \bibfield  {author} {\bibinfo {author} {\bibfnamefont {B.~P.}\ \bibnamefont
  {Abbott}} \emph {et~al.} (\bibinfo {collaboration} {LIGO Scientific,
  Virgo}),\ }\bibfield  {title} {\enquote {\bibinfo {title} {{Binary Black Hole
  Mergers in the first Advanced LIGO Observing Run}},}\ }\href {\doibase
  10.1103/PhysRevX.6.041015, 10.1103/PhysRevX.8.039903} {\bibfield  {journal}
  {\bibinfo  {journal} {Phys. Rev.}\ }\textbf {\bibinfo {volume} {X6}},\
  \bibinfo {pages} {041015} (\bibinfo {year} {2016})},\ \bibinfo {note}
  {[erratum: Phys. Rev.X8,no.3,039903(2018)]},\ \Eprint
  {http://arxiv.org/abs/1606.04856} {arXiv:1606.04856 [gr-qc]} \BibitemShut
  {NoStop}%
\bibitem [{\citenamefont {Nitz}\ \emph
  {et~al.}(2019{\natexlab{a}})\citenamefont {Nitz}, \citenamefont {Capano},
  \citenamefont {Nielsen}, \citenamefont {Reyes}, \citenamefont {White},
  \citenamefont {Brown},\ and\ \citenamefont {Krishnan}}]{Nitz:2018imz}%
  \BibitemOpen
  \bibfield  {author} {\bibinfo {author} {\bibfnamefont {Alexander~H.}\
  \bibnamefont {Nitz}}, \bibinfo {author} {\bibfnamefont {Collin}\ \bibnamefont
  {Capano}}, \bibinfo {author} {\bibfnamefont {Alex~B.}\ \bibnamefont
  {Nielsen}}, \bibinfo {author} {\bibfnamefont {Steven}\ \bibnamefont {Reyes}},
  \bibinfo {author} {\bibfnamefont {Rebecca}\ \bibnamefont {White}}, \bibinfo
  {author} {\bibfnamefont {Duncan~A.}\ \bibnamefont {Brown}}, \ and\ \bibinfo
  {author} {\bibfnamefont {Badri}\ \bibnamefont {Krishnan}},\ }\bibfield
  {title} {\enquote {\bibinfo {title} {{1-OGC: The first open
  gravitational-wave catalog of binary mergers from analysis of public Advanced
  LIGO data}},}\ }\href {\doibase 10.3847/1538-4357/ab0108} {\bibfield
  {journal} {\bibinfo  {journal} {Astrophys. J.}\ }\textbf {\bibinfo {volume}
  {872}},\ \bibinfo {pages} {195} (\bibinfo {year} {2019}{\natexlab{a}})},\
  \Eprint {http://arxiv.org/abs/1811.01921} {arXiv:1811.01921 [gr-qc]}
  \BibitemShut {NoStop}%
\bibitem [{\citenamefont {Nitz}\ \emph
  {et~al.}(2019{\natexlab{b}})\citenamefont {Nitz}, \citenamefont {Dent},
  \citenamefont {Davies}, \citenamefont {Kumar}, \citenamefont {Capano},
  \citenamefont {Harry}, \citenamefont {Mozzon}, \citenamefont {Nuttall},
  \citenamefont {Lundgren},\ and\ \citenamefont {Tápai}}]{Nitz:2019hdf}%
  \BibitemOpen
  \bibfield  {author} {\bibinfo {author} {\bibfnamefont {Alexander~H.}\
  \bibnamefont {Nitz}}, \bibinfo {author} {\bibfnamefont {Thomas}\ \bibnamefont
  {Dent}}, \bibinfo {author} {\bibfnamefont {Gareth~S.}\ \bibnamefont
  {Davies}}, \bibinfo {author} {\bibfnamefont {Sumit}\ \bibnamefont {Kumar}},
  \bibinfo {author} {\bibfnamefont {Collin~D.}\ \bibnamefont {Capano}},
  \bibinfo {author} {\bibfnamefont {Ian}\ \bibnamefont {Harry}}, \bibinfo
  {author} {\bibfnamefont {Simone}\ \bibnamefont {Mozzon}}, \bibinfo {author}
  {\bibfnamefont {Laura}\ \bibnamefont {Nuttall}}, \bibinfo {author}
  {\bibfnamefont {Andrew}\ \bibnamefont {Lundgren}}, \ and\ \bibinfo {author}
  {\bibfnamefont {Márton}\ \bibnamefont {Tápai}},\ }\bibfield  {title}
  {\enquote {\bibinfo {title} {{2-OGC: Open Gravitational-wave Catalog of
  binary mergers from analysis of public Advanced LIGO and Virgo data}},}\
  }\href {\doibase 10.3847/1538-4357/ab733f} {\bibfield  {journal} {\bibinfo
  {journal} {Astrophys. J.}\ }\textbf {\bibinfo {volume} {891}},\ \bibinfo
  {pages} {123} (\bibinfo {year} {2019}{\natexlab{b}})},\ \Eprint
  {http://arxiv.org/abs/1910.05331} {arXiv:1910.05331 [astro-ph.HE]}
  \BibitemShut {NoStop}%
\bibitem [{\citenamefont {Venumadhav}\ \emph {et~al.}(2019)\citenamefont
  {Venumadhav}, \citenamefont {Zackay}, \citenamefont {Roulet}, \citenamefont
  {Dai},\ and\ \citenamefont {Zaldarriaga}}]{Venumadhav:2019lyq}%
  \BibitemOpen
  \bibfield  {author} {\bibinfo {author} {\bibfnamefont {Tejaswi}\ \bibnamefont
  {Venumadhav}}, \bibinfo {author} {\bibfnamefont {Barak}\ \bibnamefont
  {Zackay}}, \bibinfo {author} {\bibfnamefont {Javier}\ \bibnamefont {Roulet}},
  \bibinfo {author} {\bibfnamefont {Liang}\ \bibnamefont {Dai}}, \ and\
  \bibinfo {author} {\bibfnamefont {Matias}\ \bibnamefont {Zaldarriaga}},\
  }\bibfield  {title} {\enquote {\bibinfo {title} {{New Binary Black Hole
  Mergers in the Second Observing Run of Advanced LIGO and Advanced Virgo}},}\
  }\href@noop {} {\  (\bibinfo {year} {2019})},\ \Eprint
  {http://arxiv.org/abs/1904.07214} {arXiv:1904.07214 [astro-ph.HE]}
  \BibitemShut {NoStop}%
\bibitem [{\citenamefont {Zackay}\ \emph {et~al.}(2019)\citenamefont {Zackay},
  \citenamefont {Venumadhav}, \citenamefont {Dai}, \citenamefont {Roulet},\
  and\ \citenamefont {Zaldarriaga}}]{Zackay:2019tzo}%
  \BibitemOpen
  \bibfield  {author} {\bibinfo {author} {\bibfnamefont {Barak}\ \bibnamefont
  {Zackay}}, \bibinfo {author} {\bibfnamefont {Tejaswi}\ \bibnamefont
  {Venumadhav}}, \bibinfo {author} {\bibfnamefont {Liang}\ \bibnamefont {Dai}},
  \bibinfo {author} {\bibfnamefont {Javier}\ \bibnamefont {Roulet}}, \ and\
  \bibinfo {author} {\bibfnamefont {Matias}\ \bibnamefont {Zaldarriaga}},\
  }\bibfield  {title} {\enquote {\bibinfo {title} {{Highly spinning and aligned
  binary black hole merger in the Advanced LIGO first observing run}},}\ }\href
  {\doibase 10.1103/PhysRevD.100.023007} {\bibfield  {journal} {\bibinfo
  {journal} {Phys. Rev. D}\ }\textbf {\bibinfo {volume} {100}},\ \bibinfo
  {pages} {023007} (\bibinfo {year} {2019})},\ \Eprint
  {http://arxiv.org/abs/1902.10331} {arXiv:1902.10331 [astro-ph.HE]}
  \BibitemShut {NoStop}%
\bibitem [{\citenamefont {Matzner}\ \emph {et~al.}(1995)\citenamefont
  {Matzner}, \citenamefont {Seidel}, \citenamefont {Shapiro}, \citenamefont
  {Smarr}, \citenamefont {Suen}, \citenamefont {Teukolsky},\ and\ \citenamefont
  {Winicour}}]{Matzner:1995ib}%
  \BibitemOpen
  \bibfield  {author} {\bibinfo {author} {\bibfnamefont {R.~A.}\ \bibnamefont
  {Matzner}}, \bibinfo {author} {\bibfnamefont {H.~E.}\ \bibnamefont {Seidel}},
  \bibinfo {author} {\bibfnamefont {Stuart~L.}\ \bibnamefont {Shapiro}},
  \bibinfo {author} {\bibfnamefont {L.}~\bibnamefont {Smarr}}, \bibinfo
  {author} {\bibfnamefont {W.~M.}\ \bibnamefont {Suen}}, \bibinfo {author}
  {\bibfnamefont {Saul~A.}\ \bibnamefont {Teukolsky}}, \ and\ \bibinfo {author}
  {\bibfnamefont {J.}~\bibnamefont {Winicour}},\ }\bibfield  {title} {\enquote
  {\bibinfo {title} {{Geometry of a black hole collision}},}\ }\href {\doibase
  10.1126/science.270.5238.941} {\bibfield  {journal} {\bibinfo  {journal}
  {Science}\ }\textbf {\bibinfo {volume} {270}},\ \bibinfo {pages} {941--947}
  (\bibinfo {year} {1995})}\BibitemShut {NoStop}%
\bibitem [{\citenamefont {Brill}\ and\ \citenamefont
  {Lindquist}(1963)}]{Brill:1963yv}%
  \BibitemOpen
  \bibfield  {author} {\bibinfo {author} {\bibfnamefont {Dieter~R.}\
  \bibnamefont {Brill}}\ and\ \bibinfo {author} {\bibfnamefont {Richard~W.}\
  \bibnamefont {Lindquist}},\ }\bibfield  {title} {\enquote {\bibinfo {title}
  {{Interaction energy in geometrostatics}},}\ }\href {\doibase
  10.1103/PhysRev.131.471} {\bibfield  {journal} {\bibinfo  {journal} {Phys.
  Rev.}\ }\textbf {\bibinfo {volume} {131}},\ \bibinfo {pages} {471--476}
  (\bibinfo {year} {1963})}\BibitemShut {NoStop}%
\bibitem [{\citenamefont {Pook-Kolb}\ \emph
  {et~al.}(2019{\natexlab{a}})\citenamefont {Pook-Kolb}, \citenamefont
  {Birnholtz}, \citenamefont {Krishnan},\ and\ \citenamefont
  {Schnetter}}]{Pook-Kolb:2018igu}%
  \BibitemOpen
  \bibfield  {author} {\bibinfo {author} {\bibfnamefont {Daniel}\ \bibnamefont
  {Pook-Kolb}}, \bibinfo {author} {\bibfnamefont {Ofek}\ \bibnamefont
  {Birnholtz}}, \bibinfo {author} {\bibfnamefont {Badri}\ \bibnamefont
  {Krishnan}}, \ and\ \bibinfo {author} {\bibfnamefont {Erik}\ \bibnamefont
  {Schnetter}},\ }\bibfield  {title} {\enquote {\bibinfo {title} {Existence and
  stability of marginally trapped surfaces in black-hole spacetimes},}\ }\href
  {\doibase 10.1103/PhysRevD.99.064005} {\bibfield  {journal} {\bibinfo
  {journal} {Phys. Rev. D}\ }\textbf {\bibinfo {volume} {99}},\ \bibinfo
  {pages} {064005} (\bibinfo {year} {2019}{\natexlab{a}})}\BibitemShut
  {NoStop}%
\bibitem [{\citenamefont {M{\"o}sta}\ \emph {et~al.}(2015)\citenamefont
  {M{\"o}sta}, \citenamefont {Andersson}, \citenamefont {Metzger},
  \citenamefont {Szil{\'a}gyi},\ and\ \citenamefont
  {Winicour}}]{Mosta:2015sga}%
  \BibitemOpen
  \bibfield  {author} {\bibinfo {author} {\bibfnamefont {P.}~\bibnamefont
  {M{\"o}sta}}, \bibinfo {author} {\bibfnamefont {L.}~\bibnamefont
  {Andersson}}, \bibinfo {author} {\bibfnamefont {J.}~\bibnamefont {Metzger}},
  \bibinfo {author} {\bibfnamefont {B.}~\bibnamefont {Szil{\'a}gyi}}, \ and\
  \bibinfo {author} {\bibfnamefont {J.}~\bibnamefont {Winicour}},\ }\bibfield
  {title} {\enquote {\bibinfo {title} {{The Merger of Small and Large Black
  Holes}},}\ }\href {\doibase 10.1088/0264-9381/32/23/235003} {\bibfield
  {journal} {\bibinfo  {journal} {Class. Quant. Grav.}\ }\textbf {\bibinfo
  {volume} {32}},\ \bibinfo {pages} {235003} (\bibinfo {year} {2015})},\
  \Eprint {http://arxiv.org/abs/1501.05358} {arXiv:1501.05358 [gr-qc]}
  \BibitemShut {NoStop}%
\bibitem [{\citenamefont {Evans}\ \emph {et~al.}(2020)\citenamefont {Evans},
  \citenamefont {Ferguson}, \citenamefont {Khamesra}, \citenamefont {Laguna},\
  and\ \citenamefont {Shoemaker}}]{Evans:2020lbq}%
  \BibitemOpen
  \bibfield  {author} {\bibinfo {author} {\bibfnamefont {Christopher}\
  \bibnamefont {Evans}}, \bibinfo {author} {\bibfnamefont {Deborah}\
  \bibnamefont {Ferguson}}, \bibinfo {author} {\bibfnamefont {Bhavesh}\
  \bibnamefont {Khamesra}}, \bibinfo {author} {\bibfnamefont {Pablo}\
  \bibnamefont {Laguna}}, \ and\ \bibinfo {author} {\bibfnamefont {Deirdre}\
  \bibnamefont {Shoemaker}},\ }\bibfield  {title} {\enquote {\bibinfo {title}
  {{Inside the Final Black Hole: Puncture and Trapped Surface Dynamics}},}\
  }\href@noop {} {\  (\bibinfo {year} {2020})},\ \Eprint
  {http://arxiv.org/abs/2004.11979} {arXiv:2004.11979 [gr-qc]} \BibitemShut
  {NoStop}%
\bibitem [{\citenamefont {Gannon}(1975)}]{Gan75}%
  \BibitemOpen
  \bibfield  {author} {\bibinfo {author} {\bibfnamefont {D.}~\bibnamefont
  {Gannon}},\ }\bibfield  {title} {\enquote {\bibinfo {title} {Singularities in
  nonsimply connected space-times},}\ }\href@noop {} {\bibfield  {journal}
  {\bibinfo  {journal} {J. Math. Phys.}\ }\textbf {\bibinfo {volume} {16}},\
  \bibinfo {pages} {2364} (\bibinfo {year} {1975})}\BibitemShut {NoStop}%
\bibitem [{\citenamefont {Gannon}(1976)}]{Gan76}%
  \BibitemOpen
  \bibfield  {author} {\bibinfo {author} {\bibfnamefont {D.}~\bibnamefont
  {Gannon}},\ }\bibfield  {title} {\enquote {\bibinfo {title} {On the topology
  of spacelike hypersurfaces, singularities and black holes},}\ }\href@noop {}
  {\bibfield  {journal} {\bibinfo  {journal} {Gen. Rel. Grav.}\ }\textbf
  {\bibinfo {volume} {7}},\ \bibinfo {pages} {219} (\bibinfo {year}
  {1976})}\BibitemShut {NoStop}%
\bibitem [{\citenamefont {Pook-Kolb}\ \emph
  {et~al.}(2019{\natexlab{b}})\citenamefont {Pook-Kolb}, \citenamefont
  {Birnholtz}, \citenamefont {Krishnan},\ and\ \citenamefont
  {Schnetter}}]{PhysRevLett.123.171102}%
  \BibitemOpen
  \bibfield  {author} {\bibinfo {author} {\bibfnamefont {Daniel}\ \bibnamefont
  {Pook-Kolb}}, \bibinfo {author} {\bibfnamefont {Ofek}\ \bibnamefont
  {Birnholtz}}, \bibinfo {author} {\bibfnamefont {Badri}\ \bibnamefont
  {Krishnan}}, \ and\ \bibinfo {author} {\bibfnamefont {Erik}\ \bibnamefont
  {Schnetter}},\ }\bibfield  {title} {\enquote {\bibinfo {title} {Interior of a
  binary black hole merger},}\ }\href {\doibase 10.1103/PhysRevLett.123.171102}
  {\bibfield  {journal} {\bibinfo  {journal} {Phys. Rev. Lett.}\ }\textbf
  {\bibinfo {volume} {123}},\ \bibinfo {pages} {171102} (\bibinfo {year}
  {2019}{\natexlab{b}})}\BibitemShut {NoStop}%
\bibitem [{\citenamefont {Booth}\ \emph {et~al.}(2020)\citenamefont {Booth},
  \citenamefont {Hennigar},\ and\ \citenamefont {Mondal}}]{Booth:2020qhb}%
  \BibitemOpen
  \bibfield  {author} {\bibinfo {author} {\bibfnamefont {Ivan}\ \bibnamefont
  {Booth}}, \bibinfo {author} {\bibfnamefont {Robie}\ \bibnamefont {Hennigar}},
  \ and\ \bibinfo {author} {\bibfnamefont {Saikat}\ \bibnamefont {Mondal}},\
  }\bibfield  {title} {\enquote {\bibinfo {title} {{Marginally outer trapped
  (open) surfaces and extreme mass ratio mergers}},}\ }\href@noop {} {\
  (\bibinfo {year} {2020})},\ \Eprint {http://arxiv.org/abs/2005.05350}
  {arXiv:2005.05350 [gr-qc]} \BibitemShut {NoStop}%
\bibitem [{\citenamefont {Hayward}(2000)}]{Hayward:2000ca}%
  \BibitemOpen
  \bibfield  {author} {\bibinfo {author} {\bibfnamefont {Sean~A.}\ \bibnamefont
  {Hayward}},\ }\bibfield  {title} {\enquote {\bibinfo {title} {{Black holes:
  New horizons}},}\ }in\ \href@noop {} {\emph {\bibinfo {booktitle} {{Recent
  developments in theoretical and experimental general relativity, gravitation
  and relativistic field theories. Proceedings, 9th Marcel Grossmann Meeting,
  MG'9, Rome, Italy, July 2-8, 2000. Pts. A-C}}}}\ (\bibinfo {year} {2000})\
  pp.\ \bibinfo {pages} {568--580},\ \Eprint
  {http://arxiv.org/abs/gr-qc/0008071} {arXiv:gr-qc/0008071 [gr-qc]}
  \BibitemShut {NoStop}%
\bibitem [{\citenamefont {Booth}(2005)}]{Booth:2005qc}%
  \BibitemOpen
  \bibfield  {author} {\bibinfo {author} {\bibfnamefont {Ivan}\ \bibnamefont
  {Booth}},\ }\bibfield  {title} {\enquote {\bibinfo {title} {{Black hole
  boundaries}},}\ }\href {\doibase 10.1139/p05-063} {\bibfield  {journal}
  {\bibinfo  {journal} {Can. J. Phys.}\ }\textbf {\bibinfo {volume} {83}},\
  \bibinfo {pages} {1073--1099} (\bibinfo {year} {2005})},\ \Eprint
  {http://arxiv.org/abs/gr-qc/0508107} {arXiv:gr-qc/0508107} \BibitemShut
  {NoStop}%
\bibitem [{\citenamefont {Ashtekar}\ and\ \citenamefont
  {Krishnan}(2004)}]{Ashtekar:2004cn}%
  \BibitemOpen
  \bibfield  {author} {\bibinfo {author} {\bibfnamefont {Abhay}\ \bibnamefont
  {Ashtekar}}\ and\ \bibinfo {author} {\bibfnamefont {Badri}\ \bibnamefont
  {Krishnan}},\ }\bibfield  {title} {\enquote {\bibinfo {title} {{Isolated and
  dynamical horizons and their applications}},}\ }\href@noop {} {\bibfield
  {journal} {\bibinfo  {journal} {Living Rev. Rel.}\ }\textbf {\bibinfo
  {volume} {7}},\ \bibinfo {pages} {10} (\bibinfo {year} {2004})},\ \Eprint
  {http://arxiv.org/abs/gr-qc/0407042} {arXiv:gr-qc/0407042} \BibitemShut
  {NoStop}%
\bibitem [{\citenamefont {Gourgoulhon}\ and\ \citenamefont
  {Jaramillo}(2006{\natexlab{a}})}]{Gourgoulhon:2005ng}%
  \BibitemOpen
  \bibfield  {author} {\bibinfo {author} {\bibfnamefont {Eric}\ \bibnamefont
  {Gourgoulhon}}\ and\ \bibinfo {author} {\bibfnamefont {Jose~Luis}\
  \bibnamefont {Jaramillo}},\ }\bibfield  {title} {\enquote {\bibinfo {title}
  {{A 3+1 perspective on null hypersurfaces and isolated horizons}},}\ }\href
  {\doibase 10.1016/j.physrep.2005.10.005} {\bibfield  {journal} {\bibinfo
  {journal} {Phys. Rept.}\ }\textbf {\bibinfo {volume} {423}},\ \bibinfo
  {pages} {159--294} (\bibinfo {year} {2006}{\natexlab{a}})},\ \Eprint
  {http://arxiv.org/abs/gr-qc/0503113} {arXiv:gr-qc/0503113} \BibitemShut
  {NoStop}%
\bibitem [{\citenamefont {Visser}(2008)}]{Visser:2009xp}%
  \BibitemOpen
  \bibfield  {author} {\bibinfo {author} {\bibfnamefont {Matt}\ \bibnamefont
  {Visser}},\ }\bibfield  {title} {\enquote {\bibinfo {title} {{Black holes in
  general relativity}},}\ }\href@noop {} {\bibfield  {journal} {\bibinfo
  {journal} {PoS}\ }\textbf {\bibinfo {volume} {BHSGRANDSTRINGS2008}},\
  \bibinfo {pages} {001} (\bibinfo {year} {2008})},\ \Eprint
  {http://arxiv.org/abs/0901.4365} {arXiv:0901.4365 [gr-qc]} \BibitemShut
  {NoStop}%
\bibitem [{\citenamefont {Jaramillo}(2011)}]{Jaramillo:2011zw}%
  \BibitemOpen
  \bibfield  {author} {\bibinfo {author} {\bibfnamefont {Jose~Luis}\
  \bibnamefont {Jaramillo}},\ }\bibfield  {title} {\enquote {\bibinfo {title}
  {{An introduction to local Black Hole horizons in the 3+1 approach to General
  Relativity}},}\ }\href {\doibase 10.1142/S0218271811020366} {\bibfield
  {journal} {\bibinfo  {journal} {Int. J. Mod. Phys.}\ }\textbf {\bibinfo
  {volume} {D20}},\ \bibinfo {pages} {2169} (\bibinfo {year} {2011})},\ \Eprint
  {http://arxiv.org/abs/1108.2408} {arXiv:1108.2408 [gr-qc]} \BibitemShut
  {NoStop}%
\bibitem [{\citenamefont {Hayward}(2013)}]{hayward2013black}%
  \BibitemOpen
  \bibfield  {author} {\bibinfo {author} {\bibfnamefont {S.A.}\ \bibnamefont
  {Hayward}},\ }\href {https://books.google.es/books?id=Joe6CgAAQBAJ} {\emph
  {\bibinfo {title} {Black Holes: New Horizons}}},\ New horizons\ (\bibinfo
  {publisher} {World Scientific},\ \bibinfo {year} {2013})\BibitemShut
  {NoStop}%
\bibitem [{\citenamefont {Faraoni}\ and\ \citenamefont
  {Prain}(2015)}]{Faraoni:2015pmn}%
  \BibitemOpen
  \bibfield  {author} {\bibinfo {author} {\bibfnamefont {Valerio}\ \bibnamefont
  {Faraoni}}\ and\ \bibinfo {author} {\bibfnamefont {Angus}\ \bibnamefont
  {Prain}},\ }\bibfield  {title} {\enquote {\bibinfo {title} {{Understanding
  dynamical black hole apparent horizons}},}\ }\href@noop {} {\bibfield
  {journal} {\bibinfo  {journal} {Lecture Notes in Physics}\ }\textbf {\bibinfo
  {volume} {907}},\ \bibinfo {pages} {1--199} (\bibinfo {year} {2015})},\
  \Eprint {http://arxiv.org/abs/1511.07775} {arXiv:1511.07775 [gr-qc]}
  \BibitemShut {NoStop}%
\bibitem [{\citenamefont {Newman}(1987)}]{Newman:1987}%
  \BibitemOpen
  \bibfield  {author} {\bibinfo {author} {\bibfnamefont {R.P.A.C.}\
  \bibnamefont {Newman}},\ }\bibfield  {title} {\enquote {\bibinfo {title}
  {Topology and stability of marginal 2-surfaces},}\ }\href@noop {} {\bibfield
  {journal} {\bibinfo  {journal} {Class. and Quant. Grav.}\ }\textbf {\bibinfo
  {volume} {4}},\ \bibinfo {pages} {277--290} (\bibinfo {year}
  {1987})}\BibitemShut {NoStop}%
\bibitem [{\citenamefont {Pook-Kolb}\ \emph
  {et~al.}(2019{\natexlab{c}})\citenamefont {Pook-Kolb}, \citenamefont
  {Birnholtz}, \citenamefont {Krishnan},\ and\ \citenamefont
  {Schnetter}}]{PhysRevD.100.084044}%
  \BibitemOpen
  \bibfield  {author} {\bibinfo {author} {\bibfnamefont {Daniel}\ \bibnamefont
  {Pook-Kolb}}, \bibinfo {author} {\bibfnamefont {Ofek}\ \bibnamefont
  {Birnholtz}}, \bibinfo {author} {\bibfnamefont {Badri}\ \bibnamefont
  {Krishnan}}, \ and\ \bibinfo {author} {\bibfnamefont {Erik}\ \bibnamefont
  {Schnetter}},\ }\bibfield  {title} {\enquote {\bibinfo {title}
  {Self-intersecting marginally outer trapped surfaces},}\ }\href {\doibase
  10.1103/PhysRevD.100.084044} {\bibfield  {journal} {\bibinfo  {journal}
  {Phys. Rev. D}\ }\textbf {\bibinfo {volume} {100}},\ \bibinfo {pages}
  {084044} (\bibinfo {year} {2019}{\natexlab{c}})}\BibitemShut {NoStop}%
\bibitem [{\citenamefont {Pook-Kolb}\ \emph
  {et~al.}(2019{\natexlab{d}})\citenamefont {Pook-Kolb}, \citenamefont
  {Birnholtz}, \citenamefont {Krishnan},\ and\ \citenamefont
  {Schnetter}}]{pook_kolb_daniel_2019_2591105}%
  \BibitemOpen
  \bibfield  {author} {\bibinfo {author} {\bibfnamefont {Daniel}\ \bibnamefont
  {Pook-Kolb}}, \bibinfo {author} {\bibfnamefont {Ofek}\ \bibnamefont
  {Birnholtz}}, \bibinfo {author} {\bibfnamefont {Badri}\ \bibnamefont
  {Krishnan}}, \ and\ \bibinfo {author} {\bibfnamefont {Erik}\ \bibnamefont
  {Schnetter}},\ }\href {\doibase 10.5281/zenodo.2591105} {\enquote {\bibinfo
  {title} {{MOTS Finder} version 1.1},}\ } (\bibinfo {year}
  {2019}{\natexlab{d}}),\ \bibinfo {note} {10.5281/zenodo.2591105}\BibitemShut
  {NoStop}%
\bibitem [{\citenamefont {Thornburg}(2004)}]{Thornburg:2003sf}%
  \BibitemOpen
  \bibfield  {author} {\bibinfo {author} {\bibfnamefont {Jonathan}\
  \bibnamefont {Thornburg}},\ }\bibfield  {title} {\enquote {\bibinfo {title}
  {{A Fast Apparent-Horizon Finder for 3-Dimensional Cartesian Grids in
  Numerical Relativity}},}\ }\href {\doibase 10.1088/0264-9381/21/2/026}
  {\bibfield  {journal} {\bibinfo  {journal} {Class. Quant. Grav.}\ }\textbf
  {\bibinfo {volume} {21}},\ \bibinfo {pages} {743--766} (\bibinfo {year}
  {2004})},\ \Eprint {http://arxiv.org/abs/gr-qc/0306056} {arXiv:gr-qc/0306056}
  \BibitemShut {NoStop}%
\bibitem [{\citenamefont {Thornburg}(2007)}]{Thornburg:2006zb}%
  \BibitemOpen
  \bibfield  {author} {\bibinfo {author} {\bibfnamefont {Jonathan}\
  \bibnamefont {Thornburg}},\ }\bibfield  {title} {\enquote {\bibinfo {title}
  {{Event and Apparent Horizon Finders for $3+1$ Numerical Relativity}},}\
  }\href@noop {} {\bibfield  {journal} {\bibinfo  {journal} {Living Rev. Rel.}\
  }\textbf {\bibinfo {volume} {10}},\ \bibinfo {pages} {3} (\bibinfo {year}
  {2007})},\ \Eprint {http://arxiv.org/abs/gr-qc/0512169} {arXiv:gr-qc/0512169}
  \BibitemShut {NoStop}%
\bibitem [{\citenamefont {Thornburg}(1996)}]{Thornburg:1995cp}%
  \BibitemOpen
  \bibfield  {author} {\bibinfo {author} {\bibfnamefont {Jonathan}\
  \bibnamefont {Thornburg}},\ }\bibfield  {title} {\enquote {\bibinfo {title}
  {{Finding apparent horizons in numerical relativity}},}\ }\href {\doibase
  10.1103/PhysRevD.54.4899} {\bibfield  {journal} {\bibinfo  {journal} {Phys.
  Rev. D}\ }\textbf {\bibinfo {volume} {54}},\ \bibinfo {pages} {4899--4918}
  (\bibinfo {year} {1996})},\ \Eprint
  {http://arxiv.org/abs/arXiv:gr-qc/9508014} {arXiv:gr-qc/9508014} \BibitemShut
  {NoStop}%
\bibitem [{\citenamefont {Shoemaker}\ \emph {et~al.}(2000)\citenamefont
  {Shoemaker}, \citenamefont {Huq},\ and\ \citenamefont
  {Matzner}}]{Shoemaker:2000ye}%
  \BibitemOpen
  \bibfield  {author} {\bibinfo {author} {\bibfnamefont {Deirdre~M.}\
  \bibnamefont {Shoemaker}}, \bibinfo {author} {\bibfnamefont {Mijan~F.}\
  \bibnamefont {Huq}}, \ and\ \bibinfo {author} {\bibfnamefont {Richard~A.}\
  \bibnamefont {Matzner}},\ }\bibfield  {title} {\enquote {\bibinfo {title}
  {{Generic Tracking of Multiple Apparent Horizons with Level Flow}},}\ }\href
  {\doibase 10.1103/PhysRevD.62.124005} {\bibfield  {journal} {\bibinfo
  {journal} {Phys. Rev.}\ }\textbf {\bibinfo {volume} {D62}},\ \bibinfo {pages}
  {124005} (\bibinfo {year} {2000})},\ \Eprint
  {http://arxiv.org/abs/gr-qc/0004062} {arXiv:gr-qc/0004062} \BibitemShut
  {NoStop}%
\bibitem [{\citenamefont {Lin}\ and\ \citenamefont {Novak}(2007)}]{Lin:2007cd}%
  \BibitemOpen
  \bibfield  {author} {\bibinfo {author} {\bibfnamefont {Lap-Ming}\
  \bibnamefont {Lin}}\ and\ \bibinfo {author} {\bibfnamefont {Jerome}\
  \bibnamefont {Novak}},\ }\bibfield  {title} {\enquote {\bibinfo {title} {{A
  new spectral apparent horizon finder for 3D numerical relativity}},}\ }\href
  {\doibase 10.1088/0264-9381/24/10/012} {\bibfield  {journal} {\bibinfo
  {journal} {Class. Quant. Grav.}\ }\textbf {\bibinfo {volume} {24}},\ \bibinfo
  {pages} {2665--2676} (\bibinfo {year} {2007})},\ \Eprint
  {http://arxiv.org/abs/gr-qc/0702038} {arXiv:gr-qc/0702038} \BibitemShut
  {NoStop}%
\bibitem [{\citenamefont {Jaramillo}\ \emph {et~al.}(2009)\citenamefont
  {Jaramillo}, \citenamefont {Ansorg},\ and\ \citenamefont
  {Vasset}}]{Jaramillo:2009zz}%
  \BibitemOpen
  \bibfield  {author} {\bibinfo {author} {\bibfnamefont {Jose~Luis}\
  \bibnamefont {Jaramillo}}, \bibinfo {author} {\bibfnamefont {Marcus}\
  \bibnamefont {Ansorg}}, \ and\ \bibinfo {author} {\bibfnamefont {Nicolas}\
  \bibnamefont {Vasset}},\ }\bibfield  {title} {\enquote {\bibinfo {title}
  {{Application of initial data sequences to the study of black hole dynamical
  trapping horizons}},}\ }\bibfield  {booktitle} {\emph {\bibinfo {booktitle}
  {{Physics and mathematical of gravitation. Proceedings, Spanish Relativity
  Meeting, Salamanca, Spain, September 15-19, 2008}}},\ }\href {\doibase
  10.1063/1.3141305} {\bibfield  {journal} {\bibinfo  {journal} {AIP Conf.
  Proc.}\ }\textbf {\bibinfo {volume} {1122}},\ \bibinfo {pages} {308--311}
  (\bibinfo {year} {2009})},\ \Eprint {http://arxiv.org/abs/1103.6180}
  {arXiv:1103.6180 [gr-qc]} \BibitemShut {NoStop}%
\bibitem [{\citenamefont {L{\"{o}}ffler}\ \emph {et~al.}(2012)\citenamefont
  {L{\"{o}}ffler}, \citenamefont {Faber}, \citenamefont {Bentivegna},
  \citenamefont {Bode}, \citenamefont {Diener}, \citenamefont {Haas},
  \citenamefont {Hinder}, \citenamefont {Mundim}, \citenamefont {Ott},
  \citenamefont {Schnetter}, \citenamefont {Allen}, \citenamefont
  {Campanelli},\ and\ \citenamefont {Laguna}}]{Loffler:2011ay}%
  \BibitemOpen
  \bibfield  {author} {\bibinfo {author} {\bibfnamefont {Frank}\ \bibnamefont
  {L{\"{o}}ffler}}, \bibinfo {author} {\bibfnamefont {Joshua}\ \bibnamefont
  {Faber}}, \bibinfo {author} {\bibfnamefont {Eloisa}\ \bibnamefont
  {Bentivegna}}, \bibinfo {author} {\bibfnamefont {Tanja}\ \bibnamefont
  {Bode}}, \bibinfo {author} {\bibfnamefont {Peter}\ \bibnamefont {Diener}},
  \bibinfo {author} {\bibfnamefont {Roland}\ \bibnamefont {Haas}}, \bibinfo
  {author} {\bibfnamefont {Ian}\ \bibnamefont {Hinder}}, \bibinfo {author}
  {\bibfnamefont {Bruno~C.}\ \bibnamefont {Mundim}}, \bibinfo {author}
  {\bibfnamefont {Christian~D.}\ \bibnamefont {Ott}}, \bibinfo {author}
  {\bibfnamefont {Erik}\ \bibnamefont {Schnetter}}, \bibinfo {author}
  {\bibfnamefont {Gabrielle}\ \bibnamefont {Allen}}, \bibinfo {author}
  {\bibfnamefont {Manuela}\ \bibnamefont {Campanelli}}, \ and\ \bibinfo
  {author} {\bibfnamefont {Pablo}\ \bibnamefont {Laguna}},\ }\bibfield  {title}
  {\enquote {\bibinfo {title} {{{T}he {E}instein {T}oolkit: {A} {C}ommunity
  {C}omputational {I}nfrastructure for {R}elativistic {A}strophysics}},}\
  }\href {\doibase doi:10.1088/0264-9381/29/11/115001} {\bibfield  {journal}
  {\bibinfo  {journal} {Class. Quantum Grav.}\ }\textbf {\bibinfo {volume}
  {29}},\ \bibinfo {pages} {115001} (\bibinfo {year} {2012})},\ \Eprint
  {http://arxiv.org/abs/arXiv:1111.3344 [gr-qc]} {arXiv:1111.3344 [gr-qc]}
  \BibitemShut {NoStop}%
\bibitem [{EinsteinToolkit()}]{EinsteinToolkit:web}%
  \BibitemOpen
  EinsteinToolkit,\ \href@noop {} {\enquote {\bibinfo {title} {{Einstein
  Toolkit}: Open software for relativistic astrophysics},}\ }\bibinfo {note}
  {\url{http://einsteintoolkit.org/}}\BibitemShut {NoStop}%
\bibitem [{\citenamefont {Ansorg}\ \emph {et~al.}(2004)\citenamefont {Ansorg},
  \citenamefont {Br{\"u}gmann},\ and\ \citenamefont {Tichy}}]{Ansorg:2004ds}%
  \BibitemOpen
  \bibfield  {author} {\bibinfo {author} {\bibfnamefont {Marcus}\ \bibnamefont
  {Ansorg}}, \bibinfo {author} {\bibfnamefont {Bernd}\ \bibnamefont
  {Br{\"u}gmann}}, \ and\ \bibinfo {author} {\bibfnamefont {Wolfgang}\
  \bibnamefont {Tichy}},\ }\bibfield  {title} {\enquote {\bibinfo {title} {A
  single-domain spectral method for black hole puncture data},}\ }\href
  {\doibase 10.1103/PhysRevD.70.064011} {\bibfield  {journal} {\bibinfo
  {journal} {Phys. Rev. D}\ }\textbf {\bibinfo {volume} {70}},\ \bibinfo
  {pages} {064011} (\bibinfo {year} {2004})},\ \Eprint
  {http://arxiv.org/abs/arXiv:gr-qc/0404056} {arXiv:gr-qc/0404056} \BibitemShut
  {NoStop}%
\bibitem [{\citenamefont {Ansorg}(2005)}]{Ansorg:2005bp}%
  \BibitemOpen
  \bibfield  {author} {\bibinfo {author} {\bibfnamefont {Marcus}\ \bibnamefont
  {Ansorg}},\ }\bibfield  {title} {\enquote {\bibinfo {title} {{Double-domain
  spectral method for black hole excision data}},}\ }\href {\doibase
  10.1103/PhysRevD.72.024018} {\bibfield  {journal} {\bibinfo  {journal} {Phys.
  Rev.}\ }\textbf {\bibinfo {volume} {D72}},\ \bibinfo {pages} {024018}
  (\bibinfo {year} {2005})},\ \Eprint {http://arxiv.org/abs/gr-qc/0505059}
  {arXiv:gr-qc/0505059 [gr-qc]} \BibitemShut {NoStop}%
\bibitem [{\citenamefont {Brown}\ \emph {et~al.}(2009)\citenamefont {Brown},
  \citenamefont {Diener}, \citenamefont {Sarbach}, \citenamefont {Schnetter},\
  and\ \citenamefont {Tiglio}}]{Brown:2008sb}%
  \BibitemOpen
  \bibfield  {author} {\bibinfo {author} {\bibfnamefont {J.~David}\
  \bibnamefont {Brown}}, \bibinfo {author} {\bibfnamefont {Peter}\ \bibnamefont
  {Diener}}, \bibinfo {author} {\bibfnamefont {Olivier}\ \bibnamefont
  {Sarbach}}, \bibinfo {author} {\bibfnamefont {Erik}\ \bibnamefont
  {Schnetter}}, \ and\ \bibinfo {author} {\bibfnamefont {Manuel}\ \bibnamefont
  {Tiglio}},\ }\bibfield  {title} {\enquote {\bibinfo {title} {{Turduckening
  black holes: an analytical and computational study}},}\ }\href {\doibase
  10.1103/PhysRevD.79.044023} {\bibfield  {journal} {\bibinfo  {journal} {Phys.
  Rev. D}\ }\textbf {\bibinfo {volume} {79}},\ \bibinfo {pages} {044023}
  (\bibinfo {year} {2009})},\ \Eprint {http://arxiv.org/abs/arXiv:0809.3533
  [gr-qc]} {arXiv:0809.3533 [gr-qc]} \BibitemShut {NoStop}%
\bibitem [{\citenamefont {Husa}\ \emph {et~al.}(2006)\citenamefont {Husa},
  \citenamefont {Hinder},\ and\ \citenamefont {Lechner}}]{Husa:2004ip}%
  \BibitemOpen
  \bibfield  {author} {\bibinfo {author} {\bibfnamefont {Sascha}\ \bibnamefont
  {Husa}}, \bibinfo {author} {\bibfnamefont {Ian}\ \bibnamefont {Hinder}}, \
  and\ \bibinfo {author} {\bibfnamefont {Christiane}\ \bibnamefont {Lechner}},\
  }\bibfield  {title} {\enquote {\bibinfo {title} {{Kranc: a Mathematica
  application to generate numerical codes for tensorial evolution
  equations}},}\ }\href@noop {} {\bibfield  {journal} {\bibinfo  {journal}
  {Comput. Phys. Commun.}\ }\textbf {\bibinfo {volume} {174}},\ \bibinfo
  {pages} {983--1004} (\bibinfo {year} {2006})},\ \Eprint
  {http://arxiv.org/abs/arXiv:gr-qc/0404023} {arXiv:gr-qc/0404023} \BibitemShut
  {NoStop}%
\bibitem [{Kranc()}]{Kranc:web}%
  \BibitemOpen
  Kranc,\ \href {http://kranccode.org/} {\enquote {\bibinfo {title} {{Kranc}:
  {Kranc} assembles numerical code},}\ }\BibitemShut {NoStop}%
\bibitem [{\citenamefont {Andersson}\ \emph {et~al.}(2005)\citenamefont
  {Andersson}, \citenamefont {Mars},\ and\ \citenamefont
  {Simon}}]{Andersson:2005gq}%
  \BibitemOpen
  \bibfield  {author} {\bibinfo {author} {\bibfnamefont {Lars}\ \bibnamefont
  {Andersson}}, \bibinfo {author} {\bibfnamefont {Marc}\ \bibnamefont {Mars}},
  \ and\ \bibinfo {author} {\bibfnamefont {Walter}\ \bibnamefont {Simon}},\
  }\bibfield  {title} {\enquote {\bibinfo {title} {{Local existence of
  dynamical and trapping horizons}},}\ }\href {\doibase
  10.1103/PhysRevLett.95.111102} {\bibfield  {journal} {\bibinfo  {journal}
  {Phys.Rev.Lett.}\ }\textbf {\bibinfo {volume} {95}},\ \bibinfo {pages}
  {111102} (\bibinfo {year} {2005})},\ \Eprint
  {http://arxiv.org/abs/gr-qc/0506013} {arXiv:gr-qc/0506013 [gr-qc]}
  \BibitemShut {NoStop}%
\bibitem [{\citenamefont {Andersson}\ \emph {et~al.}(2008)\citenamefont
  {Andersson}, \citenamefont {Mars},\ and\ \citenamefont
  {Simon}}]{Andersson:2007fh}%
  \BibitemOpen
  \bibfield  {author} {\bibinfo {author} {\bibfnamefont {Lars}\ \bibnamefont
  {Andersson}}, \bibinfo {author} {\bibfnamefont {Marc}\ \bibnamefont {Mars}},
  \ and\ \bibinfo {author} {\bibfnamefont {Walter}\ \bibnamefont {Simon}},\
  }\bibfield  {title} {\enquote {\bibinfo {title} {{Stability of marginally
  outer trapped surfaces and existence of marginally outer trapped tubes}},}\
  }\href@noop {} {\bibfield  {journal} {\bibinfo  {journal}
  {Adv.Theor.Math.Phys.}\ }\textbf {\bibinfo {volume} {12}} (\bibinfo {year}
  {2008})},\ \Eprint {http://arxiv.org/abs/0704.2889} {arXiv:0704.2889 [gr-qc]}
  \BibitemShut {NoStop}%
\bibitem [{\citenamefont {Andersson}\ \emph {et~al.}(2009)\citenamefont
  {Andersson}, \citenamefont {Mars}, \citenamefont {Metzger},\ and\
  \citenamefont {Simon}}]{Andersson:2008up}%
  \BibitemOpen
  \bibfield  {author} {\bibinfo {author} {\bibfnamefont {Lars}\ \bibnamefont
  {Andersson}}, \bibinfo {author} {\bibfnamefont {Marc}\ \bibnamefont {Mars}},
  \bibinfo {author} {\bibfnamefont {Jan}\ \bibnamefont {Metzger}}, \ and\
  \bibinfo {author} {\bibfnamefont {Walter}\ \bibnamefont {Simon}},\ }\bibfield
   {title} {\enquote {\bibinfo {title} {{The Time evolution of marginally
  trapped surfaces}},}\ }\href {\doibase 10.1088/0264-9381/26/8/085018}
  {\bibfield  {journal} {\bibinfo  {journal} {Class.Quant.Grav.}\ }\textbf
  {\bibinfo {volume} {26}},\ \bibinfo {pages} {085018} (\bibinfo {year}
  {2009})},\ \Eprint {http://arxiv.org/abs/0811.4721} {arXiv:0811.4721 [gr-qc]}
  \BibitemShut {NoStop}%
\bibitem [{\citenamefont {Booth}\ and\ \citenamefont
  {Fairhurst}(2007)}]{Booth:2006bn}%
  \BibitemOpen
  \bibfield  {author} {\bibinfo {author} {\bibfnamefont {Ivan}\ \bibnamefont
  {Booth}}\ and\ \bibinfo {author} {\bibfnamefont {Stephen}\ \bibnamefont
  {Fairhurst}},\ }\bibfield  {title} {\enquote {\bibinfo {title} {{Isolated,
  slowly evolving, and dynamical trapping horizons: geometry and mechanics from
  surface deformations}},}\ }\href {\doibase 10.1103/PhysRevD.75.084019}
  {\bibfield  {journal} {\bibinfo  {journal} {Phys. Rev.}\ }\textbf {\bibinfo
  {volume} {D75}},\ \bibinfo {pages} {084019} (\bibinfo {year} {2007})},\
  \Eprint {http://arxiv.org/abs/gr-qc/0610032} {arXiv:gr-qc/0610032}
  \BibitemShut {NoStop}%
\bibitem [{\citenamefont {Gao}\ \emph {et~al.}(2008)\citenamefont {Gao},
  \citenamefont {Chen}, \citenamefont {Faraoni},\ and\ \citenamefont
  {Shen}}]{Gao:2008jv}%
  \BibitemOpen
  \bibfield  {author} {\bibinfo {author} {\bibfnamefont {Changjun}\
  \bibnamefont {Gao}}, \bibinfo {author} {\bibfnamefont {Xuelei}\ \bibnamefont
  {Chen}}, \bibinfo {author} {\bibfnamefont {Valerio}\ \bibnamefont {Faraoni}},
  \ and\ \bibinfo {author} {\bibfnamefont {You-Gen}\ \bibnamefont {Shen}},\
  }\bibfield  {title} {\enquote {\bibinfo {title} {{Does the mass of a black
  hole decrease due to the accretion of phantom energy}},}\ }\href {\doibase
  10.1103/PhysRevD.78.024008} {\bibfield  {journal} {\bibinfo  {journal} {Phys.
  Rev.}\ }\textbf {\bibinfo {volume} {D78}},\ \bibinfo {pages} {024008}
  (\bibinfo {year} {2008})},\ \Eprint {http://arxiv.org/abs/0802.1298}
  {arXiv:0802.1298 [gr-qc]} \BibitemShut {NoStop}%
\bibitem [{\citenamefont {Jaramillo}(2014)}]{Jaramillo:2013rda}%
  \BibitemOpen
  \bibfield  {author} {\bibinfo {author} {\bibfnamefont {Jos\'e~Luis}\
  \bibnamefont {Jaramillo}},\ }\bibfield  {title} {\enquote {\bibinfo {title}
  {{A Young-Laplace law for black hole horizons}},}\ }\href {\doibase
  10.1103/PhysRevD.89.021502} {\bibfield  {journal} {\bibinfo  {journal} {Phys.
  Rev.}\ }\textbf {\bibinfo {volume} {D89}},\ \bibinfo {pages} {021502}
  (\bibinfo {year} {2014})},\ \Eprint {http://arxiv.org/abs/1309.6593}
  {arXiv:1309.6593 [gr-qc]} \BibitemShut {NoStop}%
\bibitem [{\citenamefont {Jaramillo}(2015{\natexlab{a}})}]{Jaramillo:2014oha}%
  \BibitemOpen
  \bibfield  {author} {\bibinfo {author} {\bibfnamefont {Jos\'e~Luis}\
  \bibnamefont {Jaramillo}},\ }\bibfield  {title} {\enquote {\bibinfo {title}
  {{Black hole horizons and quantum charged particles}},}\ }\href {\doibase
  10.1088/0264-9381/32/13/132001} {\bibfield  {journal} {\bibinfo  {journal}
  {Class. Quant. Grav.}\ }\textbf {\bibinfo {volume} {32}},\ \bibinfo {pages}
  {132001} (\bibinfo {year} {2015}{\natexlab{a}})},\ \Eprint
  {http://arxiv.org/abs/1410.0509} {arXiv:1410.0509 [gr-qc]} \BibitemShut
  {NoStop}%
\bibitem [{\citenamefont {Jaramillo}(2015{\natexlab{b}})}]{Jaramillo:2015twa}%
  \BibitemOpen
  \bibfield  {author} {\bibinfo {author} {\bibfnamefont {José~Luis}\
  \bibnamefont {Jaramillo}},\ }\bibfield  {title} {\enquote {\bibinfo {title}
  {{A perspective on Black Hole Horizons from the Quantum Charged Particle}},}\
  }\bibfield  {booktitle} {\emph {\bibinfo {booktitle} {{Proceedings, Spanish
  Relativity Meeting: Almost 100 years after Einstein Revolution (ERE 2014):
  Valencia, Spain, September 1-5, 2014}}},\ }\href {\doibase
  10.1088/1742-6596/600/1/012037} {\bibfield  {journal} {\bibinfo  {journal}
  {J. Phys. Conf. Ser.}\ }\textbf {\bibinfo {volume} {600}},\ \bibinfo {pages}
  {012037} (\bibinfo {year} {2015}{\natexlab{b}})},\ \Eprint
  {http://arxiv.org/abs/1608.05963} {arXiv:1608.05963 [gr-qc]} \BibitemShut
  {NoStop}%
\bibitem [{\citenamefont {Vaidya}(1999)}]{Vaidya:1999zz}%
  \BibitemOpen
  \bibfield  {author} {\bibinfo {author} {\bibfnamefont {P.C.}\ \bibnamefont
  {Vaidya}},\ }\bibfield  {title} {\enquote {\bibinfo {title} {{The External
  Field of a Radiating Star in Relativity}},}\ }\href {\doibase
  10.1023/A:1018871522880} {\bibfield  {journal} {\bibinfo  {journal} {Gen.
  Rel. Grav.}\ }\textbf {\bibinfo {volume} {31}},\ \bibinfo {pages} {119--120}
  (\bibinfo {year} {1999})}\BibitemShut {NoStop}%
\bibitem [{\citenamefont {Oppenheimer}\ and\ \citenamefont
  {Snyder}(1939)}]{Oppenheimer:1939ue}%
  \BibitemOpen
  \bibfield  {author} {\bibinfo {author} {\bibfnamefont {J.R.}\ \bibnamefont
  {Oppenheimer}}\ and\ \bibinfo {author} {\bibfnamefont {H.}~\bibnamefont
  {Snyder}},\ }\bibfield  {title} {\enquote {\bibinfo {title} {{On Continued
  gravitational contraction}},}\ }\href {\doibase 10.1103/PhysRev.56.455}
  {\bibfield  {journal} {\bibinfo  {journal} {Phys. Rev.}\ }\textbf {\bibinfo
  {volume} {56}},\ \bibinfo {pages} {455--459} (\bibinfo {year}
  {1939})}\BibitemShut {NoStop}%
\bibitem [{\citenamefont {Booth}\ \emph {et~al.}(2006)\citenamefont {Booth},
  \citenamefont {Brits}, \citenamefont {Gonzalez},\ and\ \citenamefont {Van
  Den~Broeck}}]{Booth:2005ng}%
  \BibitemOpen
  \bibfield  {author} {\bibinfo {author} {\bibfnamefont {Ivan}\ \bibnamefont
  {Booth}}, \bibinfo {author} {\bibfnamefont {Lionel}\ \bibnamefont {Brits}},
  \bibinfo {author} {\bibfnamefont {Jose~A.}\ \bibnamefont {Gonzalez}}, \ and\
  \bibinfo {author} {\bibfnamefont {Chris}\ \bibnamefont {Van Den~Broeck}},\
  }\bibfield  {title} {\enquote {\bibinfo {title} {{Marginally trapped tubes
  and dynamical horizons}},}\ }\href {\doibase 10.1088/0264-9381/23/2/009}
  {\bibfield  {journal} {\bibinfo  {journal} {Class. Quant. Grav.}\ }\textbf
  {\bibinfo {volume} {23}},\ \bibinfo {pages} {413--440} (\bibinfo {year}
  {2006})},\ \Eprint {http://arxiv.org/abs/gr-qc/0506119} {arXiv:gr-qc/0506119
  [gr-qc]} \BibitemShut {NoStop}%
\bibitem [{\citenamefont {Helou}\ \emph {et~al.}(2017)\citenamefont {Helou},
  \citenamefont {Musco},\ and\ \citenamefont {Miller}}]{Helou:2016xyu}%
  \BibitemOpen
  \bibfield  {author} {\bibinfo {author} {\bibfnamefont {Alexis}\ \bibnamefont
  {Helou}}, \bibinfo {author} {\bibfnamefont {Ilia}\ \bibnamefont {Musco}}, \
  and\ \bibinfo {author} {\bibfnamefont {John~C.}\ \bibnamefont {Miller}},\
  }\bibfield  {title} {\enquote {\bibinfo {title} {{Causal Nature and Dynamics
  of Trapping Horizons in Black Hole Collapse}},}\ }\href {\doibase
  10.1088/1361-6382/aa6d8f} {\bibfield  {journal} {\bibinfo  {journal} {Class.
  Quant. Grav.}\ }\textbf {\bibinfo {volume} {34}},\ \bibinfo {pages} {135012}
  (\bibinfo {year} {2017})},\ \Eprint {http://arxiv.org/abs/1601.05109}
  {arXiv:1601.05109 [gr-qc]} \BibitemShut {NoStop}%
\bibitem [{\citenamefont {Chatterjee}\ \emph {et~al.}(2020)\citenamefont
  {Chatterjee}, \citenamefont {Ghosh},\ and\ \citenamefont
  {Jaryal}}]{Chatterjee:2020khj}%
  \BibitemOpen
  \bibfield  {author} {\bibinfo {author} {\bibfnamefont {Ayan}\ \bibnamefont
  {Chatterjee}}, \bibinfo {author} {\bibfnamefont {Amit}\ \bibnamefont
  {Ghosh}}, \ and\ \bibinfo {author} {\bibfnamefont {Suresh}\ \bibnamefont
  {Jaryal}},\ }\bibfield  {title} {\enquote {\bibinfo {title} {{Marginally
  Trapped Surfaces in Spherical Gravitational Collapse}},}\ }\href@noop {} {\
  (\bibinfo {year} {2020})},\ \Eprint {http://arxiv.org/abs/2004.11266}
  {arXiv:2004.11266 [gr-qc]} \BibitemShut {NoStop}%
\bibitem [{\citenamefont {Booth}(2013)}]{Booth:2012xm}%
  \BibitemOpen
  \bibfield  {author} {\bibinfo {author} {\bibfnamefont {Ivan}\ \bibnamefont
  {Booth}},\ }\bibfield  {title} {\enquote {\bibinfo {title} {{Spacetime near
  isolated and dynamical trapping horizons}},}\ }\href {\doibase
  10.1103/PhysRevD.87.024008} {\bibfield  {journal} {\bibinfo  {journal} {Phys.
  Rev. D}\ }\textbf {\bibinfo {volume} {87}},\ \bibinfo {pages} {024008}
  (\bibinfo {year} {2013})},\ \Eprint {http://arxiv.org/abs/1207.6955}
  {arXiv:1207.6955 [gr-qc]} \BibitemShut {NoStop}%
\bibitem [{\citenamefont {Schnetter}\ \emph {et~al.}(2006)\citenamefont
  {Schnetter}, \citenamefont {Krishnan},\ and\ \citenamefont
  {Beyer}}]{Schnetter:2006yt}%
  \BibitemOpen
  \bibfield  {author} {\bibinfo {author} {\bibfnamefont {Erik}\ \bibnamefont
  {Schnetter}}, \bibinfo {author} {\bibfnamefont {Badri}\ \bibnamefont
  {Krishnan}}, \ and\ \bibinfo {author} {\bibfnamefont {Florian}\ \bibnamefont
  {Beyer}},\ }\bibfield  {title} {\enquote {\bibinfo {title} {{Introduction to
  dynamical horizons in numerical relativity}},}\ }\href {\doibase
  10.1103/PhysRevD.74.024028} {\bibfield  {journal} {\bibinfo  {journal} {Phys.
  Rev.}\ }\textbf {\bibinfo {volume} {D74}},\ \bibinfo {pages} {024028}
  (\bibinfo {year} {2006})},\ \Eprint {http://arxiv.org/abs/gr-qc/0604015}
  {arXiv:gr-qc/0604015} \BibitemShut {NoStop}%
\bibitem [{\citenamefont {Gupta}\ \emph {et~al.}(2018)\citenamefont {Gupta},
  \citenamefont {Krishnan}, \citenamefont {Nielsen},\ and\ \citenamefont
  {Schnetter}}]{Gupta:2018znn}%
  \BibitemOpen
  \bibfield  {author} {\bibinfo {author} {\bibfnamefont {Anshu}\ \bibnamefont
  {Gupta}}, \bibinfo {author} {\bibfnamefont {Badri}\ \bibnamefont {Krishnan}},
  \bibinfo {author} {\bibfnamefont {Alex}\ \bibnamefont {Nielsen}}, \ and\
  \bibinfo {author} {\bibfnamefont {Erik}\ \bibnamefont {Schnetter}},\
  }\bibfield  {title} {\enquote {\bibinfo {title} {{Dynamics of marginally
  trapped surfaces in a binary black hole merger: Growth and approach to
  equilibrium}},}\ }\href {\doibase 10.1103/PhysRevD.97.084028} {\bibfield
  {journal} {\bibinfo  {journal} {Phys. Rev.}\ }\textbf {\bibinfo {volume}
  {D97}},\ \bibinfo {pages} {084028} (\bibinfo {year} {2018})},\ \Eprint
  {http://arxiv.org/abs/1801.07048} {arXiv:1801.07048 [gr-qc]} \BibitemShut
  {NoStop}%
\bibitem [{\citenamefont {Ashtekar}\ and\ \citenamefont
  {Galloway}(2005)}]{Ashtekar:2005ez}%
  \BibitemOpen
  \bibfield  {author} {\bibinfo {author} {\bibfnamefont {Abhay}\ \bibnamefont
  {Ashtekar}}\ and\ \bibinfo {author} {\bibfnamefont {Gregory~J.}\ \bibnamefont
  {Galloway}},\ }\bibfield  {title} {\enquote {\bibinfo {title} {{Some
  uniqueness results for dynamical horizons}},}\ }\href@noop {} {\bibfield
  {journal} {\bibinfo  {journal} {Adv. Theor. Math. Phys.}\ }\textbf {\bibinfo
  {volume} {9}},\ \bibinfo {pages} {1--30} (\bibinfo {year} {2005})},\ \Eprint
  {http://arxiv.org/abs/gr-qc/0503109} {arXiv:gr-qc/0503109} \BibitemShut
  {NoStop}%
\bibitem [{\citenamefont {Bousso}\ and\ \citenamefont
  {Engelhardt}(2015{\natexlab{a}})}]{Bousso:2015qqa}%
  \BibitemOpen
  \bibfield  {author} {\bibinfo {author} {\bibfnamefont {Raphael}\ \bibnamefont
  {Bousso}}\ and\ \bibinfo {author} {\bibfnamefont {Netta}\ \bibnamefont
  {Engelhardt}},\ }\bibfield  {title} {\enquote {\bibinfo {title} {{Proof of a
  New Area Law in General Relativity}},}\ }\href {\doibase
  10.1103/PhysRevD.92.044031} {\bibfield  {journal} {\bibinfo  {journal} {Phys.
  Rev.}\ }\textbf {\bibinfo {volume} {D92}},\ \bibinfo {pages} {044031}
  (\bibinfo {year} {2015}{\natexlab{a}})},\ \Eprint
  {http://arxiv.org/abs/1504.07660} {arXiv:1504.07660 [gr-qc]} \BibitemShut
  {NoStop}%
\bibitem [{\citenamefont {Bousso}\ and\ \citenamefont
  {Engelhardt}(2015{\natexlab{b}})}]{Bousso:2015mqa}%
  \BibitemOpen
  \bibfield  {author} {\bibinfo {author} {\bibfnamefont {Raphael}\ \bibnamefont
  {Bousso}}\ and\ \bibinfo {author} {\bibfnamefont {Netta}\ \bibnamefont
  {Engelhardt}},\ }\bibfield  {title} {\enquote {\bibinfo {title} {{New Area
  Law in General Relativity}},}\ }\href {\doibase
  10.1103/PhysRevLett.115.081301} {\bibfield  {journal} {\bibinfo  {journal}
  {Phys. Rev. Lett.}\ }\textbf {\bibinfo {volume} {115}},\ \bibinfo {pages}
  {081301} (\bibinfo {year} {2015}{\natexlab{b}})},\ \Eprint
  {http://arxiv.org/abs/1504.07627} {arXiv:1504.07627 [hep-th]} \BibitemShut
  {NoStop}%
\bibitem [{\citenamefont {Sanches}\ and\ \citenamefont
  {Weinberg}(2016)}]{Sanches:2016pga}%
  \BibitemOpen
  \bibfield  {author} {\bibinfo {author} {\bibfnamefont {Fabio}\ \bibnamefont
  {Sanches}}\ and\ \bibinfo {author} {\bibfnamefont {Sean~J.}\ \bibnamefont
  {Weinberg}},\ }\bibfield  {title} {\enquote {\bibinfo {title} {{Refinement of
  the Bousso-Engelhardt Area Law}},}\ }\href {\doibase
  10.1103/PhysRevD.94.021502} {\bibfield  {journal} {\bibinfo  {journal} {Phys.
  Rev. D}\ }\textbf {\bibinfo {volume} {94}},\ \bibinfo {pages} {021502}
  (\bibinfo {year} {2016})},\ \Eprint {http://arxiv.org/abs/1604.04919}
  {arXiv:1604.04919 [hep-th]} \BibitemShut {NoStop}%
\bibitem [{\citenamefont {Ashtekar}\ \emph
  {et~al.}(2000{\natexlab{a}})\citenamefont {Ashtekar}, \citenamefont
  {Fairhurst},\ and\ \citenamefont {Krishnan}}]{Ashtekar:2000hw}%
  \BibitemOpen
  \bibfield  {author} {\bibinfo {author} {\bibfnamefont {Abhay}\ \bibnamefont
  {Ashtekar}}, \bibinfo {author} {\bibfnamefont {Stephen}\ \bibnamefont
  {Fairhurst}}, \ and\ \bibinfo {author} {\bibfnamefont {Badri}\ \bibnamefont
  {Krishnan}},\ }\bibfield  {title} {\enquote {\bibinfo {title} {{Isolated
  horizons: Hamiltonian evolution and the first law}},}\ }\href {\doibase
  10.1103/PhysRevD.62.104025} {\bibfield  {journal} {\bibinfo  {journal} {Phys.
  Rev.}\ }\textbf {\bibinfo {volume} {D62}},\ \bibinfo {pages} {104025}
  (\bibinfo {year} {2000}{\natexlab{a}})},\ \Eprint
  {http://arxiv.org/abs/gr-qc/0005083} {arXiv:gr-qc/0005083} \BibitemShut
  {NoStop}%
\bibitem [{\citenamefont {Krishnan}(2012)}]{Krishnan:2012bt}%
  \BibitemOpen
  \bibfield  {author} {\bibinfo {author} {\bibfnamefont {Badri}\ \bibnamefont
  {Krishnan}},\ }\bibfield  {title} {\enquote {\bibinfo {title} {{The spacetime
  in the neighborhood of a general isolated black hole}},}\ }\href {\doibase
  10.1088/0264-9381/29/20/205006} {\bibfield  {journal} {\bibinfo  {journal}
  {Class.Quant.Grav.}\ }\textbf {\bibinfo {volume} {29}},\ \bibinfo {pages}
  {205006} (\bibinfo {year} {2012})},\ \Eprint {http://arxiv.org/abs/1204.4345}
  {arXiv:1204.4345 [gr-qc]} \BibitemShut {NoStop}%
\bibitem [{\citenamefont {Ashtekar}\ \emph {et~al.}(2001)\citenamefont
  {Ashtekar}, \citenamefont {Beetle},\ and\ \citenamefont
  {Lewandowski}}]{Ashtekar:2001is}%
  \BibitemOpen
  \bibfield  {author} {\bibinfo {author} {\bibfnamefont {Abhay}\ \bibnamefont
  {Ashtekar}}, \bibinfo {author} {\bibfnamefont {Christopher}\ \bibnamefont
  {Beetle}}, \ and\ \bibinfo {author} {\bibfnamefont {Jerzy}\ \bibnamefont
  {Lewandowski}},\ }\bibfield  {title} {\enquote {\bibinfo {title} {{Mechanics
  of Rotating Isolated Horizons}},}\ }\href {\doibase
  10.1103/PhysRevD.64.044016} {\bibfield  {journal} {\bibinfo  {journal} {Phys.
  Rev.}\ }\textbf {\bibinfo {volume} {D64}},\ \bibinfo {pages} {044016}
  (\bibinfo {year} {2001})},\ \Eprint {http://arxiv.org/abs/gr-qc/0103026}
  {arXiv:gr-qc/0103026} \BibitemShut {NoStop}%
\bibitem [{\citenamefont {Ashtekar}\ \emph {et~al.}(2002)\citenamefont
  {Ashtekar}, \citenamefont {Beetle},\ and\ \citenamefont
  {Lewandowski}}]{Ashtekar:2001jb}%
  \BibitemOpen
  \bibfield  {author} {\bibinfo {author} {\bibfnamefont {Abhay}\ \bibnamefont
  {Ashtekar}}, \bibinfo {author} {\bibfnamefont {Christopher}\ \bibnamefont
  {Beetle}}, \ and\ \bibinfo {author} {\bibfnamefont {Jerzy}\ \bibnamefont
  {Lewandowski}},\ }\bibfield  {title} {\enquote {\bibinfo {title} {{Geometry
  of Generic Isolated Horizons}},}\ }\href {\doibase
  10.1088/0264-9381/19/6/311} {\bibfield  {journal} {\bibinfo  {journal}
  {Class. Quant. Grav.}\ }\textbf {\bibinfo {volume} {19}},\ \bibinfo {pages}
  {1195--1225} (\bibinfo {year} {2002})},\ \Eprint
  {http://arxiv.org/abs/gr-qc/0111067} {arXiv:gr-qc/0111067} \BibitemShut
  {NoStop}%
\bibitem [{\citenamefont {Lewandowski}(2000)}]{Lewandowski:1999zs}%
  \BibitemOpen
  \bibfield  {author} {\bibinfo {author} {\bibfnamefont {Jerzy}\ \bibnamefont
  {Lewandowski}},\ }\bibfield  {title} {\enquote {\bibinfo {title} {{Spacetimes
  Admitting Isolated Horizons}},}\ }\href {\doibase 10.1088/0264-9381/17/4/101}
  {\bibfield  {journal} {\bibinfo  {journal} {Class. Quant. Grav.}\ }\textbf
  {\bibinfo {volume} {17}},\ \bibinfo {pages} {L53--L59} (\bibinfo {year}
  {2000})},\ \Eprint {http://arxiv.org/abs/gr-qc/9907058} {arXiv:gr-qc/9907058}
  \BibitemShut {NoStop}%
\bibitem [{\citenamefont {Gunasekaran}\ and\ \citenamefont
  {Booth}(2019)}]{Gunasekaran:2019jrq}%
  \BibitemOpen
  \bibfield  {author} {\bibinfo {author} {\bibfnamefont {Sharmila}\
  \bibnamefont {Gunasekaran}}\ and\ \bibinfo {author} {\bibfnamefont {Ivan}\
  \bibnamefont {Booth}},\ }\bibfield  {title} {\enquote {\bibinfo {title}
  {{Horizons as boundary conditions in spherical symmetry}},}\ }\href {\doibase
  10.1103/PhysRevD.100.064019} {\bibfield  {journal} {\bibinfo  {journal}
  {Phys. Rev. D}\ }\textbf {\bibinfo {volume} {100}},\ \bibinfo {pages}
  {064019} (\bibinfo {year} {2019})},\ \Eprint
  {http://arxiv.org/abs/1905.02748} {arXiv:1905.02748 [gr-qc]} \BibitemShut
  {NoStop}%
\bibitem [{\citenamefont {G{\"u}rlebeck}(2015)}]{Gurlebeck:2015xpa}%
  \BibitemOpen
  \bibfield  {author} {\bibinfo {author} {\bibfnamefont {Norman}\ \bibnamefont
  {G{\"u}rlebeck}},\ }\bibfield  {title} {\enquote {\bibinfo {title} {{No-hair
  theorem for Black Holes in Astrophysical Environments}},}\ }\href {\doibase
  10.1103/PhysRevLett.114.151102} {\bibfield  {journal} {\bibinfo  {journal}
  {Phys. Rev. Lett.}\ }\textbf {\bibinfo {volume} {114}},\ \bibinfo {pages}
  {151102} (\bibinfo {year} {2015})},\ \Eprint
  {http://arxiv.org/abs/1503.03240} {arXiv:1503.03240 [gr-qc]} \BibitemShut
  {NoStop}%
\bibitem [{\citenamefont {Flandera}(2016)}]{Flandera:2016qwg}%
  \BibitemOpen
  \bibfield  {author} {\bibinfo {author} {\bibfnamefont {Ale\v~s}\ \bibnamefont
  {Flandera}},\ }\bibfield  {title} {\enquote {\bibinfo {title} {{Geometry of
  isolated horizons}},}\ }\href@noop {} {\  (\bibinfo {year} {2016})},\ \Eprint
  {http://arxiv.org/abs/1611.02215} {arXiv:1611.02215 [gr-qc]} \BibitemShut
  {NoStop}%
\bibitem [{\citenamefont {Lewandowski}\ and\ \citenamefont
  {Pawlowski}(2014)}]{Lewandowski:2014nta}%
  \BibitemOpen
  \bibfield  {author} {\bibinfo {author} {\bibfnamefont {Jerzy}\ \bibnamefont
  {Lewandowski}}\ and\ \bibinfo {author} {\bibfnamefont {Tomasz}\ \bibnamefont
  {Pawlowski}},\ }\bibfield  {title} {\enquote {\bibinfo {title}
  {{Neighborhoods of isolated horizons and their stationarity}},}\ }\href
  {\doibase 10.1088/0264-9381/31/17/175012} {\bibfield  {journal} {\bibinfo
  {journal} {Class. Quant. Grav.}\ }\textbf {\bibinfo {volume} {31}},\ \bibinfo
  {pages} {175012} (\bibinfo {year} {2014})},\ \Eprint
  {http://arxiv.org/abs/1404.7836} {arXiv:1404.7836 [gr-qc]} \BibitemShut
  {NoStop}%
\bibitem [{\citenamefont {Lewandowski}\ and\ \citenamefont
  {Li}(2018)}]{Lewandowski:2018khe}%
  \BibitemOpen
  \bibfield  {author} {\bibinfo {author} {\bibfnamefont {Jerzy}\ \bibnamefont
  {Lewandowski}}\ and\ \bibinfo {author} {\bibfnamefont {Carmen}\ \bibnamefont
  {Li}},\ }\bibfield  {title} {\enquote {\bibinfo {title} {{Spacetime near Kerr
  isolated horizon}},}\ }\href@noop {} {\  (\bibinfo {year} {2018})},\ \Eprint
  {http://arxiv.org/abs/1809.04715} {arXiv:1809.04715 [gr-qc]} \BibitemShut
  {NoStop}%
\bibitem [{\citenamefont {Scholtz}\ \emph {et~al.}(2017)\citenamefont
  {Scholtz}, \citenamefont {Flandera},\ and\ \citenamefont
  {Guerlebeck}}]{Scholtz:2017ttf}%
  \BibitemOpen
  \bibfield  {author} {\bibinfo {author} {\bibfnamefont {Martin}\ \bibnamefont
  {Scholtz}}, \bibinfo {author} {\bibfnamefont {Ales}\ \bibnamefont
  {Flandera}}, \ and\ \bibinfo {author} {\bibfnamefont {Norman}\ \bibnamefont
  {Guerlebeck}},\ }\bibfield  {title} {\enquote {\bibinfo {title} {{Kerr-Newman
  black hole in the formalism of isolated horizons}},}\ }\href {\doibase
  10.1103/PhysRevD.96.064024} {\bibfield  {journal} {\bibinfo  {journal} {Phys.
  Rev.}\ }\textbf {\bibinfo {volume} {D96}},\ \bibinfo {pages} {064024}
  (\bibinfo {year} {2017})},\ \Eprint {http://arxiv.org/abs/1708.06383}
  {arXiv:1708.06383 [gr-qc]} \BibitemShut {NoStop}%
\bibitem [{\citenamefont {Geroch}\ and\ \citenamefont
  {Hartle}(1982)}]{Geroch:1982bv}%
  \BibitemOpen
  \bibfield  {author} {\bibinfo {author} {\bibfnamefont {Robert~P.}\
  \bibnamefont {Geroch}}\ and\ \bibinfo {author} {\bibfnamefont {J.~B.}\
  \bibnamefont {Hartle}},\ }\bibfield  {title} {\enquote {\bibinfo {title}
  {{Distorted black holes}},}\ }\href {\doibase 10.1063/1.525384} {\bibfield
  {journal} {\bibinfo  {journal} {J. Math. Phys.}\ }\textbf {\bibinfo {volume}
  {23}},\ \bibinfo {pages} {680} (\bibinfo {year} {1982})}\BibitemShut
  {NoStop}%
\bibitem [{\citenamefont {Fairhurst}\ and\ \citenamefont
  {Krishnan}(2001)}]{Fairhurst:2000xh}%
  \BibitemOpen
  \bibfield  {author} {\bibinfo {author} {\bibfnamefont {Stephen}\ \bibnamefont
  {Fairhurst}}\ and\ \bibinfo {author} {\bibfnamefont {Badri}\ \bibnamefont
  {Krishnan}},\ }\bibfield  {title} {\enquote {\bibinfo {title} {{Distorted
  black holes with charge}},}\ }\href {\doibase 10.1142/S0218271801001086}
  {\bibfield  {journal} {\bibinfo  {journal} {Int. J. Mod. Phys.}\ }\textbf
  {\bibinfo {volume} {D10}},\ \bibinfo {pages} {691--710} (\bibinfo {year}
  {2001})},\ \Eprint {http://arxiv.org/abs/gr-qc/0010088} {arXiv:gr-qc/0010088}
  \BibitemShut {NoStop}%
\bibitem [{\citenamefont {Pilkington}\ \emph {et~al.}(2011)\citenamefont
  {Pilkington}, \citenamefont {Melanson}, \citenamefont {Fitzgerald},\ and\
  \citenamefont {Booth}}]{Pilkington:2011aj}%
  \BibitemOpen
  \bibfield  {author} {\bibinfo {author} {\bibfnamefont {Terry}\ \bibnamefont
  {Pilkington}}, \bibinfo {author} {\bibfnamefont {Alexandre}\ \bibnamefont
  {Melanson}}, \bibinfo {author} {\bibfnamefont {Joseph}\ \bibnamefont
  {Fitzgerald}}, \ and\ \bibinfo {author} {\bibfnamefont {Ivan}\ \bibnamefont
  {Booth}},\ }\bibfield  {title} {\enquote {\bibinfo {title} {{Trapped and
  marginally trapped surfaces in Weyl-distorted Schwarzschild solutions}},}\
  }\href {\doibase 10.1088/0264-9381/28/12/125018} {\bibfield  {journal}
  {\bibinfo  {journal} {Class. Quant. Grav.}\ }\textbf {\bibinfo {volume}
  {28}},\ \bibinfo {pages} {125018} (\bibinfo {year} {2011})},\ \Eprint
  {http://arxiv.org/abs/1102.0999} {arXiv:1102.0999 [gr-qc]} \BibitemShut
  {NoStop}%
\bibitem [{\citenamefont {Ashtekar}\ and\ \citenamefont
  {Krishnan}(2002)}]{Ashtekar:2002ag}%
  \BibitemOpen
  \bibfield  {author} {\bibinfo {author} {\bibfnamefont {Abhay}\ \bibnamefont
  {Ashtekar}}\ and\ \bibinfo {author} {\bibfnamefont {Badri}\ \bibnamefont
  {Krishnan}},\ }\bibfield  {title} {\enquote {\bibinfo {title} {{Dynamical
  horizons: Energy, angular momentum, fluxes and balance laws}},}\ }\href
  {\doibase 10.1103/PhysRevLett.89.261101} {\bibfield  {journal} {\bibinfo
  {journal} {Phys. Rev. Lett.}\ }\textbf {\bibinfo {volume} {89}},\ \bibinfo
  {pages} {261101} (\bibinfo {year} {2002})},\ \Eprint
  {http://arxiv.org/abs/gr-qc/0207080} {arXiv:gr-qc/0207080} \BibitemShut
  {NoStop}%
\bibitem [{\citenamefont {Ashtekar}\ and\ \citenamefont
  {Krishnan}(2003)}]{Ashtekar:2003hk}%
  \BibitemOpen
  \bibfield  {author} {\bibinfo {author} {\bibfnamefont {Abhay}\ \bibnamefont
  {Ashtekar}}\ and\ \bibinfo {author} {\bibfnamefont {Badri}\ \bibnamefont
  {Krishnan}},\ }\bibfield  {title} {\enquote {\bibinfo {title} {{Dynamical
  horizons and their properties}},}\ }\href {\doibase
  10.1103/PhysRevD.68.104030} {\bibfield  {journal} {\bibinfo  {journal} {Phys.
  Rev.}\ }\textbf {\bibinfo {volume} {D68}},\ \bibinfo {pages} {104030}
  (\bibinfo {year} {2003})},\ \Eprint {http://arxiv.org/abs/gr-qc/0308033}
  {arXiv:gr-qc/0308033} \BibitemShut {NoStop}%
\bibitem [{\citenamefont {Hayward}(1994{\natexlab{a}})}]{Hayward:1993wb}%
  \BibitemOpen
  \bibfield  {author} {\bibinfo {author} {\bibfnamefont {S.A.}\ \bibnamefont
  {Hayward}},\ }\bibfield  {title} {\enquote {\bibinfo {title} {{General laws
  of black hole dynamics}},}\ }\href {\doibase 10.1103/PhysRevD.49.6467}
  {\bibfield  {journal} {\bibinfo  {journal} {Phys.Rev.}\ }\textbf {\bibinfo
  {volume} {D49}},\ \bibinfo {pages} {6467--6474} (\bibinfo {year}
  {1994}{\natexlab{a}})}\BibitemShut {NoStop}%
\bibitem [{\citenamefont {Hayward}(1994{\natexlab{b}})}]{Hayward:1994yy}%
  \BibitemOpen
  \bibfield  {author} {\bibinfo {author} {\bibfnamefont {Sean~A.}\ \bibnamefont
  {Hayward}},\ }\bibfield  {title} {\enquote {\bibinfo {title} {{Spin
  coefficient form of the new laws of black hole dynamics}},}\ }\href {\doibase
  10.1088/0264-9381/11/12/016} {\bibfield  {journal} {\bibinfo  {journal}
  {Class. Quant. Grav.}\ }\textbf {\bibinfo {volume} {11}},\ \bibinfo {pages}
  {3025--3036} (\bibinfo {year} {1994}{\natexlab{b}})},\ \Eprint
  {http://arxiv.org/abs/gr-qc/9406033} {arXiv:gr-qc/9406033} \BibitemShut
  {NoStop}%
\bibitem [{\citenamefont {Hayward}(2004)}]{Hayward:2004fz}%
  \BibitemOpen
  \bibfield  {author} {\bibinfo {author} {\bibfnamefont {Sean~A.}\ \bibnamefont
  {Hayward}},\ }\bibfield  {title} {\enquote {\bibinfo {title} {{Energy and
  entropy conservation for dynamical black holes}},}\ }\href {\doibase
  10.1103/PhysRevD.70.104027} {\bibfield  {journal} {\bibinfo  {journal} {Phys.
  Rev.}\ }\textbf {\bibinfo {volume} {D70}},\ \bibinfo {pages} {104027}
  (\bibinfo {year} {2004})},\ \Eprint {http://arxiv.org/abs/gr-qc/0408008}
  {arXiv:gr-qc/0408008} \BibitemShut {NoStop}%
\bibitem [{\citenamefont {Hayward}(2006)}]{Hayward:2006ss}%
  \BibitemOpen
  \bibfield  {author} {\bibinfo {author} {\bibfnamefont {Sean~A.}\ \bibnamefont
  {Hayward}},\ }\bibfield  {title} {\enquote {\bibinfo {title} {{Angular
  momentum conservation for dynamical black holes}},}\ }\href {\doibase
  10.1103/PhysRevD.74.104013} {\bibfield  {journal} {\bibinfo  {journal} {Phys.
  Rev.}\ }\textbf {\bibinfo {volume} {D74}},\ \bibinfo {pages} {104013}
  (\bibinfo {year} {2006})},\ \Eprint {http://arxiv.org/abs/gr-qc/0609008}
  {arXiv:gr-qc/0609008} \BibitemShut {NoStop}%
\bibitem [{\citenamefont {Booth}\ and\ \citenamefont
  {Fairhurst}(2004)}]{Booth:2003ji}%
  \BibitemOpen
  \bibfield  {author} {\bibinfo {author} {\bibfnamefont {Ivan}\ \bibnamefont
  {Booth}}\ and\ \bibinfo {author} {\bibfnamefont {Stephen}\ \bibnamefont
  {Fairhurst}},\ }\bibfield  {title} {\enquote {\bibinfo {title} {{The first
  law for slowly evolving horizons}},}\ }\href {\doibase
  10.1103/PhysRevLett.92.011102} {\bibfield  {journal} {\bibinfo  {journal}
  {Phys. Rev. Lett.}\ }\textbf {\bibinfo {volume} {92}},\ \bibinfo {pages}
  {011102} (\bibinfo {year} {2004})},\ \Eprint
  {http://arxiv.org/abs/gr-qc/0307087} {arXiv:gr-qc/0307087} \BibitemShut
  {NoStop}%
\bibitem [{\citenamefont {Prasad}\ \emph {et~al.}(2020)\citenamefont {Prasad},
  \citenamefont {Gupta}, \citenamefont {Bose}, \citenamefont {Krishnan},\ and\
  \citenamefont {Schnetter}}]{Prasad:2020xgr}%
  \BibitemOpen
  \bibfield  {author} {\bibinfo {author} {\bibfnamefont {Vaishak}\ \bibnamefont
  {Prasad}}, \bibinfo {author} {\bibfnamefont {Anshu}\ \bibnamefont {Gupta}},
  \bibinfo {author} {\bibfnamefont {Sukanta}\ \bibnamefont {Bose}}, \bibinfo
  {author} {\bibfnamefont {Badri}\ \bibnamefont {Krishnan}}, \ and\ \bibinfo
  {author} {\bibfnamefont {Erik}\ \bibnamefont {Schnetter}},\ }\bibfield
  {title} {\enquote {\bibinfo {title} {{News from horizons in binary black hole
  mergers}},}\ }\href@noop {} {\  (\bibinfo {year} {2020})},\ \Eprint
  {http://arxiv.org/abs/2003.06215} {arXiv:2003.06215 [gr-qc]} \BibitemShut
  {NoStop}%
\bibitem [{\citenamefont {Damour}(1979)}]{Damour:membrane}%
  \BibitemOpen
  \bibfield  {author} {\bibinfo {author} {\bibfnamefont {T}~\bibnamefont
  {Damour}},\ }\emph {\bibinfo {title} {Quelques propri\'et\'es m\'ecaniques,
  \'electromagn\'etiques, thermodynamiques et quantiques des trous noirs}},\
  \href@noop {} {Ph.D. thesis},\ \bibinfo  {school} {University of Paris}
  (\bibinfo {year} {1979})\BibitemShut {NoStop}%
\bibitem [{\citenamefont {Damour}(1982)}]{damour1982surface}%
  \BibitemOpen
  \bibfield  {author} {\bibinfo {author} {\bibfnamefont {Thibaut}\ \bibnamefont
  {Damour}},\ }\href@noop {} {\emph {\bibinfo {title} {Surface effects in black
  hole physics}}}\ (\bibinfo  {publisher} {na},\ \bibinfo {year}
  {1982})\BibitemShut {NoStop}%
\bibitem [{\citenamefont {Price}\ and\ \citenamefont
  {Thorne}(1986)}]{Price:1986yy}%
  \BibitemOpen
  \bibfield  {author} {\bibinfo {author} {\bibfnamefont {R.~H.}\ \bibnamefont
  {Price}}\ and\ \bibinfo {author} {\bibfnamefont {K.~S.}\ \bibnamefont
  {Thorne}},\ }\bibfield  {title} {\enquote {\bibinfo {title} {{Membrane
  Viewpoint on Black Holes: Properties and Evolution of the Stretched
  Horizon}},}\ }\href {\doibase 10.1103/PhysRevD.33.915} {\bibfield  {journal}
  {\bibinfo  {journal} {Phys. Rev.}\ }\textbf {\bibinfo {volume} {D33}},\
  \bibinfo {pages} {915--941} (\bibinfo {year} {1986})}\BibitemShut {NoStop}%
\bibitem [{\citenamefont {Thorne}\ \emph {et~al.}(1986)\citenamefont {Thorne},
  \citenamefont {Price},\ and\ \citenamefont {MacDonald}}]{Thorne:1986}%
  \BibitemOpen
  \bibfield  {author} {\bibinfo {author} {\bibfnamefont {K.~S.}\ \bibnamefont
  {Thorne}}, \bibinfo {author} {\bibfnamefont {R.~H.}\ \bibnamefont {Price}}, \
  and\ \bibinfo {author} {\bibfnamefont {D.~A.}\ \bibnamefont {MacDonald}},\
  }\href@noop {} {\emph {\bibinfo {title} {Black holes: The membrane
  paradigm}}}\ (\bibinfo  {publisher} {Yale University Press},\ \bibinfo {year}
  {1986})\BibitemShut {NoStop}%
\bibitem [{\citenamefont {Gourgoulhon}(2005)}]{Gourgoulhon:2005ch}%
  \BibitemOpen
  \bibfield  {author} {\bibinfo {author} {\bibfnamefont {Eric}\ \bibnamefont
  {Gourgoulhon}},\ }\bibfield  {title} {\enquote {\bibinfo {title} {{A
  generalized Damour-Navier-Stokes equation applied to trapping horizons}},}\
  }\href {\doibase 10.1103/PhysRevD.72.104007} {\bibfield  {journal} {\bibinfo
  {journal} {Phys. Rev.}\ }\textbf {\bibinfo {volume} {D72}},\ \bibinfo {pages}
  {104007} (\bibinfo {year} {2005})},\ \Eprint
  {http://arxiv.org/abs/gr-qc/0508003} {arXiv:gr-qc/0508003} \BibitemShut
  {NoStop}%
\bibitem [{\citenamefont {Gourgoulhon}\ and\ \citenamefont
  {Jaramillo}(2006{\natexlab{b}})}]{Gourgoulhon:2006uc}%
  \BibitemOpen
  \bibfield  {author} {\bibinfo {author} {\bibfnamefont {Eric}\ \bibnamefont
  {Gourgoulhon}}\ and\ \bibinfo {author} {\bibfnamefont {Jose~Luis}\
  \bibnamefont {Jaramillo}},\ }\bibfield  {title} {\enquote {\bibinfo {title}
  {{Area evolution, bulk viscosity and entropy principles for dynamical
  horizons}},}\ }\href {\doibase 10.1103/PhysRevD.74.087502} {\bibfield
  {journal} {\bibinfo  {journal} {Phys. Rev.}\ }\textbf {\bibinfo {volume}
  {D74}},\ \bibinfo {pages} {087502} (\bibinfo {year} {2006}{\natexlab{b}})},\
  \Eprint {http://arxiv.org/abs/gr-qc/0607050} {arXiv:gr-qc/0607050}
  \BibitemShut {NoStop}%
\bibitem [{\citenamefont {Gourgoulhon}\ and\ \citenamefont
  {Jaramillo}(2008)}]{Gourgoulhon:2008pu}%
  \BibitemOpen
  \bibfield  {author} {\bibinfo {author} {\bibfnamefont {Eric}\ \bibnamefont
  {Gourgoulhon}}\ and\ \bibinfo {author} {\bibfnamefont {Jose~Luis}\
  \bibnamefont {Jaramillo}},\ }\bibfield  {title} {\enquote {\bibinfo {title}
  {{New theoretical approaches to black holes}},}\ }\href {\doibase
  10.1016/j.newar.2008.03.026} {\bibfield  {journal} {\bibinfo  {journal} {New
  Astron. Rev.}\ }\textbf {\bibinfo {volume} {51}},\ \bibinfo {pages}
  {791--798} (\bibinfo {year} {2008})},\ \Eprint
  {http://arxiv.org/abs/0803.2944} {arXiv:0803.2944 [astro-ph]} \BibitemShut
  {NoStop}%
\bibitem [{\citenamefont {Ashtekar}\ \emph {et~al.}(1999)\citenamefont
  {Ashtekar}, \citenamefont {Beetle},\ and\ \citenamefont
  {Fairhurst}}]{Ashtekar:1998sp}%
  \BibitemOpen
  \bibfield  {author} {\bibinfo {author} {\bibfnamefont {Abhay}\ \bibnamefont
  {Ashtekar}}, \bibinfo {author} {\bibfnamefont {Christopher}\ \bibnamefont
  {Beetle}}, \ and\ \bibinfo {author} {\bibfnamefont {Stephen}\ \bibnamefont
  {Fairhurst}},\ }\bibfield  {title} {\enquote {\bibinfo {title} {{Isolated
  horizons: A generalization of black hole mechanics}},}\ }\href {\doibase
  10.1088/0264-9381/16/2/027} {\bibfield  {journal} {\bibinfo  {journal}
  {Class. Quant. Grav.}\ }\textbf {\bibinfo {volume} {16}},\ \bibinfo {pages}
  {L1--L7} (\bibinfo {year} {1999})},\ \Eprint
  {http://arxiv.org/abs/gr-qc/9812065} {arXiv:gr-qc/9812065} \BibitemShut
  {NoStop}%
\bibitem [{\citenamefont {Ashtekar}\ \emph
  {et~al.}(2000{\natexlab{b}})\citenamefont {Ashtekar}, \citenamefont
  {Beetle},\ and\ \citenamefont {Fairhurst}}]{Ashtekar:1999yj}%
  \BibitemOpen
  \bibfield  {author} {\bibinfo {author} {\bibfnamefont {Abhay}\ \bibnamefont
  {Ashtekar}}, \bibinfo {author} {\bibfnamefont {Christopher}\ \bibnamefont
  {Beetle}}, \ and\ \bibinfo {author} {\bibfnamefont {Stephen}\ \bibnamefont
  {Fairhurst}},\ }\bibfield  {title} {\enquote {\bibinfo {title} {{Mechanics of
  Isolated Horizons}},}\ }\href {\doibase 10.1088/0264-9381/17/2/301}
  {\bibfield  {journal} {\bibinfo  {journal} {Class. Quant. Grav.}\ }\textbf
  {\bibinfo {volume} {17}},\ \bibinfo {pages} {253--298} (\bibinfo {year}
  {2000}{\natexlab{b}})},\ \Eprint {http://arxiv.org/abs/gr-qc/9907068}
  {arXiv:gr-qc/9907068} \BibitemShut {NoStop}%
\bibitem [{\citenamefont {Ashtekar}\ and\ \citenamefont
  {Corichi}(2000)}]{Ashtekar:1999sn}%
  \BibitemOpen
  \bibfield  {author} {\bibinfo {author} {\bibfnamefont {Abhay}\ \bibnamefont
  {Ashtekar}}\ and\ \bibinfo {author} {\bibfnamefont {Alejandro}\ \bibnamefont
  {Corichi}},\ }\bibfield  {title} {\enquote {\bibinfo {title} {{Laws governing
  isolated horizons: Inclusion of dilaton couplings}},}\ }\href {\doibase
  10.1088/0264-9381/17/6/301} {\bibfield  {journal} {\bibinfo  {journal}
  {Class. Quant. Grav.}\ }\textbf {\bibinfo {volume} {17}},\ \bibinfo {pages}
  {1317--1332} (\bibinfo {year} {2000})},\ \Eprint
  {http://arxiv.org/abs/gr-qc/9910068} {arXiv:gr-qc/9910068} \BibitemShut
  {NoStop}%
\bibitem [{\citenamefont {Bekenstein}(1973)}]{Bekenstein:1973ur}%
  \BibitemOpen
  \bibfield  {author} {\bibinfo {author} {\bibfnamefont {Jacob~D.}\
  \bibnamefont {Bekenstein}},\ }\bibfield  {title} {\enquote {\bibinfo {title}
  {{Black holes and entropy}},}\ }\href {\doibase 10.1103/PhysRevD.7.2333}
  {\bibfield  {journal} {\bibinfo  {journal} {Phys. Rev.}\ }\textbf {\bibinfo
  {volume} {D7}},\ \bibinfo {pages} {2333--2346} (\bibinfo {year}
  {1973})}\BibitemShut {NoStop}%
\bibitem [{\citenamefont {Bardeen}\ \emph {et~al.}(1973)\citenamefont
  {Bardeen}, \citenamefont {Carter},\ and\ \citenamefont
  {Hawking}}]{Bardeen:1973gs}%
  \BibitemOpen
  \bibfield  {author} {\bibinfo {author} {\bibfnamefont {James~M.}\
  \bibnamefont {Bardeen}}, \bibinfo {author} {\bibfnamefont {B.}~\bibnamefont
  {Carter}}, \ and\ \bibinfo {author} {\bibfnamefont {S.~W.}\ \bibnamefont
  {Hawking}},\ }\bibfield  {title} {\enquote {\bibinfo {title} {{The Four laws
  of black hole mechanics}},}\ }\href {\doibase 10.1007/BF01645742} {\bibfield
  {journal} {\bibinfo  {journal} {Commun. Math. Phys.}\ }\textbf {\bibinfo
  {volume} {31}},\ \bibinfo {pages} {161--170} (\bibinfo {year}
  {1973})}\BibitemShut {NoStop}%
\bibitem [{\citenamefont {Bhattacharyya}\ \emph {et~al.}(2008)\citenamefont
  {Bhattacharyya}, \citenamefont {Hubeny}, \citenamefont {Minwalla},\ and\
  \citenamefont {Rangamani}}]{Bhattacharyya:2008jc}%
  \BibitemOpen
  \bibfield  {author} {\bibinfo {author} {\bibfnamefont {Sayantani}\
  \bibnamefont {Bhattacharyya}}, \bibinfo {author} {\bibfnamefont {Veronika~E}\
  \bibnamefont {Hubeny}}, \bibinfo {author} {\bibfnamefont {Shiraz}\
  \bibnamefont {Minwalla}}, \ and\ \bibinfo {author} {\bibfnamefont {Mukund}\
  \bibnamefont {Rangamani}},\ }\bibfield  {title} {\enquote {\bibinfo {title}
  {{Nonlinear Fluid Dynamics from Gravity}},}\ }\href {\doibase
  10.1088/1126-6708/2008/02/045} {\bibfield  {journal} {\bibinfo  {journal}
  {JHEP}\ }\textbf {\bibinfo {volume} {02}},\ \bibinfo {pages} {045} (\bibinfo
  {year} {2008})},\ \Eprint {http://arxiv.org/abs/0712.2456} {arXiv:0712.2456
  [hep-th]} \BibitemShut {NoStop}%
\bibitem [{\citenamefont {Booth}\ \emph {et~al.}(2011)\citenamefont {Booth},
  \citenamefont {Heller}, \citenamefont {Plewa},\ and\ \citenamefont
  {Spalinski}}]{Booth:2011qy}%
  \BibitemOpen
  \bibfield  {author} {\bibinfo {author} {\bibfnamefont {Ivan}\ \bibnamefont
  {Booth}}, \bibinfo {author} {\bibfnamefont {Michal~P.}\ \bibnamefont
  {Heller}}, \bibinfo {author} {\bibfnamefont {Grzegorz}\ \bibnamefont
  {Plewa}}, \ and\ \bibinfo {author} {\bibfnamefont {Michal}\ \bibnamefont
  {Spalinski}},\ }\bibfield  {title} {\enquote {\bibinfo {title} {{On the
  apparent horizon in fluid-gravity duality}},}\ }\href {\doibase
  10.1103/PhysRevD.83.106005} {\bibfield  {journal} {\bibinfo  {journal} {Phys.
  Rev. D}\ }\textbf {\bibinfo {volume} {83}},\ \bibinfo {pages} {106005}
  (\bibinfo {year} {2011})},\ \Eprint {http://arxiv.org/abs/1102.2885}
  {arXiv:1102.2885 [hep-th]} \BibitemShut {NoStop}%
\bibitem [{\citenamefont {Figueras}\ \emph {et~al.}(2009)\citenamefont
  {Figueras}, \citenamefont {Hubeny}, \citenamefont {Rangamani},\ and\
  \citenamefont {Ross}}]{Figueras:2009iu}%
  \BibitemOpen
  \bibfield  {author} {\bibinfo {author} {\bibfnamefont {Pau}\ \bibnamefont
  {Figueras}}, \bibinfo {author} {\bibfnamefont {Veronika~E.}\ \bibnamefont
  {Hubeny}}, \bibinfo {author} {\bibfnamefont {Mukund}\ \bibnamefont
  {Rangamani}}, \ and\ \bibinfo {author} {\bibfnamefont {Simon~F.}\
  \bibnamefont {Ross}},\ }\bibfield  {title} {\enquote {\bibinfo {title}
  {{Dynamical black holes and expanding plasmas}},}\ }\href {\doibase
  10.1088/1126-6708/2009/04/137} {\bibfield  {journal} {\bibinfo  {journal}
  {JHEP}\ }\textbf {\bibinfo {volume} {04}},\ \bibinfo {pages} {137} (\bibinfo
  {year} {2009})},\ \Eprint {http://arxiv.org/abs/0902.4696} {arXiv:0902.4696
  [hep-th]} \BibitemShut {NoStop}%
\bibitem [{\citenamefont {Cao}(2011)}]{Cao:2010vj}%
  \BibitemOpen
  \bibfield  {author} {\bibinfo {author} {\bibfnamefont {Li-Ming}\ \bibnamefont
  {Cao}},\ }\bibfield  {title} {\enquote {\bibinfo {title} {{Deformation of
  Codimension-2 Surface and Horizon Thermodynamics}},}\ }\href {\doibase
  10.1007/JHEP03(2011)112} {\bibfield  {journal} {\bibinfo  {journal} {JHEP}\
  }\textbf {\bibinfo {volume} {03}},\ \bibinfo {pages} {112} (\bibinfo {year}
  {2011})},\ \Eprint {http://arxiv.org/abs/1009.4540} {arXiv:1009.4540 [gr-qc]}
  \BibitemShut {NoStop}%
\bibitem [{\citenamefont {Jaramillo}\ \emph {et~al.}(2012)\citenamefont
  {Jaramillo}, \citenamefont {Macedo}, \citenamefont {M{\"o}sta},\ and\
  \citenamefont {Rezzolla}}]{Jaramillo:2011rf}%
  \BibitemOpen
  \bibfield  {author} {\bibinfo {author} {\bibfnamefont {Jose~Luis}\
  \bibnamefont {Jaramillo}}, \bibinfo {author} {\bibfnamefont {Rodrigo~P.}\
  \bibnamefont {Macedo}}, \bibinfo {author} {\bibfnamefont {Philipp}\
  \bibnamefont {M{\"o}sta}}, \ and\ \bibinfo {author} {\bibfnamefont {Luciano}\
  \bibnamefont {Rezzolla}},\ }\bibfield  {title} {\enquote {\bibinfo {title}
  {{Black-hole horizons as probes of black-hole dynamics II: geometrical
  insights}},}\ }\href {\doibase 10.1103/PhysRevD.85.084031} {\bibfield
  {journal} {\bibinfo  {journal} {Phys. Rev.}\ }\textbf {\bibinfo {volume}
  {D85}},\ \bibinfo {pages} {084031} (\bibinfo {year} {2012})},\ \Eprint
  {http://arxiv.org/abs/1108.0061} {arXiv:1108.0061 [gr-qc]} \BibitemShut
  {NoStop}%
\bibitem [{\citenamefont {Jaramillo}\ \emph {et~al.}(2011)\citenamefont
  {Jaramillo}, \citenamefont {Macedo}, \citenamefont {M{\"o}sta},\ and\
  \citenamefont {Rezzolla}}]{Jaramillo:2012rr}%
  \BibitemOpen
  \bibfield  {author} {\bibinfo {author} {\bibfnamefont {J.~L.}\ \bibnamefont
  {Jaramillo}}, \bibinfo {author} {\bibfnamefont {R.~P.}\ \bibnamefont
  {Macedo}}, \bibinfo {author} {\bibfnamefont {P.}~\bibnamefont {M{\"o}sta}}, \
  and\ \bibinfo {author} {\bibfnamefont {L.}~\bibnamefont {Rezzolla}},\
  }\bibfield  {title} {\enquote {\bibinfo {title} {{Towards a cross-correlation
  approach to strong-field dynamics in Black Hole spacetimes}},}\ }\bibfield
  {booktitle} {\emph {\bibinfo {booktitle} {{Proceedings, Spanish Relativity
  Meeting : Towards new paradigms. (ERE 2011): Madrid, Spain, August
  29-September 2, 2011}}},\ }\href {\doibase 10.1063/1.4734411} {\bibfield
  {journal} {\bibinfo  {journal} {AIP Conf. Proc.}\ }\textbf {\bibinfo {volume}
  {1458}},\ \bibinfo {pages} {158--173} (\bibinfo {year} {2011})},\ \Eprint
  {http://arxiv.org/abs/1205.3902} {arXiv:1205.3902 [gr-qc]} \BibitemShut
  {NoStop}%
\bibitem [{\citenamefont {Price}\ \emph {et~al.}(2011)\citenamefont {Price},
  \citenamefont {Khanna},\ and\ \citenamefont {Hughes}}]{Price:2011fm}%
  \BibitemOpen
  \bibfield  {author} {\bibinfo {author} {\bibfnamefont {Richard~H.}\
  \bibnamefont {Price}}, \bibinfo {author} {\bibfnamefont {Gaurav}\
  \bibnamefont {Khanna}}, \ and\ \bibinfo {author} {\bibfnamefont {Scott~A.}\
  \bibnamefont {Hughes}},\ }\bibfield  {title} {\enquote {\bibinfo {title}
  {{Systematics of black hole binary inspiral kicks and the slowness
  approximation}},}\ }\href {\doibase 10.1103/PhysRevD.83.124002} {\bibfield
  {journal} {\bibinfo  {journal} {Phys.Rev.}\ }\textbf {\bibinfo {volume}
  {D83}},\ \bibinfo {pages} {124002} (\bibinfo {year} {2011})},\ \Eprint
  {http://arxiv.org/abs/1104.0387} {arXiv:1104.0387 [gr-qc]} \BibitemShut
  {NoStop}%
\bibitem [{\citenamefont {Galloway}(2000)}]{Galloway:1999ny}%
  \BibitemOpen
  \bibfield  {author} {\bibinfo {author} {\bibfnamefont {Gregory~J.}\
  \bibnamefont {Galloway}},\ }\bibfield  {title} {\enquote {\bibinfo {title}
  {{Maximum principles for null hypersurfaces and null splitting theorems}},}\
  }\href {\doibase 10.1007/s000230050006} {\bibfield  {journal} {\bibinfo
  {journal} {Annales Henri Poincare}\ }\textbf {\bibinfo {volume} {1}},\
  \bibinfo {pages} {543--567} (\bibinfo {year} {2000})},\ \Eprint
  {http://arxiv.org/abs/math/9909158} {arXiv:math/9909158} \BibitemShut
  {NoStop}%
\bibitem [{\citenamefont {Wall}(2013)}]{Wall:2013uza}%
  \BibitemOpen
  \bibfield  {author} {\bibinfo {author} {\bibfnamefont {Aron~C.}\ \bibnamefont
  {Wall}},\ }\bibfield  {title} {\enquote {\bibinfo {title} {{The Generalized
  Second Law implies a Quantum Singularity Theorem}},}\ }\href {\doibase
  10.1088/0264-9381/30/19/199501} {\bibfield  {journal} {\bibinfo  {journal}
  {Class. Quant. Grav.}\ }\textbf {\bibinfo {volume} {30}},\ \bibinfo {pages}
  {165003} (\bibinfo {year} {2013})},\ \bibinfo {note} {[Erratum:
  Class.Quant.Grav. 30, 199501 (2013)]},\ \Eprint
  {http://arxiv.org/abs/1010.5513} {arXiv:1010.5513 [gr-qc]} \BibitemShut
  {NoStop}%
\bibitem [{\citenamefont {Padmanabhan}(2011)}]{PhysRevD.83.044048}%
  \BibitemOpen
  \bibfield  {author} {\bibinfo {author} {\bibfnamefont {T.}~\bibnamefont
  {Padmanabhan}},\ }\bibfield  {title} {\enquote {\bibinfo {title} {Entropy
  density of spacetime and the navier-stokes fluid dynamics of null
  surfaces},}\ }\href {\doibase 10.1103/PhysRevD.83.044048} {\bibfield
  {journal} {\bibinfo  {journal} {Phys. Rev. D}\ }\textbf {\bibinfo {volume}
  {83}},\ \bibinfo {pages} {044048} (\bibinfo {year} {2011})}\BibitemShut
  {NoStop}%
\bibitem [{\citenamefont {Uribe-Vargas}(2004)}]{Uribe04}%
  \BibitemOpen
  \bibfield  {author} {\bibinfo {author} {\bibfnamefont {Ricardo}\ \bibnamefont
  {Uribe-Vargas}},\ }\bibfield  {title} {\enquote {\bibinfo {title} {On
  singularities, “perestroikas” and differential geometry of space
  curves},}\ }\href {\doibase 10.5169/seals-2641} {\bibfield  {journal}
  {\bibinfo  {journal} {L'Enseignement Math\'ematique. IIe S\'erie}\ }\textbf
  {\bibinfo {volume} {50}} (\bibinfo {year} {2004}),\
  10.5169/seals-2641}\BibitemShut {NoStop}%
\bibitem [{\citenamefont {Arnol{\'d}}\ and\ \citenamefont
  {Union}(1989)}]{Arnold89}%
  \BibitemOpen
  \bibfield  {author} {\bibinfo {author} {\bibfnamefont {V.I.}\ \bibnamefont
  {Arnol{\'d}}}\ and\ \bibinfo {author} {\bibfnamefont
  {International~Mathematical}\ \bibnamefont {Union}},\ }\href
  {https://books.google.es/books?id=RiYZAQAAIAAJ} {\emph {\bibinfo {title}
  {Contact Geometry and Wave Propagation: Lectures Given at the University of
  Oxford Under the Sponsorship of the International Mathematical Union}}},\
  Monographie de l{\'E}nseignement math{\'e}matique\ (\bibinfo  {publisher}
  {L'Enseignement math{\'e}matique, Universit{\'e} de Gen{\`e}ve},\ \bibinfo
  {year} {1989})\BibitemShut {NoStop}%
\bibitem [{\citenamefont {Friedrich}\ and\ \citenamefont
  {Stewart}(1983)}]{Friedrich:1983vi}%
  \BibitemOpen
  \bibfield  {author} {\bibinfo {author} {\bibfnamefont {H.}~\bibnamefont
  {Friedrich}}\ and\ \bibinfo {author} {\bibfnamefont {J.M.}\ \bibnamefont
  {Stewart}},\ }\bibfield  {title} {\enquote {\bibinfo {title} {{Characteristic
  initial data and wave front singularities in general relativity}},}\ }\href
  {\doibase 10.1098/rspa.1983.0018} {\bibfield  {journal} {\bibinfo  {journal}
  {Proc. Roy. Soc. Lond. A}\ }\textbf {\bibinfo {volume} {A385}},\ \bibinfo
  {pages} {345--371} (\bibinfo {year} {1983})}\BibitemShut {NoStop}%
\bibitem [{\citenamefont {Ehlers}\ and\ \citenamefont
  {Newman}(2000)}]{EhlNew00}%
  \BibitemOpen
  \bibfield  {author} {\bibinfo {author} {\bibfnamefont {J\"{u}rgen}\
  \bibnamefont {Ehlers}}\ and\ \bibinfo {author} {\bibfnamefont {Ezra}\
  \bibnamefont {Newman}},\ }\bibfield  {title} {\enquote {\bibinfo {title} {The
  theory of caustics and wave front singularities with physical
  applications},}\ }\href {\doibase 10.1063/1.533316} {\bibfield  {journal}
  {\bibinfo  {journal} {Journal of Mathematical Physics, v.41, 3344-3378
  (2000)}\ }\textbf {\bibinfo {volume} {41}} (\bibinfo {year} {2000}),\
  10.1063/1.533316}\BibitemShut {NoStop}%
\bibitem [{\citenamefont {Arnold}(2013)}]{Arnold13}%
  \BibitemOpen
  \bibfield  {author} {\bibinfo {author} {\bibfnamefont {V.}~\bibnamefont
  {Arnold}},\ }\href {https://books.google.es/books?id=I-9rCQAAQBAJ} {\emph
  {\bibinfo {title} {Singularities of Caustics and Wave Fronts}}},\ Mathematics
  and its Applications\ (\bibinfo  {publisher} {Springer Netherlands},\
  \bibinfo {year} {2013})\BibitemShut {NoStop}%
\end{thebibliography}%

\end{document}